\documentclass{JHEP3}

\usepackage{graphicx}
\usepackage{epsfig}
\usepackage{amsmath}
\usepackage{amssymb}
\usepackage{latexsym}

\newcommand{\smartpap}{p\hskip-7pt\hbox{$^{^{(\!-\!)}}$}}
\newcommand{\smallz}{{\scriptscriptstyle Z}} %
\newcommand{\mz}{m_\smallz}
\newcommand{\smallw}{{\scriptscriptstyle W}}
\newcommand{\mw}{m_\smallw} 
\newcommand{\gw}{\Gamma_{\smallw}}

\newcommand{\sdw}{\sin^2\theta_{\smallw}}

\newcommand{\smallh}{{\scriptscriptstyle H}}
\newcommand{\mh}{m_\smallh}

\title{Combination of electroweak and QCD corrections to 
single $W$ production at the Fermilab Tevatron and the CERN LHC}

\author{Giovanni Balossini and Guido Montagna\\
Dipartimento di Fisica
Nucleare e Teorica, Universit\`a di Pavia,
and INFN, Sezione di Pavia, via A. Bassi 6, 27100 Pavia, Italy\\
Email: \email{giovanni.balossini@pv.infn.it}\\
Email:  \email{guido.montagna@pv.infn.it}
}

\author{ Carlo Michel Carloni Calame\\
Istituto Nazionale di Fisica Nucleare, Via E.  Fermi 40, Frascati, Italy 
\\and School of Physics \& Astronomy, University of Southampton, 
Highfield, Southampton SO17 1BJ, UK\\
Email: \email{c.carloni-calame@phys.soton.ac.uk}
}

\author{Mauro Moretti\\
Dipartimento di Fisica, Universit\`a di 
Ferrara, and INFN, Sezione di Ferrara, 
via Saragat 1, 44100 Ferrara, Italy\\
Email: \email{mauro.moretti@fe.infn.it}
}

\author{Oreste Nicrosini and Fulvio Piccinini\\
INFN, Sezione di Pavia, 
via A. Bassi 6, 27100 Pavia, Italy\\
Email: \email{oreste.nicrosini@pv.infn.it}\\
Email: \email{fulvio.piccinini@pv.infn.it}
}

\author{Michele Treccani\\
Departamento de F\'{i}sica Te\'orica y del Cosmos, CAPFE,
       Universidad de Granada, Avenida de Fuentenueva, 
       E-18071 Granada, Spain \\
Email: \email{treccani@ugr.es}
}

\author{Alessandro Vicini\\
Dipartimento di Fisica, Universit\`a di Milano, and INFN, 
Sezione di Milano, Via Celoria 16, 20133 Milano, Italy\\
Email:  \email{alessandro.vicini@mi.infn.it}
}

\abstract{Precision studies of the production of a high-transverse momentum lepton  
in association with missing energy 
at hadron colliders require that electroweak and QCD higher-order contributions are simultaneously taken into 
account in theoretical predictions and data analysis. Here we present  
a detailed phenomenological study of the impact of electroweak and strong contributions, 
as well as of their combination, to all the observables 
relevant for the various facets of the 
$p\smartpap \to {\rm lepton} + X$ 
physics programme 
at hadron colliders, including luminosity 
monitoring and Parton Distribution Functions constraint, $W$ precision physics and search for new physics signals. 
We provide a theoretical recipe to carefully combine electroweak and strong corrections, 
that are mandatory in view of the challenging 
experimental accuracy already reached at the Fermilab Tevatron and aimed at the CERN LHC, and discuss the uncertainty inherent the combination. 
We conclude that the theoretical accuracy of our calculation can be conservatively estimated to be about $2\%$ for standard event selections at the Tevatron and the LHC, and about $5\%$ in the very high $W$ transverse mass/lepton transverse momentum tails. We also provide arguments for  a  more aggressive error estimate (about $1 \%$ and $3\%$, respectively) and  conclude that in order to attain a one per cent accuracy: 1) exact mixed 
${\cal O}(\alpha \alpha_s)$ corrections should be computed in addition to the already available NNLO QCD contributions and two-loop electroweak Sudakov logarithms; 2) QCD and electroweak corrections should be coherently included into a single event generator. }


\preprint{FNT/T 2009/02\\
SHEP-09-14\\
IFUM-942-FT}

\begin{document}

\section{Introduction}
\label{intro}

The electroweak gauge bosons $W^{\pm}$ and $Z^0$ were discovered at the 
SpS collider at CERN more than twenty years ago \cite{denegri}.  
Albeit a long time has passed since then, the production of electroweak gauge
bosons in hadronic collisions is still a topic of deep interest in modern particle physics. 
Actually, single $W$ and $Z$ boson production is used at the Fermilab Tevatron
collider to derive precision measurements of the $W$ boson mass and width \cite{tevewwg}, to 
extract the electroweak (EW) mixing  angle from the forward-backward asymmetry in the 
neutral current (NC) channel \cite{cdf-afb} and to severely constrain the Parton Distribution Functions (PDF) through 
the measurement of the $W$ charge asymmetry \cite{wch-exp}. These processes are also used 
for detector calibration and the measurement of their total production cross section
can be compared with the corresponding QCD prediction, in order to test the convergence
of the strong coupling expansion in perturbative QCD calculations \cite{baur,doreen}.

In the near future, at the Large Hadron Collider (LHC) at CERN, the production of $W$ and
$Z$ bosons will continue to be a relevant process, because of its large cross section and 
very clean signature \cite{lhc}, given by one isolated charged lepton with missing 
transverse energy (for $W$ production) and by two isolated charged leptons with
opposite charges (for $Z$ production). In particular, thanks to the very large statistics, a measurement of the $W$ mass with an uncertainty of about $10$~MeV should be feasible at the LHC~\cite{gk,cmsnote}. 
These processes are also
 good candidates to understand the detectors performances in the early stage of data analysis at the
 LHC \cite{fmexp}, to monitor the collider luminosity with per cent precision \cite{lumi0,lumi1,lumi2} 
 and constitute a background to new physics searches, noticeably new heavy gauge 
 bosons $Z^{\prime}$ and $W^{\prime}$ \cite{fmexp,ppr}, whose discovery is an important goal of the 
 LHC \cite{baur,doreen}.

It is important to realize that the many facets of the $W/Z$ physics programme at hadron 
colliders require, for obvious reasons, the measurement of different observables,  depending on the physics goal of interest. Correspondingly, a number of observables must be precisely
predicted and simulated, to avoid theoretical bias in data analysis. For example, for precise 
$W$ mass measurements the relevant observables are the $W$ transverse mass and the 
lepton transverse momentum, while luminosity studies require a deep understanding of the
total cross section, the $W/Z$ rapidity and lepton pseudorapidity. On the other hand, the lepton 
pair invariant mass produced in $Z$ production and the $W$ transverse mass in their high tails
are the observables to be focused on for the search for new physics signals.

To fully exploit the potential of the Tevatron and  the LHC for all the above physics goals, the 
theoretical predictions have to be of the highest standard as possible.  
In particular, the high luminosity at the LHC implies that systematic errors, including 
theoretical ones, will play a dominant role in determining the accuracy of the
cross sections. This requires to 
make available calculations of $W$ and $Z$ production cross sections including higher-order 
corrections originating from the EW and QCD sector of the Standard Model (SM). 
Furthermore, the implementation of such calculations in Monte Carlo (MC) generators 
is mandatory, in order to perform realistic studies of the impact of higher-order corrections
on the observables and to compare theory with data.

In this paper we study the impact of electroweak and strong 
corrections, as well as of their combination, to 
the production process $p\smartpap \to W \to {\rm lepton} + X$. 
We also include in our analysis photon-induced contributions to $W \to {\rm lepton} + X$
production and 
we provide results for 
$top$-pair production process,  which may give rise to the 
signature of a high-transverse momentum lepton  in association with missing energy,  in order 
to include a non-negligible contribution for a precise evaluation of the
overall background to new-physics searches. A similar study, which some of us have 
coauthored, has been recently performed for the NC Drell-Yan (DY) 
channel in the high invariant mass tail  in \cite{LH2007}, while  
an evaluation of the theoretical uncertainties in the cross sections 
of the charged current (CC) and NC 
 processes  around  the $W$/$Z$ peaks  
has been addressed in~\cite{Adam:2008pc, Adam:2008ge} by strictly following the procedure presented by our group in some conference proceedings~\cite{noiproc}. 
However, it is important to notice that at a difference 
with respect to (w.r.t.) the works~\cite{Adam:2008pc, Adam:2008ge}, where the effects of EW and QCD corrections are separately studied, in the present paper the combination of such effects is analyzed in detail, by providing different  theoretical recipes and discussing the related uncertainties.  Furthermore, we provide also predictions for the transverse mass region above 1~TeV, which is an important range for new physics searches. 

As far as complete ${\cal O}(\alpha)$ SM EW corrections to $p\smartpap  \to W \to l \nu_l$
are concerned, they have  been computed independently by various authors in~\cite{HW,bkw,dk, bw,ZYK,SANC,CMNV}. 
Also one-loop corrections in the Minimal Supersymmetric Standard Model have been calculated in~\cite{DK-new}. 
The predictions of a subset of such calculations have been 
compared, with the same input parameters and cuts, in~\cite{LesHouches,tev4lhc}, 
finding a very satisfactory agreement. 
The contributions due to the emission of multiple photon radiation for
the observables of interest for $W$ precision physics have been calculated 
in~\cite{CMNTW,CMNTZ,WINHAC,GW,HR,DK-new}. Comparisons of such 
multi-photon calculations can be found in~\cite{GW,HR,app}, showing good agreement.

Concerning QCD calculations and tools, the present situation
reveals  quite a rich structure, that includes NLO/NNLO corrections to $W$ total production rate \cite{AEM,HvNM},
fully differential cross sections known at NLO/NNLO 
QCD \cite{mcfm,ADMP,ADMP1,mp,mp1,as2}, resummation of leading/next-to-leading  
logarithms arising from soft gluon radiation \cite{BY,resbos}, 
NLO corrections merged with parton shower evolution 
\cite{MC@NLO,pnason,Alioli:2008gx}, as well as 
leading-order matrix elements generators matched with vetoed parton shower \cite{Alpgen,sherpa, madevent,phegas}.

A review of the theory of $W$ and $Z$ boson production can be found in \cite{pn,brev}, 
while recent developments about QCD and generators are described in 
\cite{ps,dko}.

On the other hand, it must be noticed that the combination of EW and QCD 
corrections is still at a very preliminary stage,  
only a few attempts being known in the literature~\cite{cy,jadach}.  
In particular, in \cite {cy} the effects of QCD resummation are combined with the leading 
part  (in the statistically dominant region of the process around the $W$ mass peak) 
of EW corrections, given by NLO 
final-state QED radiation and improved Born approximation to account for the pure weak corrections. Therefore, in the latter analysis the effect of purely (Sudakov-like) EW  
logarithms is
missing, when it is known that  the full set of EW corrections is required in addition 
to photonic corrections to reach a 
per cent theoretical precision \cite{dk,bw,CMNV}. Furthermore, 
in~\cite{ps} the treatment of QCD effects is limited to the inclusion of multiple soft-gluon corrections, leaving
room to more detailed studies of the QCD sources of contributions. The present situation 
of EW calculations and the recent development of more  sophisticated QCD generators makes feasible a more complete analysis,
both for what concerns QCD and EW effects.
 
Therefore, the aims of the present paper are the following: 
 \begin{enumerate}
 
 \item to investigate the various sources of QCD effects (PDF uncertainties, 
comparison of presently available perturbative QCD calculations, allowing for 
 variations of the factorization/renormalization scale), 
 extending the analyses of~\cite{mp,cy,fm} to consider the full set of observables
 of experimental interest and to make 
 use of computational tools which include 
 recent advances 
 in QCD phenomenology;
 
 \item to combine the results of  the state-of-the-art of EW calculations  
 with the QCD calculations and generators,  that are typically used at the Tevatron and the LHC, in order to investigate the role of EW effects in association with QCD contributions; 
 
 \item to assess the reliability of existing tools and demonstrate the need of a deep understanding of QCD and of EW-QCD combination for precision
physics studies at hadron colliders.

 \end{enumerate}

 The outline of the paper is as follows. In Section \ref{ta} we review the main features of the 
 theoretical tools used in our study, describing the 
 treatment of PDF uncertainties (Section \ref{pdf}), the implementation of QCD effects according 
 to different degrees of approximation in existing generators  (Section \ref{qcdgenerators}), the treatment 
 of NLO EW corrections and photon-induced processes  (Section \ref{horace}), as well as the combination of EW 
 effects with strong corrections (Section \ref{combining}), which is the main theoretical 
 feature of the paper. In Section \ref{numerics} we present and 
 discuss the numerical results for the observables that are grouped on the basis of their 
 specific experimental motivation, both at the energies of the Fermilab Tevatron 
proton-antiproton ($p\bar{p}$) collider and of the CERN LHC 
proton-proton ($pp$) collider. Conclusions and possible developments
 of our work are drawn in Section \ref{concl}.

\section{Theoretical ingredients}
\label{ta}

\subsection{Parton Distribution Functions (PDF)}
\label{pdf}

It is widely documented  \cite{hera4lhc} that a precise knowledge of the partonic structure of the proton will be an essential ingredient for the physics potentials of the LHC, both for what concerns discovery and precision physics. At present, there is a great deal of interest in understanding 
how 
the uncertainties on the determination of the  Parton Distribution Functions (PDF), 
both of experimental and 
theoretical origin, translate into uncertainties for those observables that will be measured at the
LHC.  A short review of recent progress in Parton Distributions and implications for LHC physics 
can be found in~\cite{pdfrev}. 
A parton distribution library, known as LHAPDF \cite{lhapdf}, is available 
and gives the possibility of comparing different PDF parameterizations. 

In particular, some sets 
allow to 
estimate 
how the errors of the experimental measurements affect the  PDF parameterization  within a certain practical tolerance. 
The
latter is defined as the maximum allowed of the $\Delta \chi^2$ variation w.r.t. the parameters
of the best PDF fit. In our analysis, among the PDFs sets available in
LHAPDF, we will make use of MRST2001E \cite{mrst2001e} and CTEQ6 \cite{cteq6}, as done in previous similar studies 
\cite{fm,tetal}.  For these two different sets of NLO PDFs the tolerance $T$ is assumed 
to be $T = \sqrt{50}$ 
(MRST2001E) and $T = 10$ (CTEQ6), resulting in PDFs uncertainties larger for a factor of 
$\sqrt{2}$ for CTEQ6. It is worth emphasizing that the uncertainties obtained according to such
a procedure are purely experimental (i.e. as due to the systematic and statistical errors of the
data used in the global fit), leaving aside other sources of uncertainty of theoretical 
origin, such as, for example, uncertainties due to the truncation of DGLAP perturbation
expansion, choice of the parameterization of non-perturbative input PDF 
or $\ln (1/x)$ 
 effects. However, they can be estimated separately and it is known from previous studies \cite{mrst2003,cteq05} that the theory-induced uncertainty is similar in magnitude to the uncertainty
 due to experimental errors at the Tevatron, but larger at the LHC.
 However, in analogy to
 previous phenomenological studies \cite{fm,tetal}, we limit to consider PDFs uncertainties
 deriving from errors on the data, just to get an idea of the order of magnitude of such effects
 in comparison with the size of perturbative corrections that are the main subject of our analysis.

Finally, we point out that,  since we include in our predictions EW corrections and, therefore,
we need to treat
collinear singularities 
due to photon radiation off the initial state quarks for the sake of
consistency at NLO EW, it is necessary a set of PDF 
 including QED effects in DGLAP evolution. The MRST group performed a 
global PDFs analysis including QED contributions, which have been incorporated
in the set of PDFs known as MRST2004QED \cite{MRST2004QED}. It turns out that the QED evolution slightly modifies the standard QCD-evolved 
PDFs for typical $x$ and $Q^2$ values probed by weak boson production at the Tevatron and
the LHC, as already argued in previous studies \cite{pdf-qed,pdf-qed1,pdf-qed2}. However, it must be stressed that the inclusion of QED corrections into PDFs 
dynamically generates a 
photon density  inside the  (anti)proton. 
This
implies that EW processes initiated by a photon (photon-induced processes) 
contribute to EW gauge boson production in hadronic collisions, as discussed 
in Section \ref{horace}. As largely discussed in the literature, the present PDF uncertainties are expected to be significantly reduced when LHC data will become available. 

The parameterizations MSTW~\cite{MSTW} and CTEQ6.6~\cite{CTEQ66}, appeared in the literature during the completion of the present work, will be discussed shortly in
Section \ref{pdfnum}.

\subsection{QCD corrections and Monte Carlo tools}
\label{qcdgenerators}

The QCD codes typically used in high-energy collider experiments are ``general purpose'' parton shower 
MC's, such as, for example, HERWIG~\cite{herwig} or PYTHIA~\cite{pythia}, 
Apacic++~\cite{apacic++} 
or MC 
programs based on a fixed-order perturbative calculation of a given process. However, 
the new challenges of the Tevatron Run II and the LHC have spurred on improved 
approaches to QCD MC's and calculations, including the matching of exact leading-order 
matrix elements for multiparticle production with parton shower, the matching of 
NLO corrections with parton shower, advances in the techniques for NNLO calculations, as
well as improvements in the parton shower algorithms \cite{ps,dko, herwig++, sjostrand2}. 

As already remarked in the Introduction, we make use in our analysis, 
for the sake of definiteness, of those QCD 
calculations and relative codes that are already used at the Tevatron or will be presumably
used at the LHC for data analysis, and, at least in our opinion, are representative of 
those advances sketched above. We therefore consider the following set of programs 

\begin{enumerate}

\item ALPGEN \cite{Alpgen}: it is a code for the generation of many multiparton 
processes in hadronic collisions, computing exact matrix elements by means of
the ALPHA algorithm \cite{ALPHA}. The leading-order matrix elements are matched 
with appropriate Sudakov form factors to add parton shower effects 
(according to
the so-called MLM prescription to match matrix 
elements and parton shower \cite{ckkw,lo,mlm,mr,mmpt}) and by vetoing shower evolution leading to multiparton final states already described by the matrix element computation.  A review of the matching algorithm implemented in ALPGEN and here used can be 
found in \cite{mmpt}, while a comparison between the predictions of the 
different procedures of matrix elements-parton shower matching is given in \cite{hkl, Alwall:2007fs}.
Additional information is available at 
{\tt http://mlm.home.cern.ch/mlm/alpgen/}

\item MCFM \cite{mcfm}: it is a program for the MC simulations of various processes at hadron colliders, 
giving predictions, for most processes (including $W/Z$ production), 
at NLO in QCD and incorporating full spin correlations. 
It is available, together with a detailed documentation, at 
{\tt http://mcfm.fnal.gov}

\item MC@NLO \cite{MC@NLO}: it is a generator incorporating NLO QCD corrections 
into the parton-shower generator HERWIG for a large class of processes.
As such, MC@NLO provides a consistent description of exact ${\cal O}(\alpha_s)$ 
emission effects together with a leading-logarithmic resummation of soft and 
collinear QCD radiation. More details can be found at
{\tt http://www.hep.phy.cam.ac.uk/theory/webber/MCatNLO/}

\item ResBos \cite{BY,resbos}: it is a MC integrator program, presently used at the
Tevatron, that computes  fully differential cross sections of various 
boson production processes in hadron-hadron collisions, including 
DY-like processes, either with NLO initial-state QCD corrections, or with soft-gluon 
initial-state resummed QCD corrections, the latter according to the Collins, Soper 
and Sterman (CSS) resummation formalism \cite{css}. Recently, Resbos has been 
improved by the inclusion of final-state NLO electromagnetic contributions to
$W$ boson production \cite{cy}, resulting in the ResBos-A code. More information 
can be found at {\tt http://hep.pa.msu.edu/resum/}. Since in our study we are interested in
both QCD and EW effects, we use, among the different ResBos versions available 
on the web, the ResBos-A  generator, including CSS resummation and final-state QED 
radiation, making use of the grids publicly available for the Tevatron energy

\item FEWZ~\cite{mp, mp1}: it is a code computing fully differential cross sections at NLO/NNLO in perturbative QCD for $W$ and $Z$ boson production in hadron collisions, including finite-width effects and full spin correlations. \\ \noindent
It is available at {\tt http://www.phys.hawaii.edu/$\tilde{\phantom{x}}$kirill/FEHiP.htm}

\end{enumerate}

As far as FEWZ is concerned, it will be used in the present paper for tuning purposes at the NLO level for fully integrated cross sections. Its predictions at the NNLO level and the phenomenological relevance of NNLO corrections have already been analyzed in~\cite{Adam:2008pc,Adam:2008ge}. 

It should also be mentioned that a recent advance in the QCD predictions for DY processes is represented by the calculation of~\cite{pnason}. This consistently combines the NLO QCD 
corrections with parton shower resummation, avoiding negative weights in the generation of events, and
is available in the MC POWHEG. 
We will not show in our analysis  predictions of POWHEG, since they are already compared with 
MC@NLO~\cite{Alioli:2008gx}, finding fair agreement for {\it jet}-inclusive quantities.

\subsection{Electroweak contributions}
\label{horace}
The radiative corrections originating from the EW sector of the SM  
are taken into account in our analysis according to the recent calculation of~\cite{CMNV}, which the reader is addressed to for more details. At a variance of QCD
corrections, we do not need to take into consideration here the results of different, independent
calculations because it is known from recent tuned-comparison studies \cite{LesHouches,tev4lhc} that
the predictions of those calculations typically used or under consideration by hadron collider 
collaborations, such as
DK \cite{dk}, SANC \cite{SANC} and WGRAD \cite{bkw,bw}, are in very good agreement 
with the results of HORACE, which is the generator implementing the theoretical 
formulation of  \cite{CMNV}, available at {\tt http://www.pv.infn.it/$\tilde{\phantom{x}}$hepcomplex/horace.html}. In HORACE, the exact one-loop EW corrections 
to the CC DY process are
consistently matched with higher-order leading logarithms due to multiple photon emission, that have been shown to be not irrelevant for the precision target of  $W$ mass measurements at hadron colliders \cite{CMNTW,CMNTZ}. 
By itself, HORACE does not include any effect due initial-state QCD parton shower, but the
events generated with it can be interfaced with standard shower MC programs.

\begin{figure}[h]
\begin{center}
\includegraphics[width=5cm]{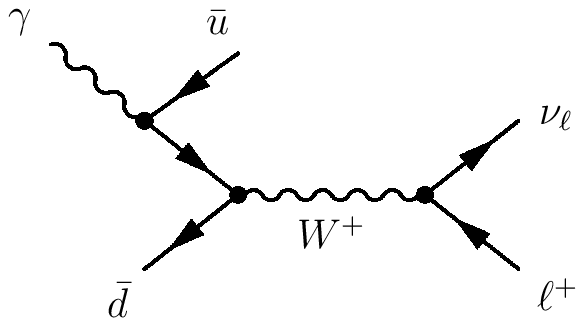}~\hskip 12pt\includegraphics[width=4cm]{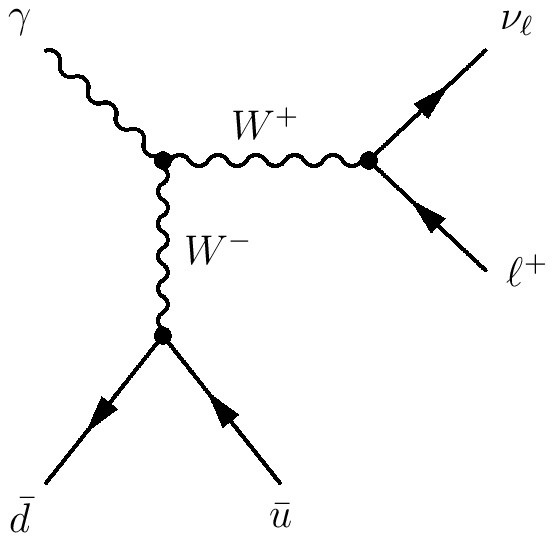}
\caption{Examples of Feynman diagrams for the photon-induced processes}
\label{ginduced}
\end{center}
\end{figure}

Since we are interested in a detailed description of EW effects too, we need to include in our calculation, for
the sake of consistency, also the contribution of the so-called photon-induced 
processes $\gamma q \to q^{\prime} l \nu_l$ and $\gamma \bar{q^{\prime}} \to \bar{q} l \nu_l$
(see Figure \ref{ginduced}). Actually, these processes are of the same EW  
perturbative order of one-loop EW corrections and give rise to the same
signature of the inclusive $W$ production process.
Furthermore, their contribution
has been shown by Dittmaier-Kr\"amer in \cite{LesHouches} to be non negligible for
$W$-boson production in the presence of realistic event selection cuts at the LHC.
The matrix elements for the photon-induced processes ($\gamma q\to
q^\prime l \nu_l$) are obtained by crossing symmetry from the
$q\bar{q^\prime}\to l \nu_l \gamma$ matrix elements, already available from the 
calculation of hard-bremsstrahlung diagrams contributing to EW corrections.

It is worth noting that the inclusion of photon-induced processes introduces in the 
${\cal O}(\alpha)$-corrected parton-level cross section mass singularities due to 
a collinear splitting $\gamma \to q \bar q$. As for photon radiation off the initial-state
quarks present in the NLO EW calculation and, again, in analogy with factorization 
in NLO QCD calculations, the collinear singularities due to $\gamma \to q \bar q$ collinear
splitting can be absorbed in the distribution functions, by replacing the (anti-)quark distribution
as described in detail, with all the necessary formulae, in \cite{ddh}. 

\subsection{Combination of electroweak and QCD corrections}
\label{combining}
The aim of this Section is to describe how the contributions of EW origin 
are combined, in our approach, with QCD 
predictions. 

A first strategy  for the combination of EW and QCD 
corrections consists in the 
following additive  formula
\begin{eqnarray}
\left[\frac{d\sigma}{d\cal O}\right]_{{\rm QCD} \& {\rm EW}} = 
\left\{\frac{d\sigma}{d\cal O}\right\}_{{\rm QCD}}
+\left\{\left[\frac{d\sigma}{d\cal O}\right]_{{\rm EW}} - 
\left[\frac{d\sigma}{d\cal O}\right]_{{\rm LO}} \right\}_{{\rm HERWIG\, \,  PS}}
\label{eq:qcd-ew}
\end{eqnarray}
where ${d\sigma/d\cal O}_{{\rm QCD}}$ stands for the prediction of the 
observable ${d\sigma/d\cal O}$ 
as  obtained  by means of one the QCD generators 
described in Section \ref{qcdgenerators}, 
${d\sigma/d\cal O}_{{\rm EW}}$ is the HORACE
prediction for the EW corrected  ${d\sigma/d\cal O}$ observable, 
and ${d\sigma/d\cal O}_{{\rm LO}}$ is the lowest-order hadron-level result. 
The label {{\rm HERWIG PS} in the second term in r.h.s. 
of eq. (\ref{eq:qcd-ew}) means that EW corrections are convoluted with QCD PS 
evolution through the HERWIG event generator, in order to (approximately) include
mixed ${\cal O}(\alpha \alpha_s)$ corrections and to obtain a more realistic 
description of the observables under study. In principle, concerning QCD, the predictions 
of any of the codes considered in Section \ref{qcdgenerators} could be used in eq. (\ref{eq:qcd-ew}).
However, because in our approach EW corrections
are treated at NLO and are convoluted with the QCD parton shower cascade by HERWIG, 
it follows that, in eq. (\ref{eq:qcd-ew}), the QCD contributions can be consistently combined with EW  
effects only by making use of the results provided, among the different 
QCD generators, 
by MC@NLO, in order 
to treat QCD NLO and shower corrections on the same ground of the EW 
counterpart.  Therefore, in the following, we will show results for the QCD and EW  
combination assuming $\left[{d\sigma} / {d\cal O}\right]_{\rm QCD} 
\equiv \left[{d\sigma} / {d\cal O}\right]_{\rm MC@NLO}$\footnote{Alternatively one could use POWHEG in association with PYTHIA PS, without spoiling the general procedure presented here. }. 
This implies, in turn, that the HORACE and MC@NLO generators need to 
 be properly tuned at the level of input parameters and PDF set, in order to provide meaningful
 predictions. Furthermore, since we are interested in consistent comparisons between the results of
 independent codes, this tuning procedure has to be applied to each generator in use, 
as discussed in Section \ref{mctuning}. 
Concerning the convolution 
of NLO EW corrections with QCD PS,  it is worth noting that, according to eq. (\ref{eq:qcd-ew}), 
 the contributions of the order of 
$\alpha \alpha_s$ are not reliable when hard non-collinear  QCD radiation 
turns out to be relevant, e.g. for the lepton and vector boson transverse momentum distributions 
in the absence of severe cuts able to exclude resonant $W$ production. In this case, a full  ${\cal O}(\alpha \alpha_s)$ calculation would be 
needed for a sound evaluation of mixed EW and QCD corrections. Full ${\cal O}(\alpha)$ 
EW corrections to the exclusive process $pp \to W + j$ (where $j$ stands for jet) have been
recently computed, in the approximation of real $W$ bosons, in \cite{hkk,kkp}, and for off-shell $W$ in~\cite{ddkm},  while 
one-loop weak corrections to $Z$ hadro-production have been computed, for on-shell 
$Z$ bosons, in \cite{mmr}. It is also 
worth stressing that in eq.~(\ref{eq:qcd-ew}) the infrared part of QCD corrections is factorized, whereas the infrared-safe matrix element residue is included in  an additive  form. 

It is otherwise possible to implement a fully factorized combination 
as follows: 
\begin{eqnarray}
\left[\frac{d\sigma}{d\cal O}\right]_{{\rm QCD} \otimes {\rm EW}} = & &
\left( 1 + \frac{
\left[{d\sigma} / {d\cal O}\right]_{\rm MC@NLO} - \left[{d\sigma}/{d\cal O}\right]_{\rm HERWIG\, \, PS}
}{
\left[{d\sigma}/{d\cal O}\right]_{\rm LO/NLO}
}
\right) \times \nonumber \\
& & \times 
\left\{ 
\frac{d\sigma}{{d\cal O}_{\rm EW}}
\right\}_{{\rm HERWIG\, \,  PS}} , 
\label{eq:qcd-ew-factor}
\end{eqnarray}
where the ingredients are the same as in eq.~(\ref{eq:qcd-ew}) but also the QCD matrix element residue in now factorized. It is worth noticing that the QCD correction factor in front of $\left\{ 
d\sigma / {d\cal O}_{\rm EW}
\right\}_{{\rm HERWIG\, \,  PS}} 
$
is defined in terms of two different normalization cross sections, namely the LO or the NLO one, respectively. As it can be easily checked, the two prescriptions differ at the order $\alpha_s^2$ by non-leading contributions. Nevertheless, eq.~(\ref{eq:qcd-ew-factor}) normalized in terms of the LO cross section can give rise to pathologically large order $\alpha_s^2$ corrections in the presence of huge NLO effects, as will be discussed in the next section. On the other hand, when NLO matrix element effects do not introduce particularly relevant corrections, the two prescriptions are substantially equivalent, as it will be demonstrated too in the discussion of numerical results. 
Some comments are in order here. Equations~(\ref{eq:qcd-ew}) and (\ref{eq:qcd-ew-factor}) in both prescriptions have the very same ${\cal O}(\alpha)$ and ${\cal O}(\alpha_s)$ content. 
While the two factorized prescriptions contain the same 
${\cal O}(\alpha \alpha_s)$ contributions, the additive prescription has 
mixed ${\cal O}(\alpha \alpha_s)$ contributions which differ from the factorized recipes. Therefore, their relative difference can be taken as an estimate of the uncertainty of QCD and EW combination and will be discussed in detail in the next section.

\section{Numerical results and discussion}
\label{numerics}

\subsection{Input parameters and event selection criteria}
\label{criteria}

The numerical results have been obtained using {\it in each code} the following values for the input parameters 

\begin{center}
\begin{tabular}{lll}
$G_{\mu} = 1.16639~10^{-5}$ GeV$^{-2}$ & 
$\mw = 80.419$~GeV&
$\mz=91.188$~GeV \\
$\gw = 2.048$~GeV & 
$\sdw = 1 - \mw^2/\mz^2$&
$\mh = 120$~GeV\\
$m_e=510.99892$~KeV &
$m_{\mu}=105.658369$~MeV &
$m_{\tau}=1.77699$~GeV \\
$m_u = 320$~MeV &
$m_c = 1.55$~GeV &
$m_t = 174.3$~GeV \\
$m_d = 320$~MeV &
$m_s = 500$~MeV &
$m_b = 4.7$~GeV \\
$V_{cd}=0.224$ &
$V_{cs}= 0.975$ &
$V_{cb}=0$ \\
$V_{ud}=\sqrt{1 - V_{cd}^2}$ &
$V_{us}=0.224$ &
$V_{ub}=0$ \\
$V_{td}=0$ &
$V_{ts}=0$ &
$V_{tb}=1$ \\
\end{tabular}
\end{center}
and adopting the $G_{\mu}$ input scheme for the calculation of EW corrections, where, in particular,
the effective electromagnetic coupling constant is given in the tree-level approximation by
\begin{eqnarray}
\alpha^{\rm tree}_{G_{\mu}} = \frac{\sqrt{2} G_{\mu} \sdw \mw^2}{\pi}
\end{eqnarray}
However, for the coupling of external photons to charged particles needed for the 
evaluation of photonic corrections we use 
$\alpha (0) =1/137.03599911$. 

We study the production process $p\smartpap \to {\rm lepton} + X$, both at the Tevatron $p \bar{p}$ collider ($\sqrt{s}$ = 1.96~TeV) and at the $pp$ LHC  ($\sqrt{s}$ = 14~TeV), presenting results only for 
final-state muons, for
the sake of definiteness. Actually, the difference between electron and muon final states
is confined at the level of pure EW and multiple photon corrections, and can be inferred from the
literature addressing such a topic in detail \cite{bkw,dk,CMNV, CMNTW, CMNTZ}. 

\begin{table}[h]
\begin{center}
\begin{tabular}{|c|}
\hline
{\large Tevatron}\\
\hline
$\! \! \! \! \! \! \! \, p_{\perp}^{l} \geq$~25 GeV $\, \, \quad $ $\rlap{\slash}{\! E_T} \geq$~25 GeV and 
$|\eta_l|< 1.2$\\
$\, \, \, \, p_{\perp}^{W} \leq 50$~GeV  $\quad$ 50~GeV $\leq M_{l \nu} \leq$~200 GeV\\
\hline
\end{tabular}
\caption{Selection criteria for the Tevatron}
\label{tab:tev}
\end{center}
\end{table}

\begin{table}[h]
\begin{center}
\begin{tabular}{|c|}
\hline
{\large LHC}\\
\hline
$\, \,$ a. $\,$ $p_{\perp}^{l} \geq$~25 GeV $\, \, $ $\rlap{\slash}{\! E_T}  \geq$~25 GeV and 
$|\eta_l|< 2.5$\\
\hline
$\!  \! \! \! \! \! \! \! \! \! \! \! \! \! \! \! \!$ b. the cuts as above $\, \, $  $\oplus$ $\, \,$ $M_\perp^W \geq 1$~TeV\\
\hline
\end{tabular}
\caption{Selection criteria for the LHC}
\label{tab:lhc}
\end{center}
\end{table}

The selection criteria used for our analysis at the Tevatron and the LHC are summarized 
in Table \ref{tab:tev} and Table \ref{tab:lhc}, respectively. In the Tables $p_{\perp}^{l}$ and
$\eta_l$ are the transverse momentum and the pseudorapidity of the muon, $\rlap{\slash}{\! E_T} $
is the missing transverse energy, which we identify with the transverse momentum of 
the neutrino, as typically done in phenomenological studies \cite{bkw,dk,bw,CMNV,cy,fm}; 
$p_{\perp}^W$ and $M_{\perp}^W$ are the transverse momentum 
and transverse mass of the $W$ boson, $M_{l\nu}$ 
is the invariant mass of the muon-neutrino pair. 
The selection criteria in the first line of
Table~\ref{tab:tev} are introduced to model the acceptance cuts used by the experimental 
collaborations at the Tevatron, while the constraints on $p_{\perp}^W$ and $M_{l\nu}$
in the second line are imposed in order to include in our study of the QCD corrections the 
predictions of ResBos-A v1.1, which makes publicly available grids for numerical integration 
corresponding to such constraints. For the LHC, we consider two different set up, labeled as a. and b.
in Table \ref{tab:lhc}. The first one 
corresponds to typical cuts used for LHC simulations, while for the second one a severe cut
on the $W$ transverse mass is superimposed to the cuts of set up a., in order to isolate the 
region of the high tail of $M_T^W$, which is interesting for new physics searches.

In order to avoid edge effects in the final phenomenological analysis, introduced by 
cuts at the generation level, we make use of the following 
loose generation cuts $p_{\perp}^{l} \geq$ 5 GeV, $\rlap{\slash}{\! E_T}  \geq$~5 GeV, $| \eta_l |< 5$ (for both
the Tevatron and set up a. at the LHC), with, additionally, $M_\perp^W \geq 0.8$~TeV 
for set up b. at the LHC. These cuts have been imposed simply to improve the 
efficiency in the generation of MC events.
For the ALPGEN code, we also use the following 
generation cuts for additional QCD partons: $p_T>$ 20 GeV, $|\eta_j|<$ 5, $\Delta R_{ij} >$ 0.7, 
where $\Delta R_{ij}$ is the separation between the $i$-th and $j$-th parton in the $\eta-\phi$ plane.
The parameters required in ALPGEN for the matching between matrix elements and
shower evolution are chosen as: $E_{T,clus.}=$ 25 GeV, $|\eta_j|=$ 5, $R_j =$ 1.05, 
as done in previous studies \cite{mmpt,hkl,Alwall:2007fs}. 
A first attempt to assess the intrinsic systematics associated 
         with the matching procedure has been presented in~\cite{Alwall:2007fs}. For our purposes we assume 
         that the up to one jet samples are well described 
         by NLO and next-next leading logarithmic codes. Therefore 
         the impact of the uncertainties of the matching procedure 
         is related to the importance of higher parton multiplicities 
         for the various distributions under consideration. 
         As it will be shown in the following, their size is at the level 
         of (or below) 1\% at the Tevatron 
         and the LHC (event selection a), while they become more relevant 
         at the LHC, event selection b, where a more thorough investigation 
         could be worthwhile. 

The set of parton density functions used in our study is CT6EQ6M~\cite{cteq6} for the 
case of the Tevatron, with factorization/renormalization scale  $\mu_R = \mu_F = 
M_{l \nu}$ for both the LO and NLO predictions. Again, this choice has been dictated by the necessity of including 
ResBos-A in our comparisons. Although it is theoretically inconsistent using a PDF set, like 
CT6EQ6M, without QED contributions when combining QCD corrections with NLO EW 
contributions, it is also known, as previously discussed, that the effect of QED evolution 
is small and translates into negligible numerical effects on the observables of
interest here within the adopted selection criteria. 

For the LHC, all the analysis is performed using the set
MRST2004QED, in order to consistently incorporate
EW corrections and photon-induced processes in association with QCD corrections. 
For the LHC, the QCD 
factorization/renormalization scale and the analogous QED scale (present 
in MRST2004QED) are chosen to be equal, as usually done in the literature \cite{bkw,dk,bw,CMNV}, and 
fixed at $\mu_R = \mu_F = \sqrt{p_{\perp W}^2 + M_{l\nu}^2}$, where $M_{l\nu}$, 
as done in previous studies \cite{fm}, both for LO and NLO results. 

The numerical results showing 
the estimate of the PDF uncertainties on the observables at the LHC energies (Section \ref{pdfnum}) have been obtained using the generator MC@NLO version 3.3 implementing the
LHAPDF package. Results for the NLO MRST2001E (with  30 error sets) and CTEQ6 (with
40 error sets) parameterizations as available in  LHAPDF  are shown, 
using as default scale choice $\mu_R = \mu_F = \sqrt{p_{\perp W}^2 +  M_{l\nu}^2}$.

For $\alpha_s(Q^2)$ we use as input parameters $\alpha_s(M_Z) = 0.118$ at the Tevatron and $\alpha_s(M_Z) = 0.121$ at the LHC. We checked that for all the codes under consideration the values of $\alpha_s$ agree at the 0.1\% level when varying $Q$ from $M_Z$ to a few TeV. 

All the numerical simulations have been done without taking into account the effects of the
underlying event and hadronization processes. 

The numerical results discussed in the following have been 
obtained according to a MC statistics of approximately 
$10^6$ - $10^8$ events for the integrated sections 
 and of about $10^6$ events for the distributions. 
For the integrated cross sections, the numbers in the 
Tables quoted in parenthesis after the last digit correspond to the $1\sigma$ MC 
errors. As for the distributions, each 
of them has been generated as an histogram consisting of 25 bins. Therefore,
in the dominant region of a given distribution (e.g. in the central
range of the $W$ rapidity and lepton pseudorapidity and around the
Jacobian peak of the $W$ transverse mass and lepton transverse momentum),
the statistical uncertainty associated to the cross section in 
a bin is at the level of some per mille. In the tails of the 
distributions, the MC error can grow up to a few per cent level. 
This statistics is sufficient for a meaningful interpretation 
of the effects of interest in the present study and discussed in the
following. Actually, whenever rather mild effects are registered (e.g.
of a few per cent), they typically show up in the dominant regions where
the statistical uncertainty is about a factor of ten smaller. On the 
other hand, in the tails of the distributions, characterized by a poorer
statistics, the contributions of interest are typically quite large 
(at the level of several or tens of per cent) and are therefore meaningful 
in comparison with the size of the statistical fluctuations. Last but not least, 
the most moderate effects generally correspond, as it will be shown in 
the following, to the difference between the results of the additive and 
factorized recipes for the combination of EW and QCD corrections. However, 
since the two recipes are defined in terms of the very same theoretical
ingredients, and use the same MC data, it turns out that predictions of the two 
 combinations are 
strongly correlated variables and, as such, the uncertainty associated to
each of these two predictions largely cancel out when considering the 
difference of the two.

\subsection{Monte Carlo tuning}
\label{mctuning}

Before discussing the numerical results, we would like  to spend some words about the effort
done in MC tuning, in order to normalize all the codes to the same LO or NLO 
cross section and, therefore, avoid possible bias in the interpretation of the physical
effects of interest.

Concerning the program ResBos-A, we switched off some EW contributions present 
inside the code in the form of Improved Born Approximation and transformed the running 
$W$ width in the $W$ propagator into a fixed width, to normalize the code at a pure
LO cross section. Moreover, we used the program without NLO QED final-state corrections
when providing pure QCD predictions. 

 To tune FEWZ against the other codes, we substituted the default choice 
of using the experimental branching ratio BR($W \to \mu \nu$) with the tree-level theoretical branching ratio 
$\Gamma(W \to \mu \nu) / \Gamma (W \to {\rm all}) = 1/9$, for consistency with the other used programs.

As far as MC@NLO is concerned, 
we tuned it at the next-to-leading-order in QCD, by comparing its predictions with the 
corresponding NLO QCD results of FEWZ and MCFM. This tuning has been performed at the level of 
completely inclusive (i.e. without cuts) cross sections both for the Tevatron and the LHC, as well as
in the presence of a cut on the invariant mass $M_{l\nu} > 0.8$~TeV at LHC energies, in order to check  
the correct normalization  in the region important for new-physics studies.

\begin{table}[h]
\begin{center}
\begin{tabular}{|c|c|c|c|c|c|}
\hline
Monte Carlo & ALPGEN & FEWZ & HORACE & MCFM & ResBos-A  \\
\hline
$\sigma_{\rm LO}$~(pb) & 906.3(3) & 905.4(2) & 905.6(1) & 905.1(1) &  905.3(2) \\
\hline
\end{tabular}
\caption{MC tuning at the Tevatron for the LO cross sections of the sum of the
processes $p\bar p \to W^+ \to \mu^+ + \nu$ and $p\bar p \to W^- \to \mu^- + \bar\nu$, according to the cuts 
of Table~1 
and using the CT6EQ6M 
%CTEQ6M 
PDFs with scale $M_{l\nu}$. }
\label{mctev}
\end{center}
\end{table}

\begin{table}[h]
\begin{center}
\begin{tabular}{|c|c|c|c|c|c|}
\hline
Monte Carlo & ALPGEN &FEWZ & HORACE&  MCFM  \\
\hline
$\sigma_{\rm LO}$~(pb) & 8310(2) & 8306(1) &  8308(1) & 8305(1)  \\
\hline
\end{tabular}
\caption{MC tuning at the LHC for the LO cross sections of the sum of the
processes $p p \to W^+ \to \mu^+ + \nu$ and $p p \to W^- \to \mu^- + \bar\nu$, according to the cuts 
of set up a. in Table 2 and using the MRST2004QED PDFs with scale  $\sqrt{p_{\perp W}^2 +  M_{l\nu}^2}$}
\label{mclhc}
\end{center}
\end{table}

\begin{table}[h]
\begin{center}
\begin{tabular}{|c|c|c|c|}
\hline
Monte Carlo &  FEWZ & MC@NLO & MCFM  \\
\hline
$\sigma_{\rm NLO}^{\rm Tevatron} ({\rm pb})$ & 2635.5(4) & 2639.1(5) &  2640(1)  \\
\hline
 $\sigma_{\rm NLO}^{\rm LHC} ({\rm pb})$ & 21058(3) & 21031(3) &  21008(2)   \\
\hline
\end{tabular}
\caption{MC tuning at the Tevatron and LHC for MC@NLO and MCFM NLO 
inclusive cross sections of the sum of the processes $p \smartpap \to W^- \to \mu^- \bar \nu$ and
$p \smartpap \to W^+ \to \mu^+ \bar \nu$, using the PDF sets CT6EQ6M 
%CTEQ6M 
at the Tevatron with scale $M_{l\nu}$ and MRST2004QED at the LHC with scale $\sqrt{p_{\perp W}^2 +  M_{l\nu}^2}$, respectively. }
\label{tuningnlo}
\end{center}
\end{table}

\begin{table}
\begin{center}
\begin{tabular}{|c|c|c|}
\hline
Monte Carlo &  MC@NLO & MCFM\\
\hline
$\sigma_{\rm NLO} ({\rm fb})$ & 50.34(1) & 50.28(2) \\
\hline
\end{tabular}
\caption{MC tuning at the LHC for MC@NLO and MCFM NLO 
cross sections of the sum of the processes $p p \to W^- \to \mu^- \bar \nu$ and
$p p \to W^+ \to \mu^+ \bar \nu$, in the presence of a cut on the invariant mass $M_{l\nu} > 0.8$~TeV 
and no other lepton selection criteria. The PDF set is MRST2004QED with scale $\sqrt{p_{\perp W}^2 +  M_{l\nu}^2}$. }
\label{tuningnlo-hm}
\end{center}
\end{table}

The results are shown in Table \ref{mctev} for the LO cross 
sections at the Tevatron, in Table \ref{mclhc} for the LO cross 
sections at the LHC, in Table \ref{tuningnlo} for the tuning between FEWZ, MCFM and MC@NLO
NLO inclusive cross section both at the Tevatron and LHC, and in Table \ref{tuningnlo-hm} 
for the MCFM and MC@NLO predictions at LHC energies when imposing the invariant mass cut 
$M_{l\nu} > 0.8$~TeV. 
 As can be seen, there is a very good agreement between the
predictions of the different programs, since all the 
relative differences are at the few per mille level (or better), i.e. much smaller than the size of the radiative corrections discussed in the following. 
As already remarked in \cite{LH2007}, the tuning procedure is essential because it validates the interpretation of the various contributions as due to physical effects and
not to a mismatch in the set up of the codes under consideration.

\subsection{Integrated cross sections (Tevatron and LHC)}
\label{xsect}

In Table \ref{comparison} and Table \ref{nlo-nnlo} we present the results for the 
integrated cross sections corresponding to
the selection criteria at the Tevatron and LHC quoted in 
Section~\ref{criteria}, as obtained according to the 
following codes and theoretical recipes

\begin{enumerate}

\item Leading order, as obtained by MCFM code in terms of the LO matrix element supplemented by the 
PDF sets defined in Section \ref{criteria}

\item ALPGEN~S$_0$: the hadron-level LO cross section of the CC DY process is convoluted with parton shower evolution through HERWIG Parton Shower. Hence, such a prediction
can be considered as analogous to that of a pure parton shower generator;

\item ALPGEN~S$_1$: the exact LO matrix elements for up to one parton radiation are interfaced
to HERWIG parton shower, according to the MLM prescription of matrix elements-PS matching;

\item ALPGEN~S$_2$: the exact LO matrix elements for up to two parton radiation are interfaced
to HERWIG parton shower, according to the MLM prescription of matrix elements-PS matching;

\item MCFM, i.e. pure NLO QCD predictions (MCFM NLO QCD), including also factorization and 
renormalization scale variation by varying them by a factor 1/2 or 2
around the central values specified in Section \ref{criteria} (see Table \ref{nlo-nnlo});

\item HORACE NLO EW: HORACE, in the presence of NLO EW corrections;

\item HORACE NLO+HERWIG PS: the NLO EW predictions of HORACE are 
interfaced with HERWIG QCD Shower evolution;

\item MC@NLO: NLO QCD combined with HERWIG parton shower;

\item ResBos-A CSS: ResBos-A, with pure QCD CSS resummation, without fixed-order perturbative QCD 
contributions and final-state QED radiation;

\item ResBos-A NLO QED: without QCD resummation, without fixed-order perturbative QCD 
results but with final-state QED radiation; 

\item ResBos-A, as far as its complete predictions, including QCD 
resummation and final-state QED radiation, but without perturbative QCD corrections, are concerned;

\item Equation (\ref{eq:qcd-ew}), obtained by summing the predictions of MC@NLO with those
of  HORACE NLO convoluted with HERWIG PS according to eq. (\ref{eq:qcd-ew}) (additive combination);

\item Equation (\ref{eq:qcd-ew-factor}), for the combination of QCD and EW contributions in factorized
form.

\end{enumerate}

\begin{table}[hbtp]
  \medskip
  \begin{center}
  \begin{tabular}{|l||c |c c|} \hline
     & \multicolumn{1}{c|}{Tevatron} 
     & \multicolumn{2}{c|}{LHC}\\ \cline{0-3}
    Prediction & & a. & b. \\
    \hline
    \hline
    Leading order                                    & 905.1(1) & 8305(1) & 7.128(3) \\
    ALPGEN~S$_0$                                         & 815(1) &
                                                                    7224(5) & 7.15(1) \\
    ALPGEN~S$_2$                               & 814(1) &
                                                                     7578(5) & 7.29(1) \\                                                                                                           
    MCFM  NLO  QCD                                            & 979.1(6) & 
                                                                      8135(2) & 8.681(2) \\
    HORACE NLO  EW                               & 881.5(2) & 
                                                                      8091(1) & 5.569(2) \\ 
     HORACE NLO+HERWIG PS           & 792(1) & 
                                                                      7008(5) & 5.55(1) \\ 
                                                                 
     MC@NLO                                           & 967(3) & 
                                                                      8254(5) & 8.64(1)\\                                                                                                                           
     ResBos-A CSS                                    & 933.4(4) &
                                                                       -- & -- \\
     ResBos-A NLO QED                        & 877.9(3) & 
                                                                       -- & -- \\
     ResBos-A                                           & 920(3) & 
                                                                       -- & -- \\   
     Eq. (\ref{eq:qcd-ew}) QCD+EW add.     & 944(3) & 8038(9)
                                                                        & 7.04(2) \\ 
      Eq. (\ref{eq:qcd-ew-factor}) QCD $\otimes$ EW fact. LO    & 925(3) & 7877(8)
                                                                        & 6.71(2)  \\         
     Eq. (\ref{eq:qcd-ew-factor}) QCD $\otimes$ EW fact. NLO    & 915(3) & 7895(8)
                                                                        & 6.50(1)  \\                                                                                                                                                                                                                                                                                                                                                                                             
    \hline
  \end{tabular}
  \caption{Integrated cross sections, relative to the selection criteria of Table 1 and Table 2, according to different approximations for the various programs considered 
  in the study. For the Tevatron  and set up a. at the LHC the cross sections are given in pb, while the units of measure
  are fb for set up b. at the LHC.} 
  \label{comparison}
  \end{center}
\end{table}

  As can be seen from Table \ref{comparison} and, in particular from the comparison between the LO results and the
  ALPGEN~S$_0$  predictions, the QCD PS effects lower the cross section by about $10 \div 15\%$  at the Tevatron and LHC set up a, while they are negligible at LHC set up b. 
The exact matrix element corrections, as present in  ALPGEN~S$_2$, are not relevant for the cross section at the
Tevatron, while they induce a some per cent (positive) contribution for the two set up at the LHC.\footnote{The results of ALPGEN~S$_1$  are not shown in Table \ref{comparison} because they are practically indistinguishable,
 within the statistical errors of the MC integration, from the predictions of  ALPGEN~S$_2$. 
 } The NLO QCD 
corrections are positive, of about 8\%, at the Tevatron,  negative  of about 2\% at the
LHC set up a and positive of about 20\% at the LHC set up b. The EW corrections are at the level of a few per cent and negative at the Tevatron and
LHC set up a., while they are quite large, of the order of $-  25\%$, for set up b. at the LHC, in agreement with
the results of various studies in the literature 
on the virtual 
EW Sudakov logarithms. 
It is worth noticing that for the considered event selection the NLO QCD and EW corrections tend to compensate at the Tevatron and LHC set up b, while they sum up coherently at the LHC set up a. 
The QCD PS, when 
convoluted with the NLO EW contributions, decreases the cross sections of about 10\%, as already noted above
at the level of pure QCD results, at the Tevatron and the LHC set up a, while it is irrelevant at the LHC set up b.  The combination of NLO QCD corrections with PS resummation, as predicted by 
MC@NLO, gives a correction at the per cent level w.r.t. the pure NLO QCD correction
obtained with MCFM. 
The soft-gluon resummation a la CSS is responsible of a few per 
cent increase of the LO cross section at the Tevatron, while NLO final-state QED radiation is of the same order but of
opposite sign, so that the two effects tend to compensate, yielding a full ResBos-A prediction of only about 1\% larger
than the LO cross section. On the other hand, the combination of EW and QCD corrections according to 
the (additive) formula of eq. (\ref{eq:qcd-ew})  predicts an enhancement of the LO cross section at the Tevatron of
about 7\%, differing of some per cent from the ResBos-A result, as a consequence of the different QCD and 
EW ingredients between eq. (\ref{eq:qcd-ew}) and the ResBos-A formulation. The calculation of the combined
EW and QCD as in the (factorized) formula  of eq. (\ref{eq:qcd-ew-factor}) differs from the additive prediction
of about 2\%, thus reducing the difference w.r.t. ResBos-A. At the LHC, the combined QCD/EW correction is negative, of the order of few per cent, both in the set up a and b. It is worth noticing, however, that in the set up a this is due to the sum of mild corrections of the same sign, whereas in the set up b it is the result of the almost complete cancellation of large QCD and EW corrections. Also in this case, the difference of additive and factorized prescriptions amounts to few per cent.

\begin{table}[h]
\begin{center}
\begin{tabular}{|c|c|c|c|}
\hline
Collider & $\sigma_{\rm NLO}^{(\mu/2)}$ & $\sigma_{\rm NLO}^{(\mu)}$& $\sigma_{\rm NLO}^{(2\mu)}$\\
\hline
Tevatron & 978.5(8) &  979.1(6) & 987.1(6) \\
\hline
LHC a. & 7875(1) & 8135(2) & 8393(2) \\
\hline
LHC b. & 8.890(5) &  8.681(2) & 8.501(4) \\
\hline
\end{tabular}
\caption{NLO cross sections, including scale variation, as obtained by means of MCFM
at the Tevatron and LHC, set up a. and b.  The cross sections are given in pb 
for the Tevatron  and set up a. at the LHC, in fb for set up b. at the LHC.}
\label{nlo-nnlo}
\end{center}
\end{table}

In Table \ref{nlo-nnlo} we show the NLO predictions of MCFM at three different fac\-to\-ri\-za\-tion/re\-nor\-ma\-li\-za\-tion
scales, $\mu_R = \mu_F = \mu/2, \mu, 2 \mu$, 
as it is customary, with
$\mu =   M_{l\nu}$ at the Tevatron and 
$\mu  = \sqrt{p_{\perp W}^2 +  M_{l\nu}^2}$ at the LHC. The scale dependence at the Tevatron is 
at the 1\% level, indicating a good stability of the NLO prediction for the assumed scales. At the
LHC set up a., the cross section varies of about 3-4\% as a function of the scale variation, while a 2\% 
variation is observed for the cross section at the LHC set up b. 
As a whole, the results shown in Table \ref{comparison} point out the following aspects: first of all, a proper combination of EW and QCD corrections is mandatory to achieve a cross section accuracy of the order of a few per cent, in particular for the set up b at the LHC where a subtle cancellation occurs; secondly, at the per cent accuracy the NNLO QCD corrections should be taken into account, being their contribution of a few per cent  as estimated by the scale variations shown in Table \ref{nlo-nnlo} and confirmed by dedicated calculations in the 
literature~\cite{mp,mp1,Adam:2008pc,Adam:2008ge,as2}; last, at such a precision level also exact mixed ${\cal O}(\alpha\alpha_s)$ contributions, at present not known, can be expected to be relevant as pointed out by the relative difference between the additive and factorized prescriptions of eqs.~(\ref{eq:qcd-ew}) and (\ref{eq:qcd-ew-factor}), respectively.

\begin{figure}[h]
\begin{center}
\includegraphics[width=5.5cm]{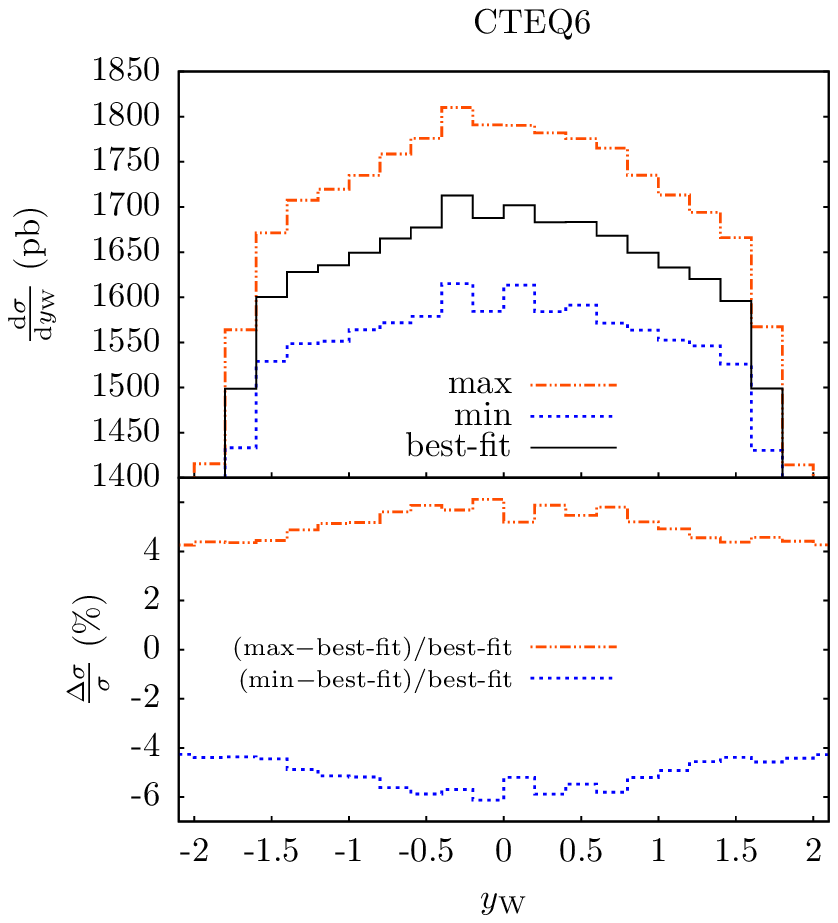}\hspace{2.cm}\includegraphics[width=5.5cm]{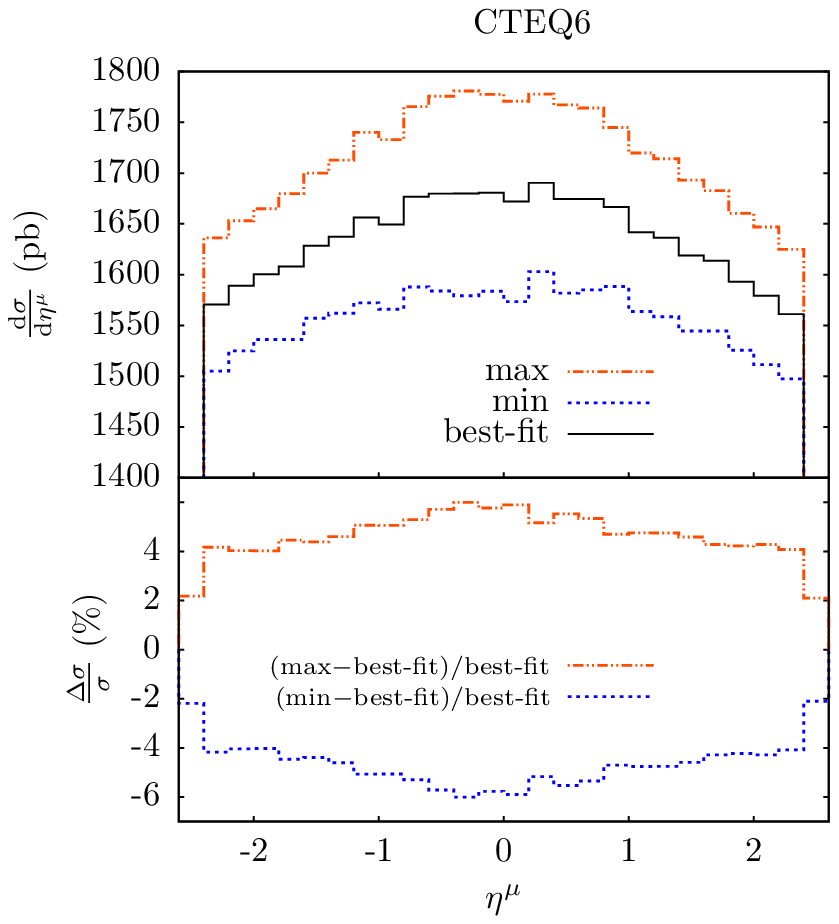}
\end{center}
\vskip 12pt
\begin{center}
\includegraphics[width=5.5cm]{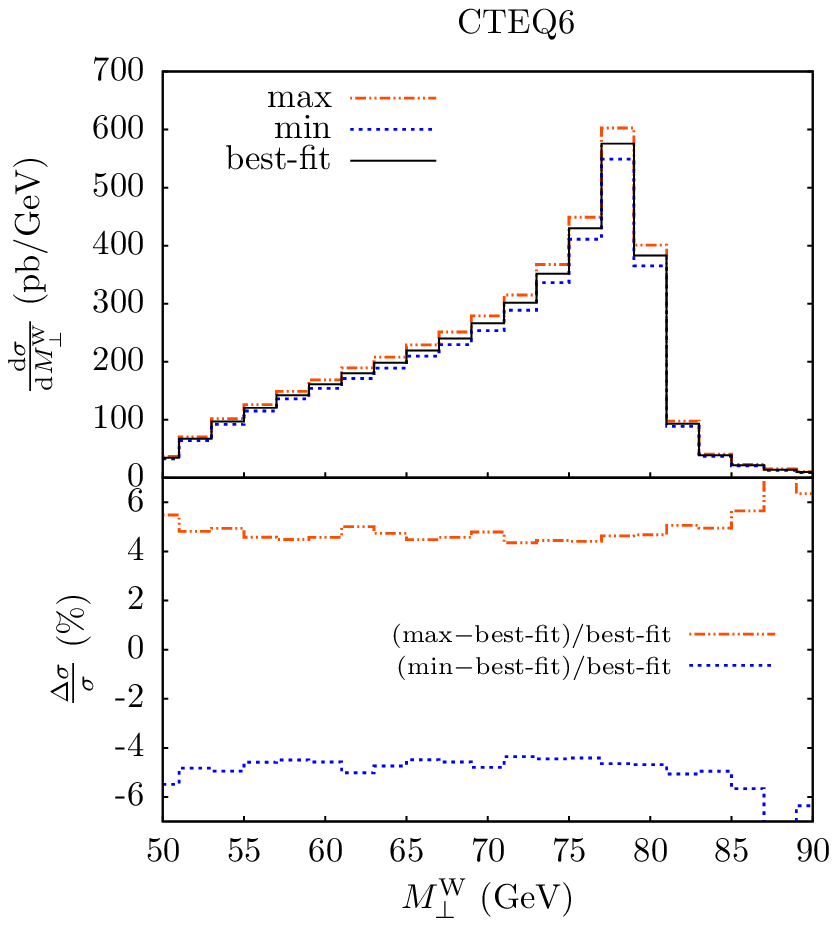}\hspace{2.cm}\includegraphics[width=5.5cm]{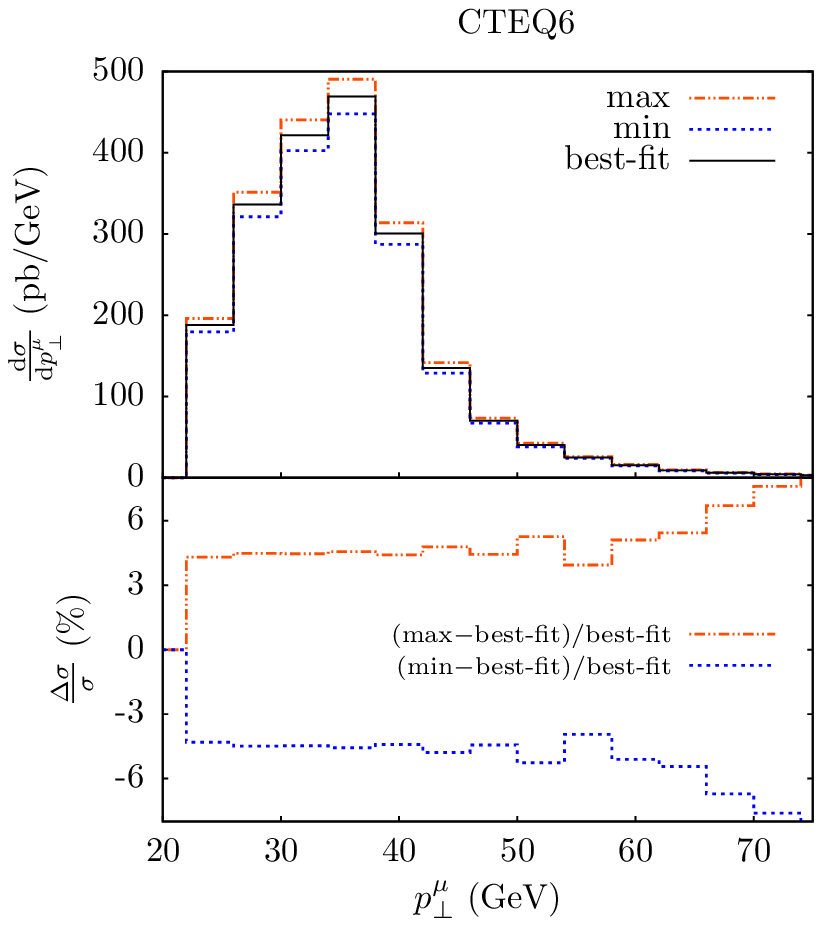}
\end{center}
\caption{CTEQ61 PDFs uncertainties for $W$ rapidity and muon pseudorapidity (upper plots), 
and for the $W$ transverse mass and muon transverse momentum (lower plots), according to
set up a. at the LHC.  In the lower panel of each plot, the relative deviations of the minimum
and maximum predicted values w.r.t. the best fit PDF are shown.}
\label{CTEQ6-nocuts}
\end{figure}

\subsection{PDF uncertainties (LHC)}
\label{pdfnum}

In this Section we discuss how the PDFs uncertainties of experimental origin, as known of today,  affect
all the observables of interest for the physics analysis of the process 
$p p \to \mu + X$ at the LHC. 

In Figure \ref{CTEQ6-nocuts} we show the results
for the $W$ rapidity and muon pseudorapidity (upper plots) and for the
$W$ transverse mass and muon transverse momentum (lower plots)  
according to set up a. specified in Table \ref{tab:lhc}, as obtained with the NLO CTEQ61 parameterization available in 
the LHAPDF package. 

For each observable, we show the predictions corresponding to the
maximum and minimum values returned by CTEQ61 PDFs, together with the result of the
best fit parton densities.

\begin{figure}[h]
\begin{center}
\includegraphics[width=5.5cm]{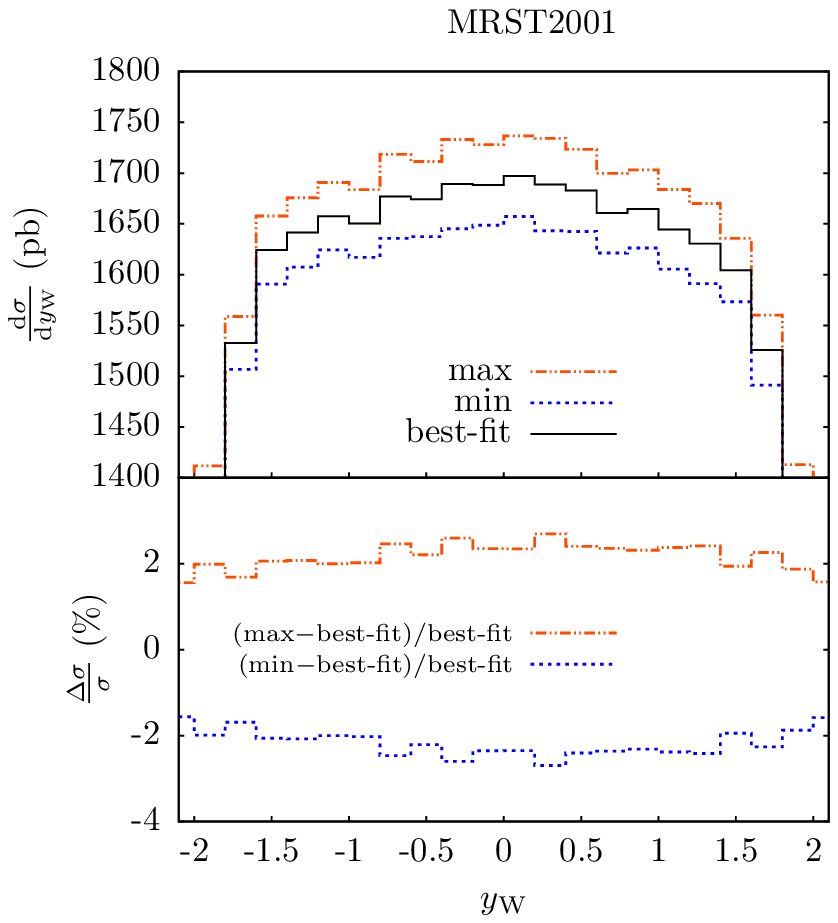}\hspace{2.cm}\includegraphics[width=5.5cm]{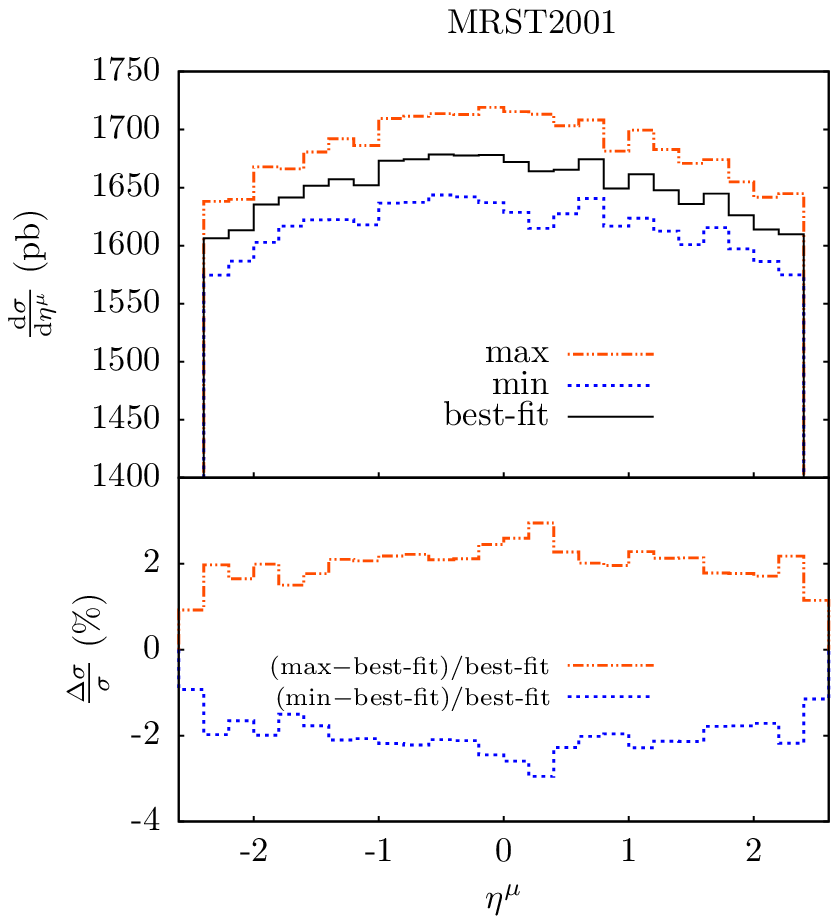}
\end{center}
\vskip 12pt
\begin{center}
\includegraphics[width=5.5cm]{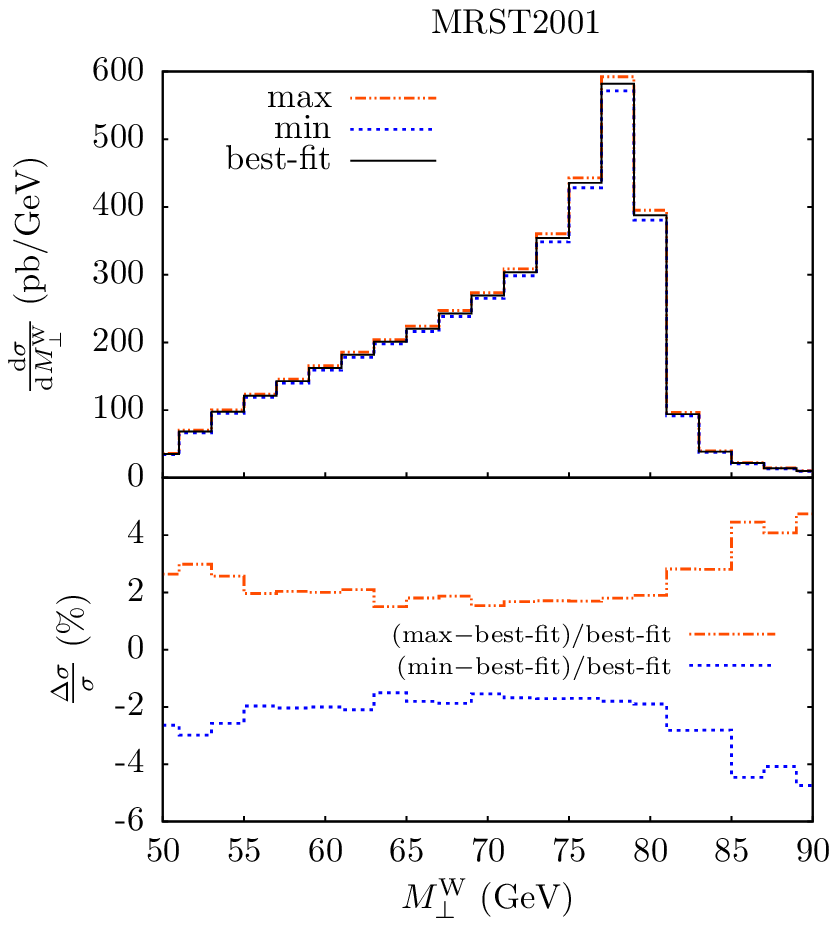}\hspace{2.cm}\includegraphics[width=5.5cm]{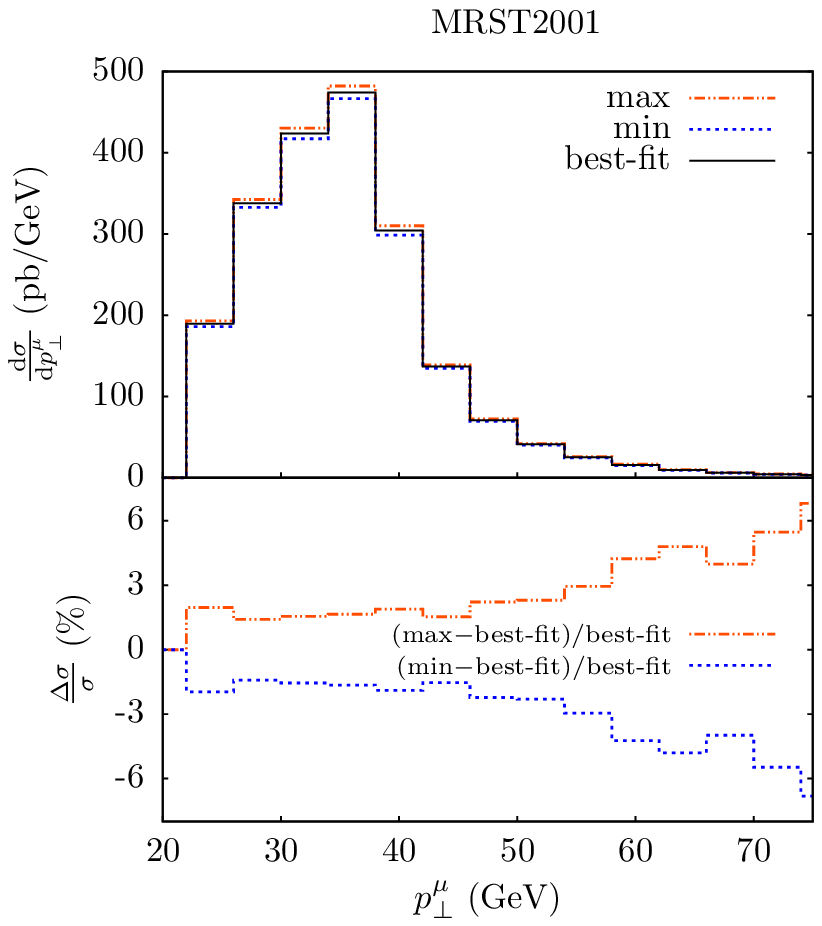}
\end{center}
\caption{The same as Figure 2 for MRST2001E parameterization.}
\label{MRST-nocuts}
\end{figure}

The PDF uncertainty for an observable O is the maximal change in O as
a function of variables $\{ z_i \}$ varying within the tolerance hypersphere~\cite{pdfuncert}:
\begin{equation}
\delta O = \sqrt{\sum_{i=1}^{n/2} \delta O_i^2} \nonumber
\end{equation}
where
\begin{equation}
\delta O_i \equiv T \frac{\partial O}{\partial z_i} \approx T \frac{O(z_i^0 + t) - O(z_i^0 - t)}{2 t} . \nonumber
\end{equation}
$n$ is the number of the error sets (30 for MRST, 40 for CTEQ), $T$ is the tolerance, 
$t$ is a small step in the space of $z_i$. Here $O(z^0_1 , . . . , z^0_i \pm t , . . . , z^0_{n/2})$
is denoted as $O (z^0_i \pm t)$.

In the lower panel of each plot, the relative deviations of the minimum
and maximum predicted values w.r.t. the best fit PDF are shown. As can be seen,
the spread of the predictions is at the level of some per cents  for all the distributions. In particular, for the $W$ rapidity and muon pseudorapidity the uncertainties vary within 2-4\%, reaching their
maximum value in the central rapidity region, while for the muon $p_\perp$ and $M_\perp^W$
the relative deviations are about 3-4\% around the jacobian peaks, reaching the 6\% level
in the hard tails of the distributions.

The same analysis is shown in Figure \ref{MRST-nocuts} for the NLO MRST2001E
parameterization. The observed uncertainties are smaller by about  a factor of two, as expected from the 
discussion addressed in Section \ref{pdf} about the different values of the tolerance parameter 
adopted in the two PDF sets.

\begin{figure}[h]
\begin{center}
\includegraphics[width=5.5cm]{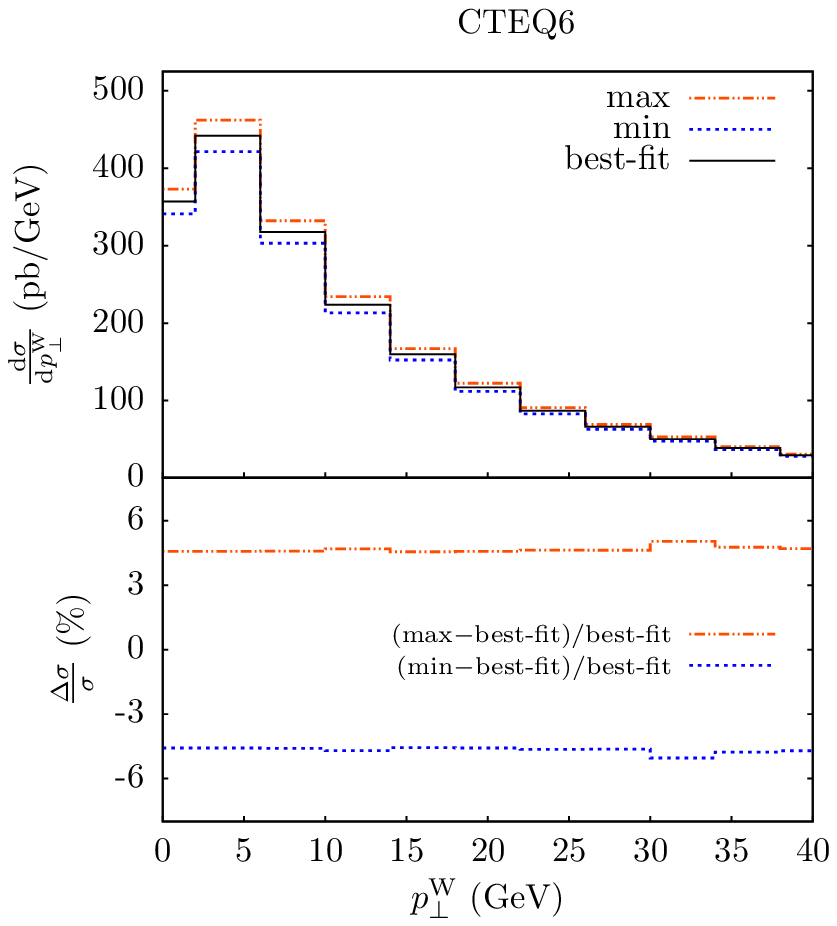}\hspace{2.cm}\includegraphics[width=5.5cm]{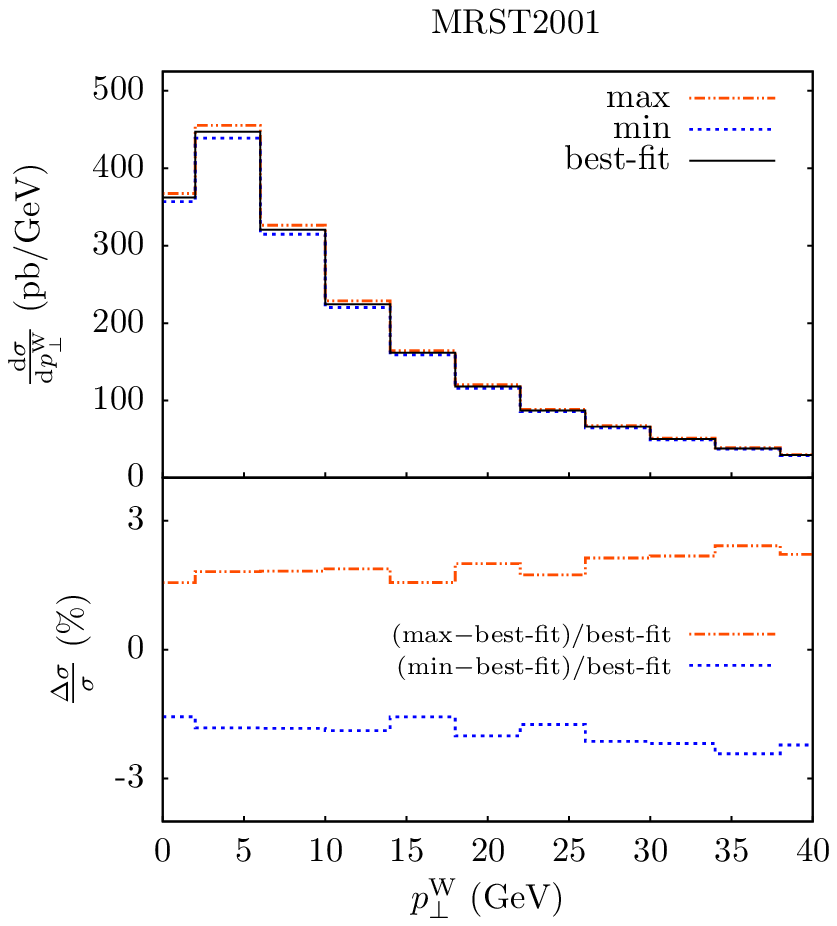}
\end{center}
\caption{CTEQ61 (left plot) and MRST2001E (right plot) PDFs uncertainties for the $W$ transverse
momentum distribution, according to
set up a. at the LHC.}
\label{CTEQ-MRST-ptw}
\end{figure}

\begin{figure}[h]
\begin{center}
\includegraphics[width=6cm]{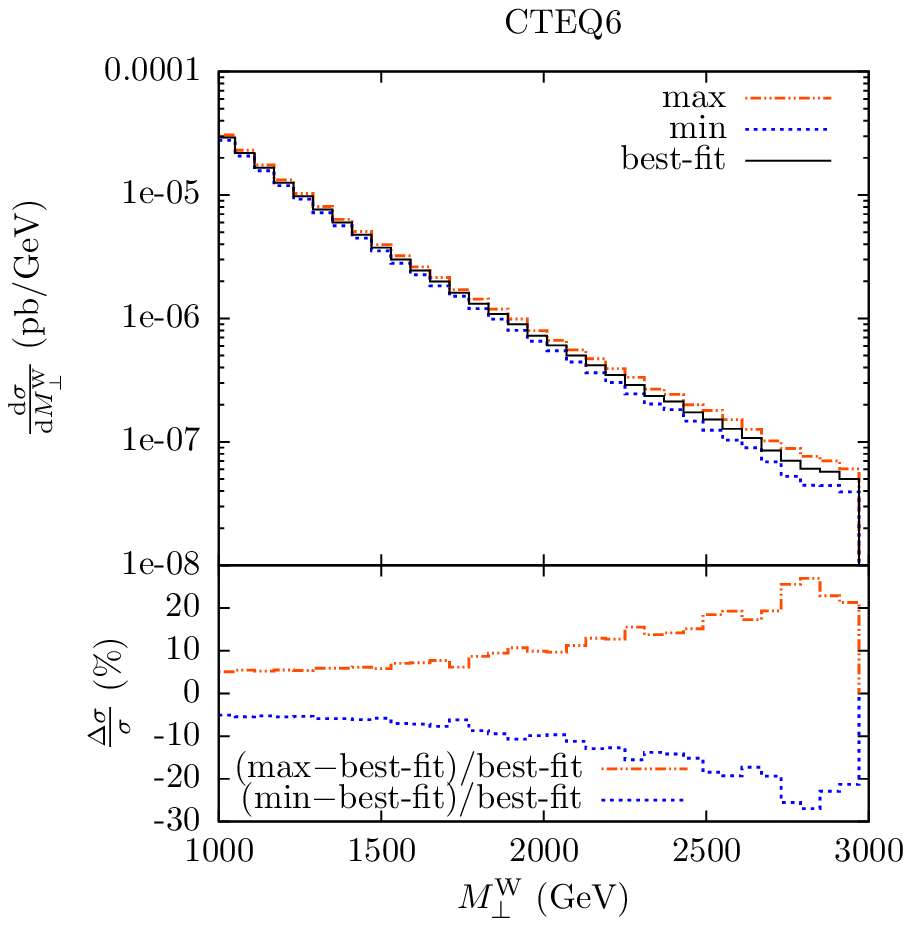}\hspace{1.5cm}\includegraphics[width=6cm]{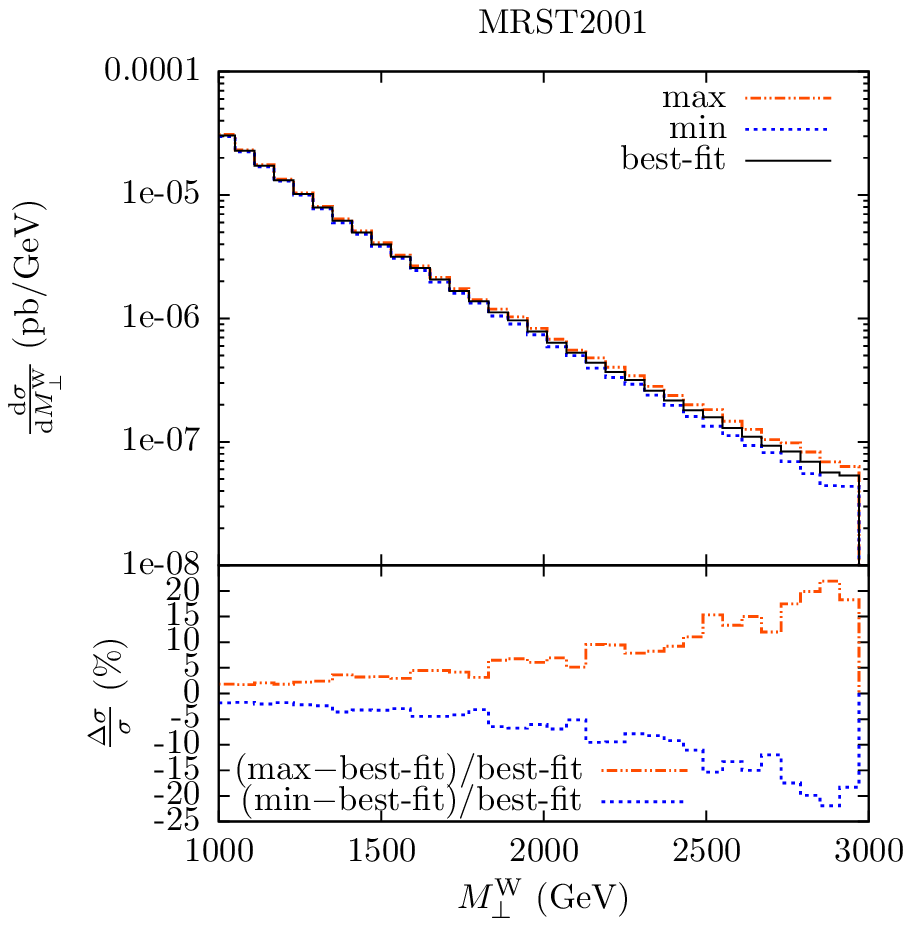}
\caption{CTEQ61 (left plot) and MRST2001E (right plot) PDFs uncertainties for the $W$ transverse mass in the high tail, according to set up b. at the LHC.}
\label{fig:CTEQ-MRST-whigh}
\end{center}
\end{figure}

The results for the $W$ transverse momentum are shown in Figure 
\ref{CTEQ-MRST-ptw}, both for CTEQ61 (left plot) and
MRST2001E (right plot). The uncertainties are almost flat in both cases, and of about 3\%  
and 1.5\% for CTEQ61 and MRST2001E parameterization, respectively.  

We also analyzed the PDF uncertainties on the transverse mass distribution when considering set up b. of Table \ref{tab:lhc} at the LHC, as shown in Figure \ref{fig:CTEQ-MRST-whigh}. We observed that both the CTEQ61 and MRST2001E predictions display deviations w.r.t to the best fit at some per cent level 
from 1 to 2 TeV, while the uncertainties are, in the average, of the order of 20\% in the vicinity 
of 3 TeV, as due to the well-known gluon PDF uncertainties for large $x$ values dominant in such a
region.

It is important to remind, as already remarked, that the estimate of PDF uncertainties obtained 
according to such a procedure are of experimental origin only, leaving aside other sources
of uncertainty due to theory.

At present the only available set which includes electromagnetic
effects in the evolution of the PDFs is MRST2004QED.
We have chosen this set to  consistently subtract the initial state
QED collinear divergences, and also to investigate the contribution
of the extra photon-induced partonic subprocesses, due to the presence
of a photon density in the proton.

%The MRST2004QED set does not contain massive quarks effects. The latter
%have been included in the most recent PDFs sets~\cite{MSTW, CTEQ66} and
%turned out to be important for DY phenomenology at the LHC,
%mostly because of the large contribution given by the $c\bar
%s$-initiated partonic process:
%in fact the CC and NC total cross-sections within cuts
%change by approximately +6\%, w.r.t. the
%predictions obtained 
%with massless
%quarks PDFs; 
The recent PDFs sets~\cite{MSTW, CTEQ66,thorne09}  have included new data and a refined treatment of 
the quark masses effects. These updates turned out to be important for DY 
phenomenology at the LHC: in fact the total cross sections within cuts 
change by approximately 3-6\% w.r.t. the predictions obtained with the older 
sets. 
On top of the overall normalization variation we
observed
a moderate (of the order of 1\%) but not constant shape change 
when comparing the
transverse mass or the lepton transverse momentum distributions
computed with or without massive quark corrections.

Our choice of using MRST2004QED, although it is not the most
up-to-date proton parameterization, has been taken to keep the
possibility of discussing the relevance of the photon-induced
processes, which might be not negligible especially in the searches
for new physics signals.
For consistency at the Tevatron we have used the Resbos grids based
on the CTEQ6.1 PDFs set, which does not include massive charm effects.

The analysis of the present Section 
is not spoiled by the use of older sets of PDFs nor it does
overestimate the uncertainties
due to the experimental error from which the PDFs are affected.
In fact we numerically checked that the 1-$\sigma$ spread of the predictions obtained 
with the most recent sets is of the same size, w.r.t. the results shown here. 

\subsection{Numerical results for the Tevatron}
\label{tev-results}

In this section, we present the results obtained for the process $p \bar{p} \to W^\pm \to \mu^\pm + X$ at the
energy of the Tevatron Run II  ($\sqrt{s}$ = 1.96~TeV), when imposing the cuts quoted 
in Table~\ref{tab:tev}. For the various observables of interest, we begin with a discussion
of QCD effects, to continue with the analysis of the combination of EW and 
QCD corrections. The latter is realized according both to the additive formula of eq. (\ref{eq:qcd-ew}) and the factorized combination as in eq. (\ref{eq:qcd-ew-factor}). 
 The numerical results presented in the following have been obtained using HORACE including exact NLO 
 EW corrections, but neglecting the contribution of higher-order effects due to  
 multiple 
 photon emission, since the phenomenological impact of these contributions has been already studied
 in detail, both at the Tevatron and LHC,  in a series of previous papers \cite{CMNV,CMNTW,CMNTZ,app}.  
 Moreover, results for the $W$ transverse 
 momentum will not be shown because of the strong sensitivity of this observable to non-perturbative 
 effects, that require a careful modeling and comparison with real data and are, therefore, beyond the scope of the present
 analysis.

\subsubsection{Observables for luminosity monitoring and PDF constraint}

After the successful MC tuning at the level of integrated cross sections discussed in Section \ref{mctuning}, we performed detailed comparisons between the predictions of QCD codes under
consideration for those observables of experimental interest to monitor the collider luminosity and
to constrain the PDFs. 

\begin{figure}[h]
\begin{center}
\includegraphics[height=5.5cm]{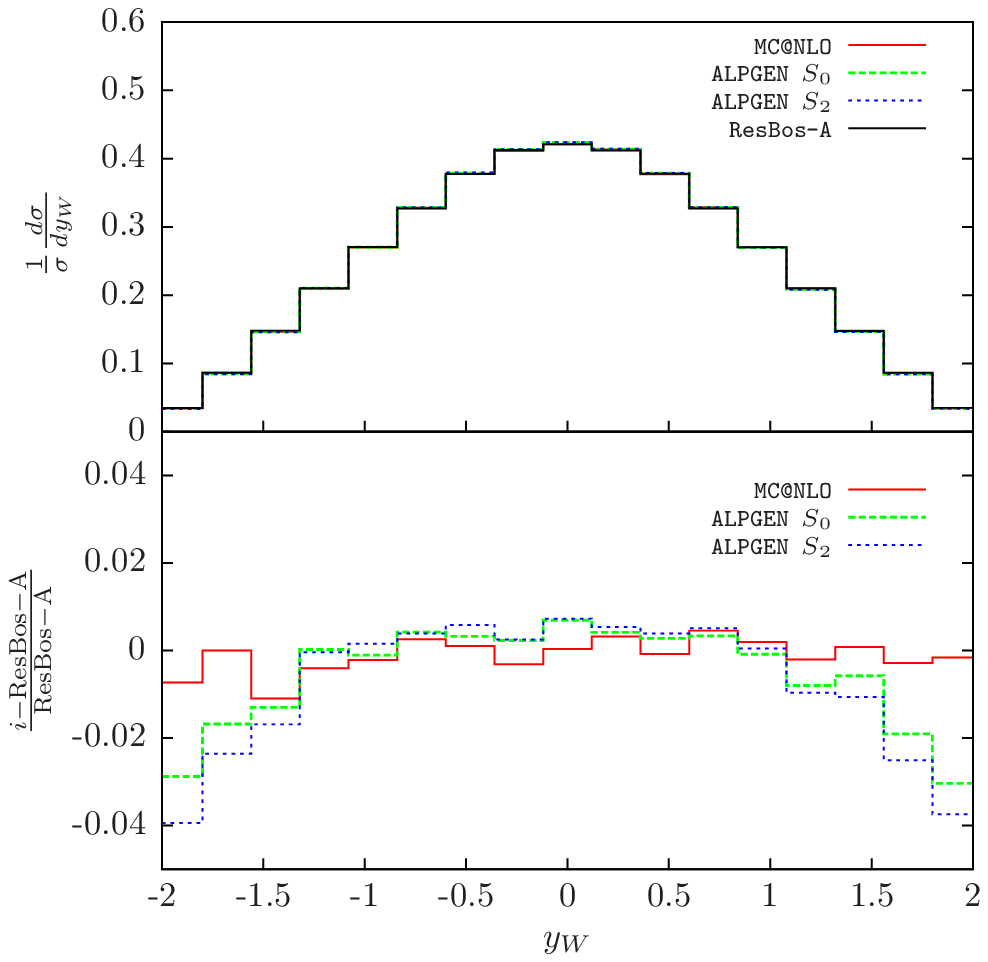}~\hskip 24pt\includegraphics[height=5.5cm]{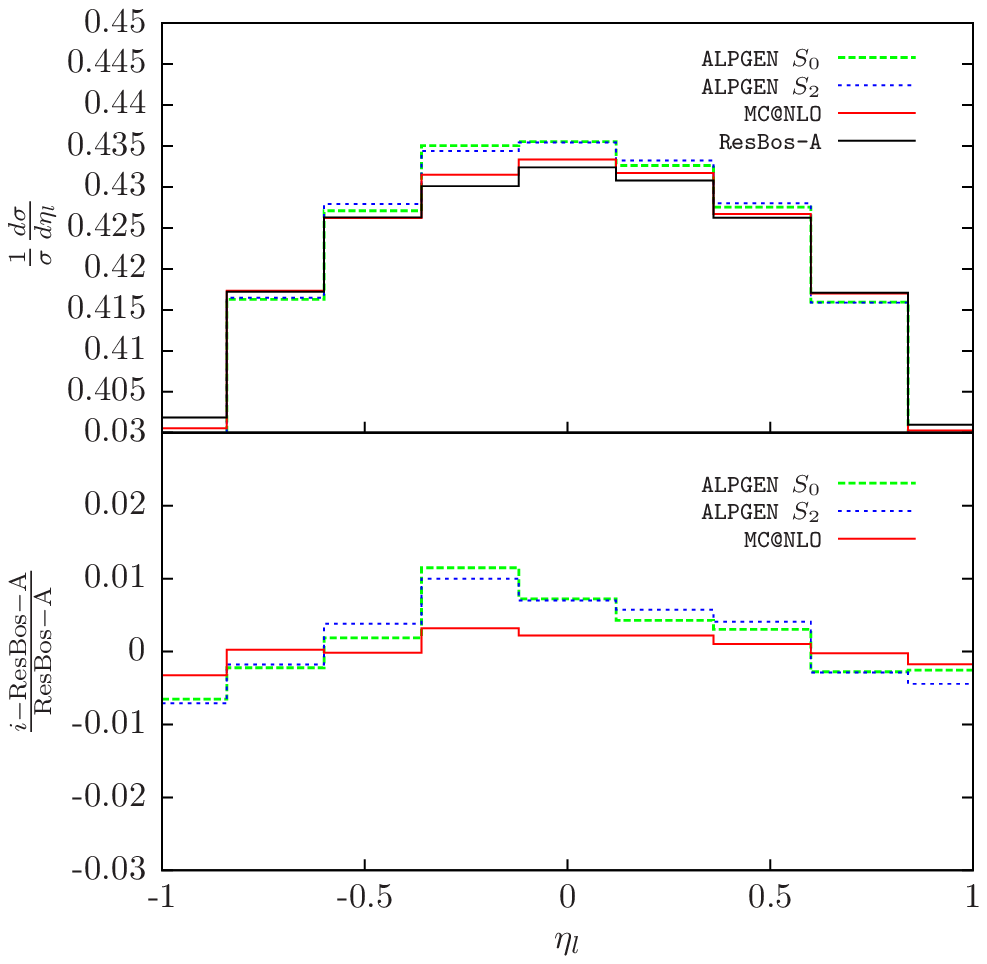}
\caption{$W$ rapidity (left plot) and muon pseudorapidity (right plot) distributions according to 
the QCD predictions of ALPGEN~S$_0$, ALPGEN~S$_2$, MC@NLO and ResBos-A. In the lower panels
the relative deviations of each code w.r.t. ResBos-A are shown.}
\label{yw-etal-tev}
\end{center}
\end{figure}

In Figure \ref{yw-etal-tev} we show the results obtained for the $W$ 
rapidity (left plot) and muon pseudorapidity (right plot) with 
ALPGEN~S$_0$, ALPGEN~S$_2$, MC@NLO and ResBos-A CSS.\footnote{As already stated for the integrated cross sections in Section \ref{xsect}, the results of ALPGEN~S$_1$  are not shown in  Figure \ref{yw-etal-tev} and in the next plots
for the Tevatron because they are practically indistinguishable from the predictions of  ALPGEN~S$_2$, 
as we explicitly checked.}
It is important to emphasize that in Figure \ref{yw-etal-tev}, and in the following  plots referring to 
distributions in the presence of QCD only, the results of all QCD programs have been normalized 
to the corresponding integrated cross section, 
in order to point out just the differences in the shape description. In the lower panels of 
 Figure \ref{yw-etal-tev} we show the relative deviations of each QCD tool w.r.t. ResBos-A, chosen as
 a benchmark because of its wide use at the Tevatron\footnote{More precisely, the code used at the Tevatron for the modeling of QCD in DY processes is ResBos. However, ResBos-A coincides with ResBos as far as QCD contributions are concerned. }. It can be seen that, in spite of the different theoretical
 ingredients, the predictions of the QCD programs agree at the  $\sim 1\%$ level almost in the whole shape.  
This can be easily understood because $y_W$ and $\eta_l$ are rather smooth distributions 
and, as such, quite insensitive to QCD shape differences, at least for the  ranges  
accessible at the Tevatron.

We also investigated the level of agreement between the QCD codes when considering 
their predictions for the $W$ charge asymmetry, which is an important quantity at the Tevatron
to derive information about the partonic contents of the proton. As can be seen from Figure 
\ref{wcharge}, the absolute differences between the various predictions reach at most the 1\% level, 
again as a consequence of the quite smooth behavior of this distribution.

\begin{figure}[h]
\begin{center}
\includegraphics[height=5.5cm]{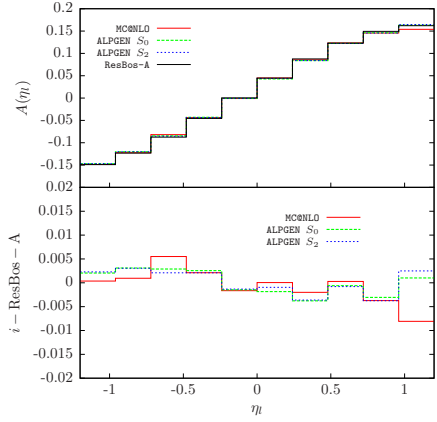}
\caption{The $W$ charge asymmetry according to 
the QCD predictions of ALPGEN~S$_0$, ALPGEN~S$_2$, MC@NLO and ResBos-A. In the lower panel
the absolute deviations of each code w.r.t. ResBos-A are shown.}
\label{wcharge}
\end{center}
\end{figure}

A first illustration of the combination of EW and QCD corrections at the Tevatron is 
shown in Figure \ref{yw-etal-tev-ewqcd} for the $y_W$ and $\eta_l$ distributions. 
The upper 
panels show the {\it absolute} predictions obtained by means of the codes ALPGEN~$S_0$, MC@NLO and according
to eq. (\ref{eq:qcd-ew}), when using MC@NLO\footnote{Instead of adopting MC@NLO, the same 
results could be obtained using ALPGEN with an overall $K$-factor (defined as 
$\sigma_{NLO}/\sigma_{LO}$), since the shape predictions of ALPGEN and MC@NLO are in 
good agreement, as previously shown.} for the simulation of QCD effects in association
with HORACE convoluted with HERWIG Parton Shower. In the lower panels, 
the relative effects in units of ALPGEN~$S_0$ due to QCD
(MC@NLO), EW ($\rm HORACE_{HERWIG}$) and the combination of electroweak and
QCD corrections (eq. (\ref{eq:qcd-ew})) are shown, as obtained by appropriate combinations of the absolute predictions shown in the upper panel. 
\begin{figure}[h]
\begin{center}
\includegraphics[height=5.5cm]{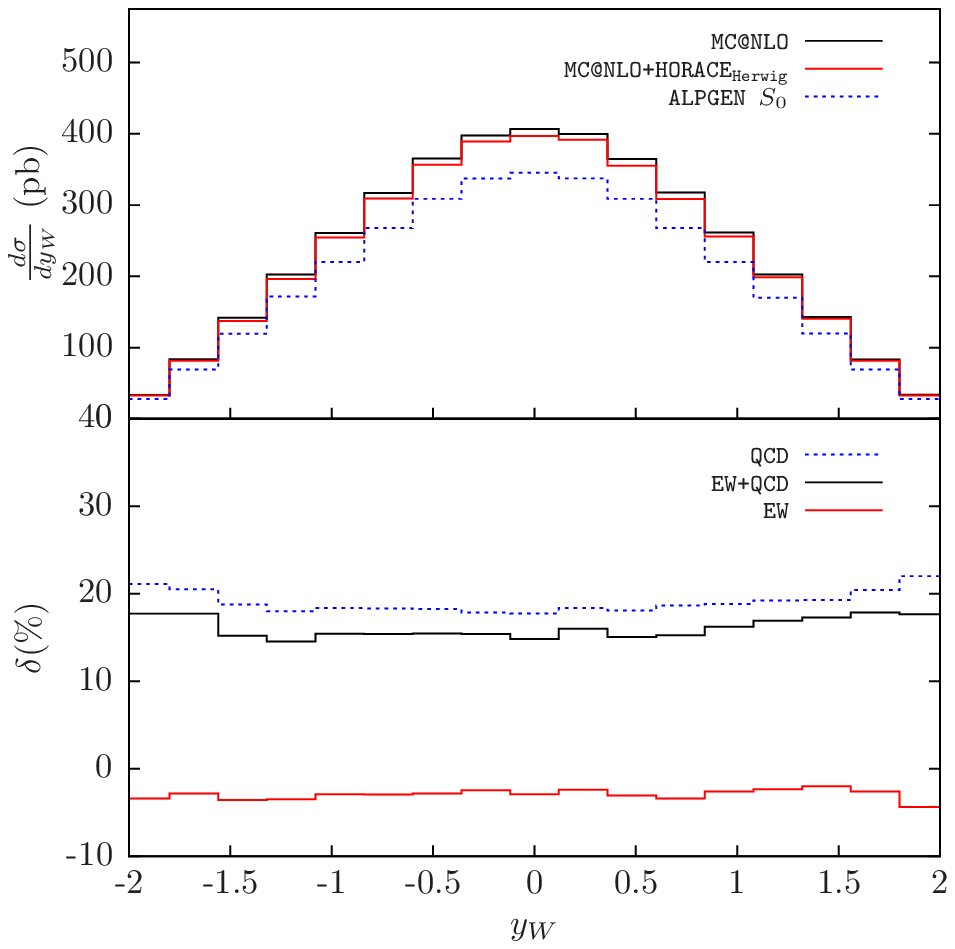}~\hskip 24pt\includegraphics[height=5.5cm]{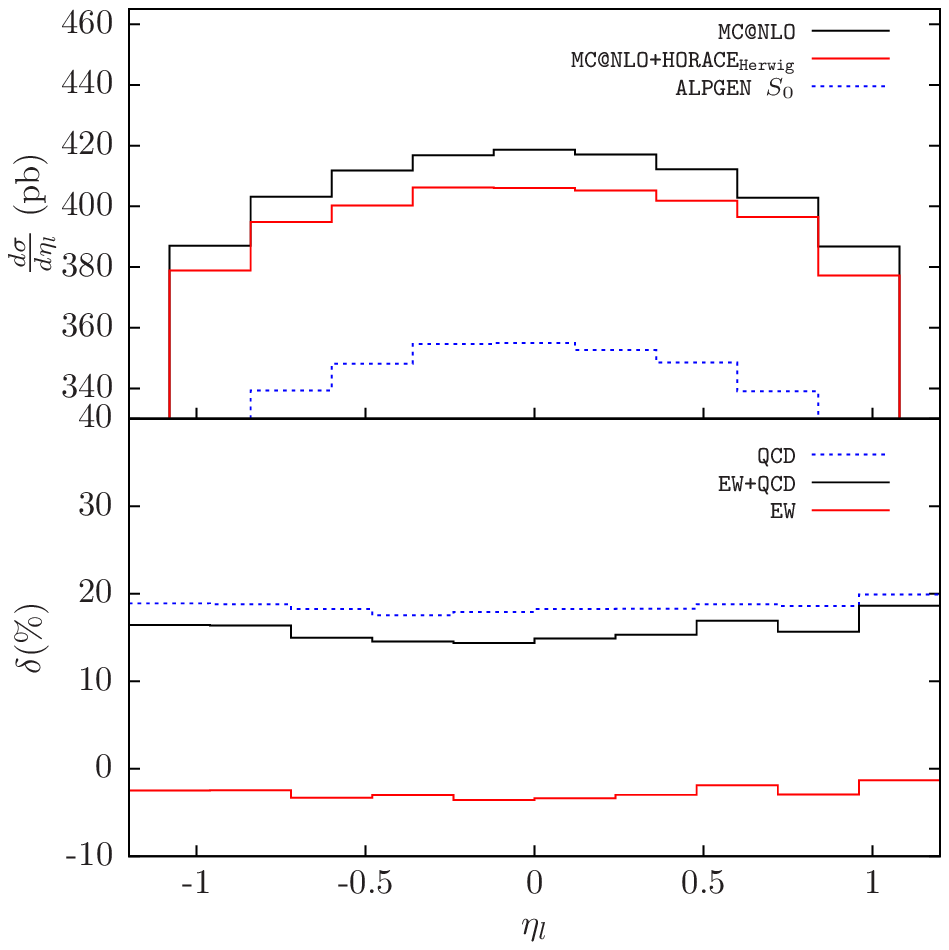}
\caption{
$W$ rapidity (left plot) and muon pseudorapidity (right plot) distributions according to 
MC@NLO, ALPGEN~$S_0$ and the additive EW+QCD combination.  In the lower panels
the relative effects due to QCD, EW and EW+QCD corrections are shown in units of ALPGEN~$S_0$. 
}
\label{yw-etal-tev-ewqcd}
\end{center}
\end{figure}
Strictly speaking, the results referring to 
MC@NLO+HORACE$_{\rm HERWIG}$ and denoted in the plots as EW correspond to the second term of 
eq. (\ref{eq:qcd-ew}), in units of  the predictions of  ALPGEN~$S_0$.  
It can be seen that EW corrections amount to a few per cent negative contribution, QCD NLO effects are of the order of 20\%, resulting in a total combined correction of about 15\%. 
We also studied the EW and QCD corrections  to the W charge asymmetry and observed that they are
 of comparable size and tend to sum up, yielding a total correction at
 the
 1\% level on the average. We also performed a comparison with ResBos-A, to notice that the predictions of our EW/QCD recipe are in agreement with those of
 ResBos-A within about 1\% in the whole $\eta_l$ range.

\begin{figure}[h]
\begin{center}
\includegraphics[height=5.5cm]{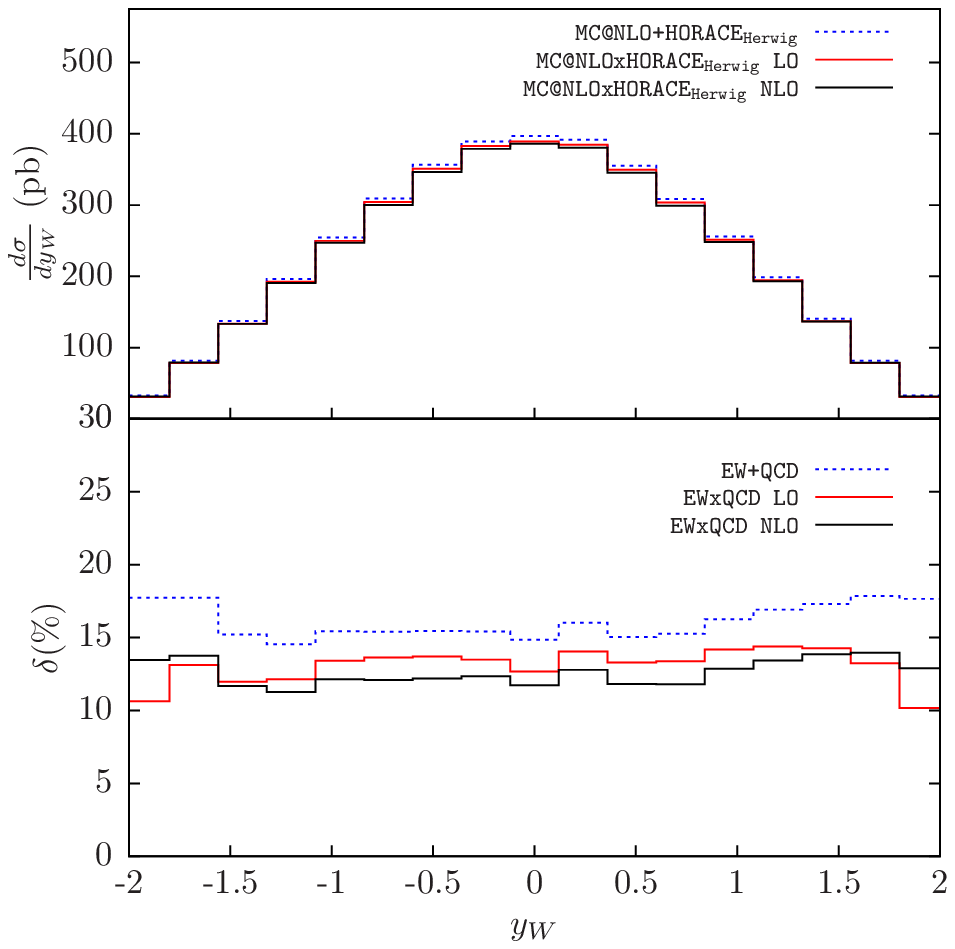}~\hskip 24pt\includegraphics[height=5.5cm]{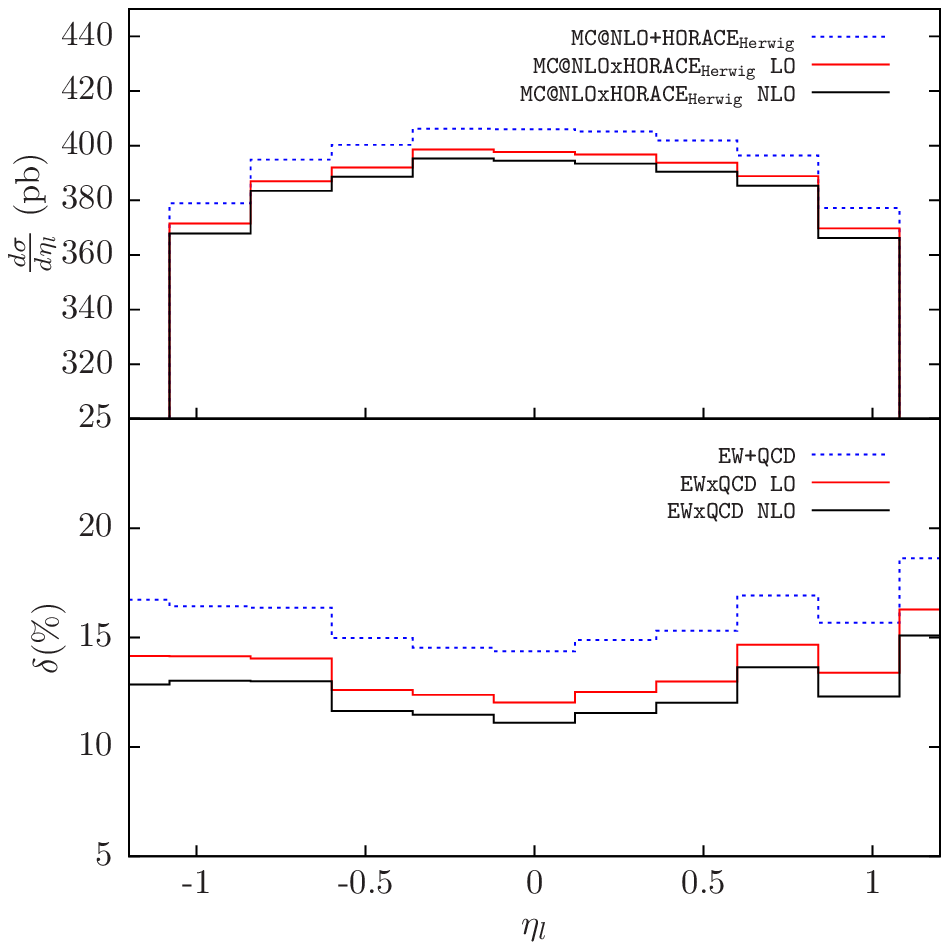}
\caption{
$W$ rapidity (left plot) and muon pseudorapidity (right plot) distributions according to 
the additive EW+QCD combination and the two EW$\otimes$QCD factorized prescriptions.  In the lower panels
the relative effects due to the different combinations are shown in units of ALPGEN~$S_0$. 
}
\label{yw-etal-tev-ewqcd-fvsa}
\end{center}
\end{figure}

Figure~\ref{yw-etal-tev-ewqcd-fvsa} shows the comparison between the additive formula of eq.~(\ref{eq:qcd-ew}) and the factorized prescriptions of (\ref{eq:qcd-ew-factor}), as normalized to the LO hadron level distributions convoluted with HERWIG PS.  The predictions lie in a few per cent band, the factorized prescriptions being quite close to one another. The relative difference between the additive and factorized recipes can be seen as an estimate of the uncertainty due to ${\cal O}(\alpha_s^2)$ and ${\cal O}(\alpha \alpha_s)$ corrections.

\begin{figure}[h]
\begin{center}
\includegraphics[height=5.5cm]{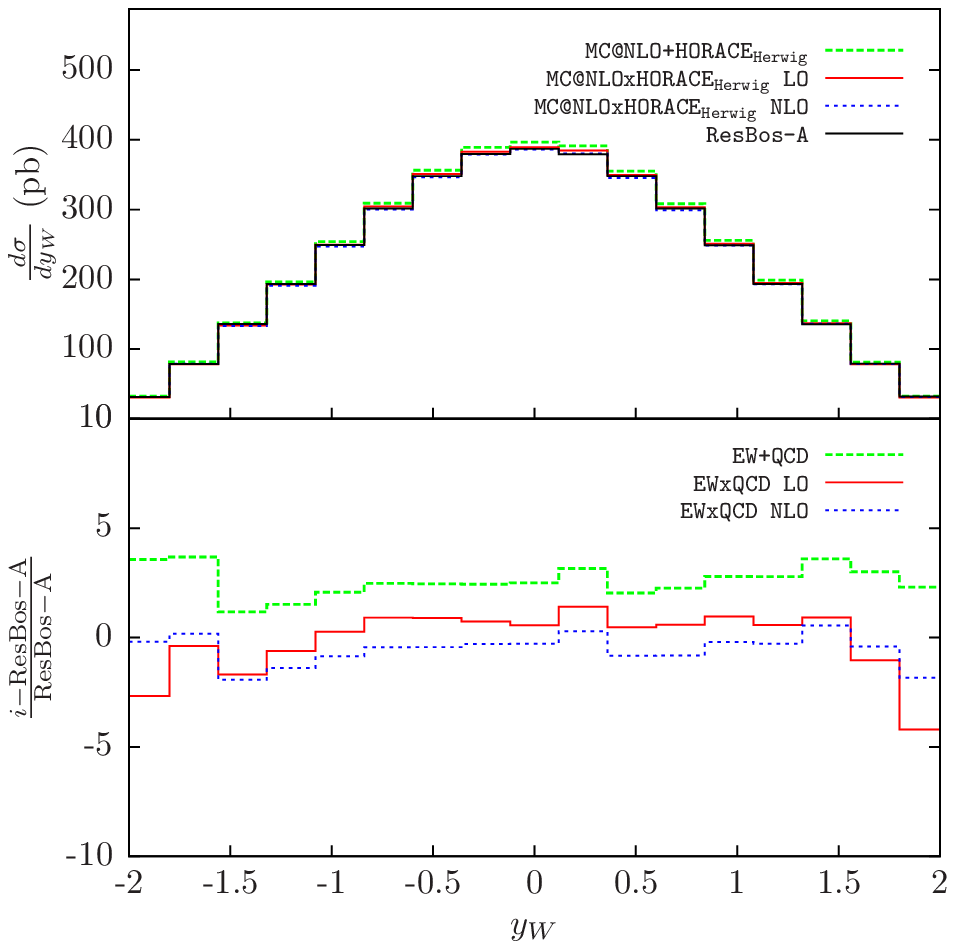}~\hskip 24pt\includegraphics[height=5.5cm]{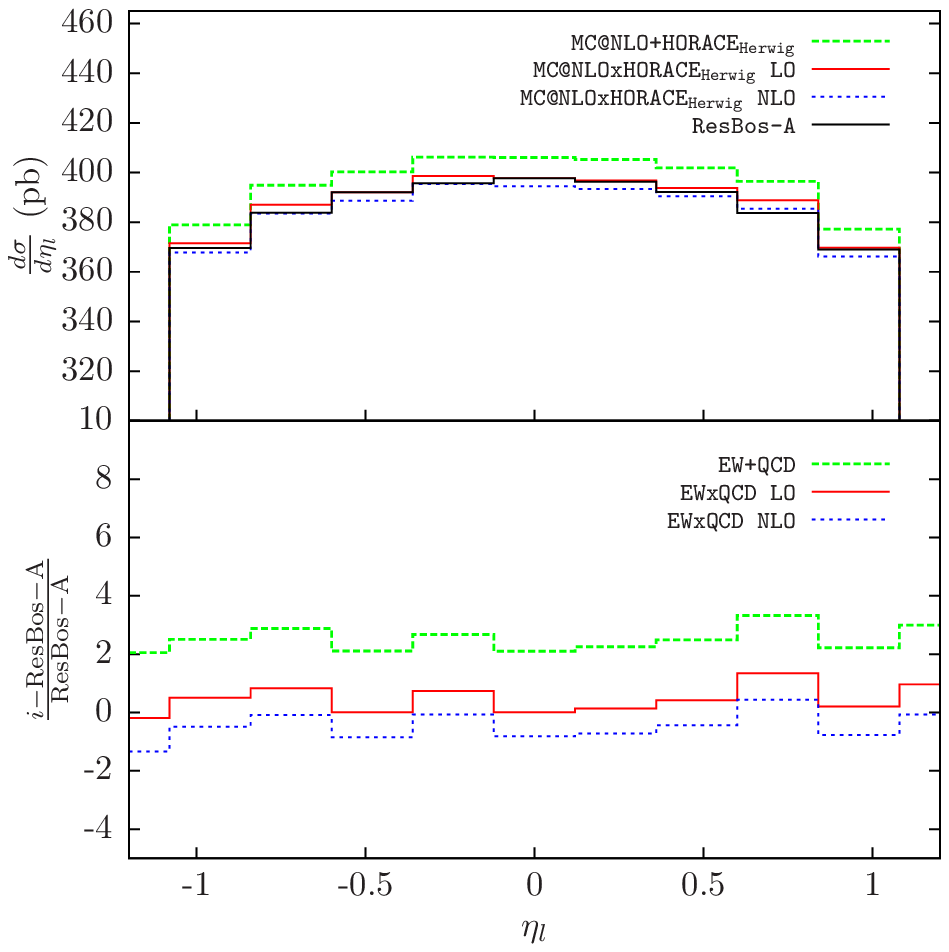}
\caption{
$W$ rapidity (left plot) and muon pseudorapidity (right plot) distributions according to 
the additive EW+QCD combination, the two EW$\otimes$QCD factorized prescriptions and ResBos-A.  In the lower panels
the relative differences  of the various combinations w.r.t. ResBos-A are shown. 
}
\label{yw-etal-tev-ewqcd-fvsa-comp}
\end{center}
\end{figure}

In figure~\ref{yw-etal-tev-ewqcd-fvsa-comp} the absolute predictions of eqs.~(\ref{eq:qcd-ew}) and (\ref{eq:qcd-ew-factor}) are compared with those of ResBos-A in the upper panel. As can be seen in the lower panels, the relative difference does not exceed the few per cent level, the agreement being better for the factorized prescriptions, consistently with the fact that ResBos-A adopts a factorized formulation as well.  

\subsubsection{Observables for $W$ precision physics}

\begin{figure}[h]
\begin{center}
\includegraphics[height=5.5cm]{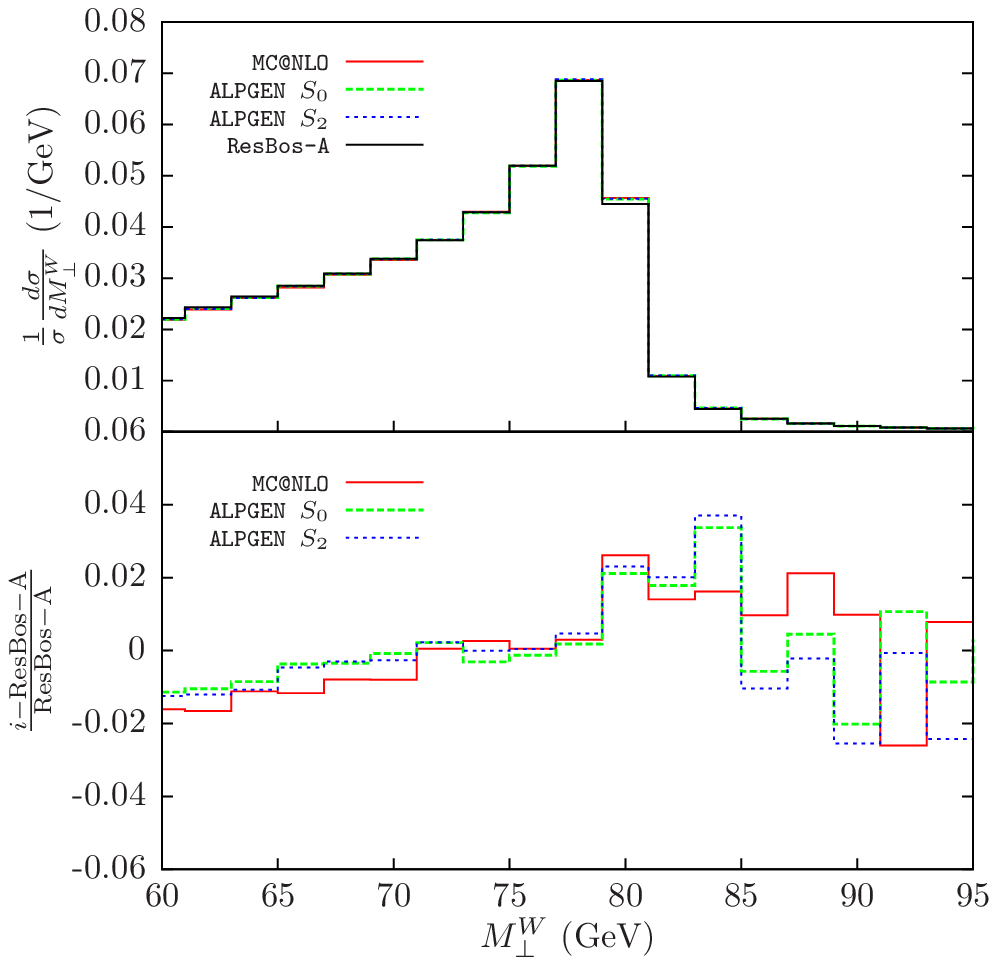}~\hskip 24pt\includegraphics[height=5.5cm]{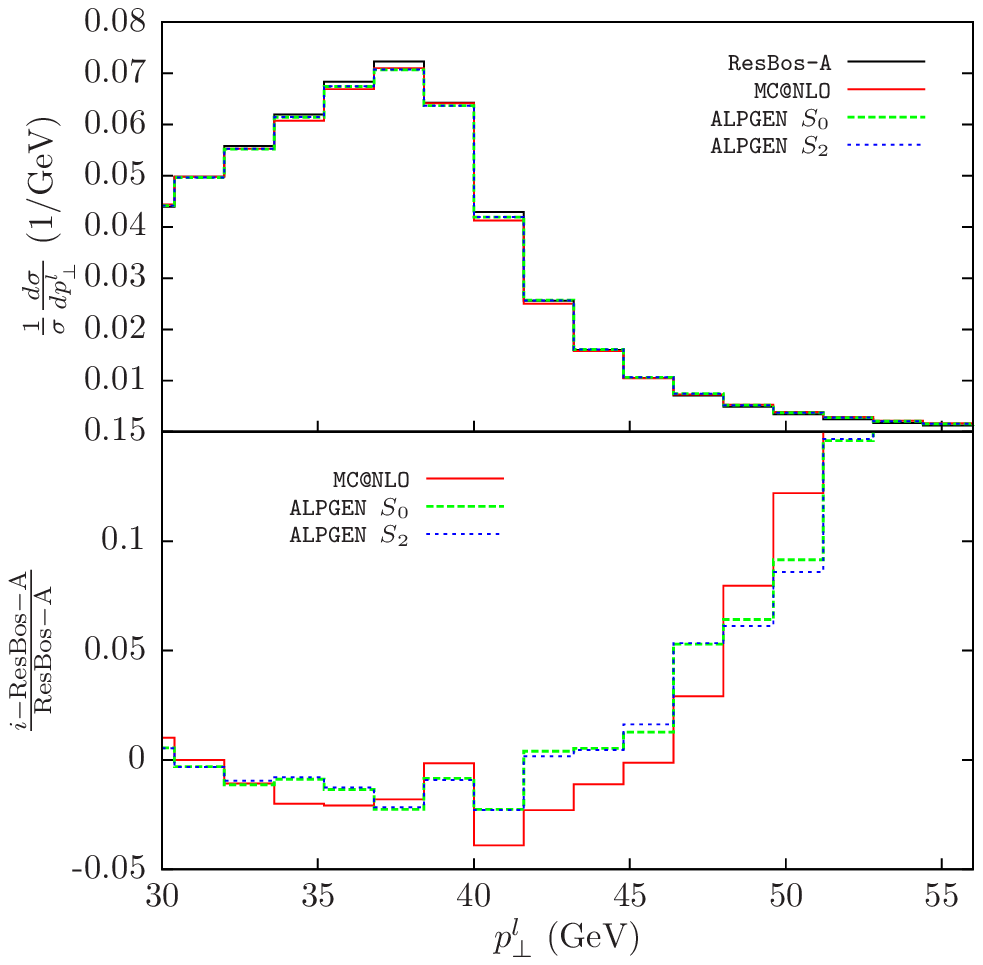}
\caption{The same as Figure 6 for the $W$ transverse mass (left plot) and muon transverse momentum (right plot) distributions.}
\label{mtw-ptl-tev}
\end{center}
\end{figure}

The status of the QCD predictions for the observables relevant for precision measurements of the
$W$-boson mass is presented in Figure \ref{mtw-ptl-tev}, showing the $W$ transverse mass (left 
plot) and the muon transverse momentum (right plot), according to the results of
ALPGEN~S$_0$, ALPGEN~S$_2$, MC@NLO and ResBos-A, as 
in Figure \ref{yw-etal-tev} and Figure \ref{wcharge}. For such strongly varying distributions,
it can be noticed that the predictions of the QCD programs differ at some per cent level around the
jacobian peak, which is  the crucial region for $M_W$ extraction. The relative differences can 
reach the 5-10\% level in the distributions tails. Presumably, the discrepancies around 50~GeV for $M_\perp^W$ and 25~GeV for $p_\perp^l$ have to be ascribed to 
soft-gluon resummation, which is accurately described in ResBos-A but not taken into account at the same precision level 
in the other codes. On the other hand, the differences observed 
around 90~GeV for $M_\perp^W$ and 50~GeV of $p_\perp^l$ are probably due to hard collinear PS effects, which are
incorporated in ALPGEN and in MC@NLO but are absent in ResBos-A. Actually, one can notice that
in this region the predictions of ALPGEN~S$_2$  and MC@NLO well agree with the
pure PS approximation of ALPGEN~S$_0$, supporting the above interpretation.

\begin{figure}[h]
\begin{center}
\includegraphics[height=5.5cm]{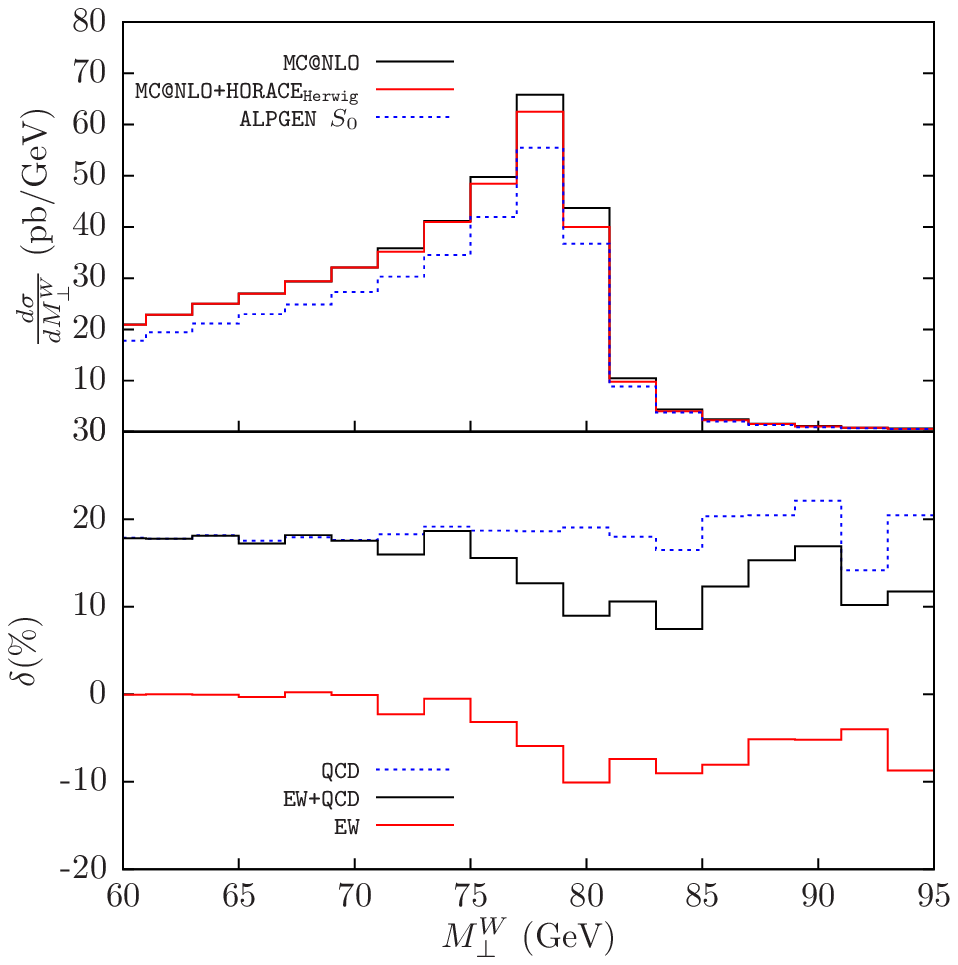}~\hskip 24pt\includegraphics[height=5.5cm]{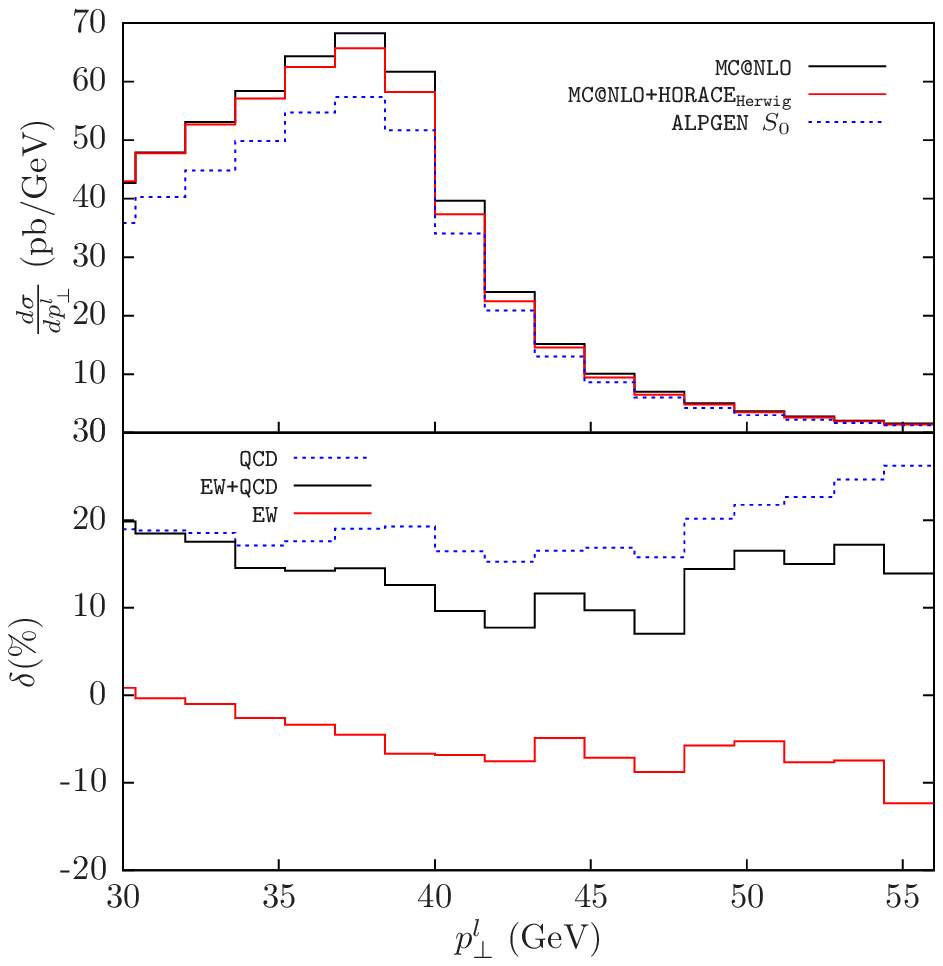}
\caption{$W$ transverse mass (left plot) and muon transverse momentum (right plot) distributions according to 
MC@NLO, ALPGEN~$S_0$ and the additive EW+QCD combination.  In the lower panels
the relative effects due to QCD, EW and EW+QCD corrections are shown in units of ALPGEN~$S_0$. }
\label{mtw-ptl-tev-ewqcd}
\end{center}
\end{figure}
A further example of the combination of EW and QCD corrections  at the Tevatron is 
shown in Figure \ref{mtw-ptl-tev-ewqcd} for the $M_\perp^W$ and $p_\perp^l$ distributions. 
Similarly to Figure \ref{yw-etal-tev-ewqcd}, the upper 
panels show the {\it absolute} predictions of 
ALPGEN~$S_0$, MC@NLO and according
to eq. (\ref{eq:qcd-ew}), 
 while the lower panels illustrate the relative effects due to QCD and EW 
corrections only and to the combination of electroweak and strong contributions. As can be observed, in the most interesting region around the jacobian peak the NLO corrections are partially compensated by the EW contributions, yielding a total correction of about 10-15\%. It is worth noticing that the convolution of EW corrections with the QCD shower changes the shape of EW corrections themselves, smearing them w.r.t. a pure parton-level calculation. 

\begin{figure}[h]
\begin{center}
\includegraphics[height=5.5cm]{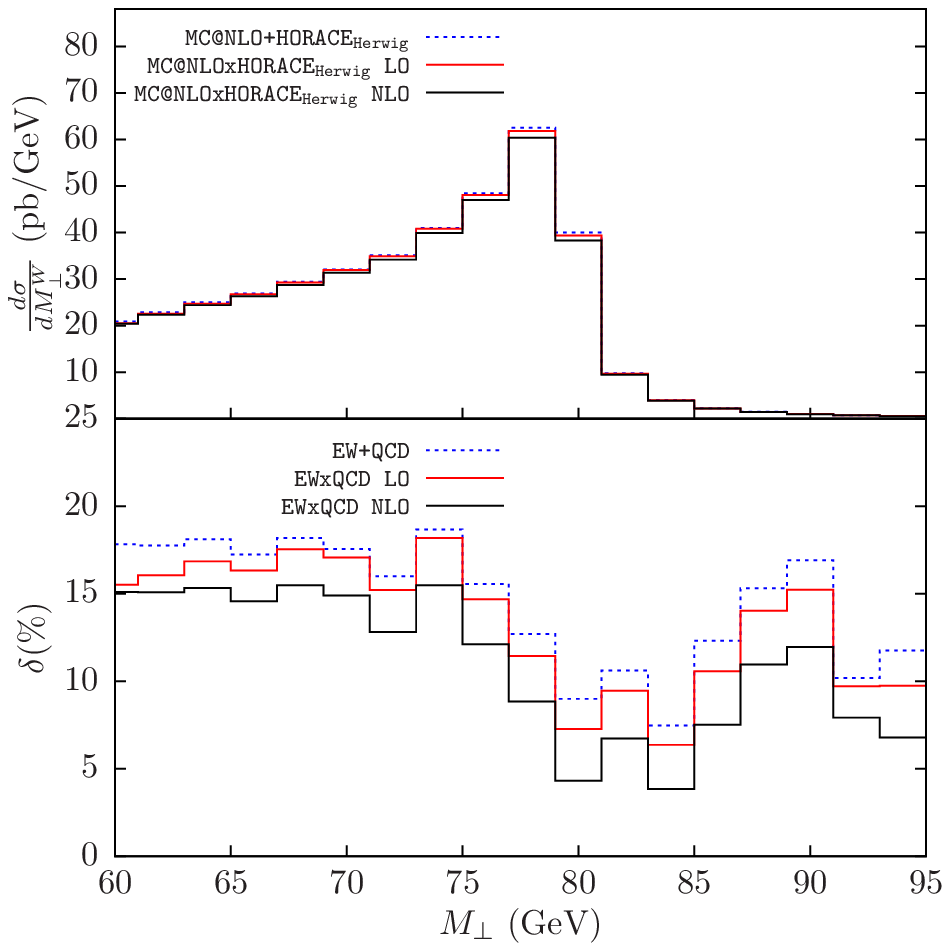}~\hskip 24pt\includegraphics[height=5.5cm]{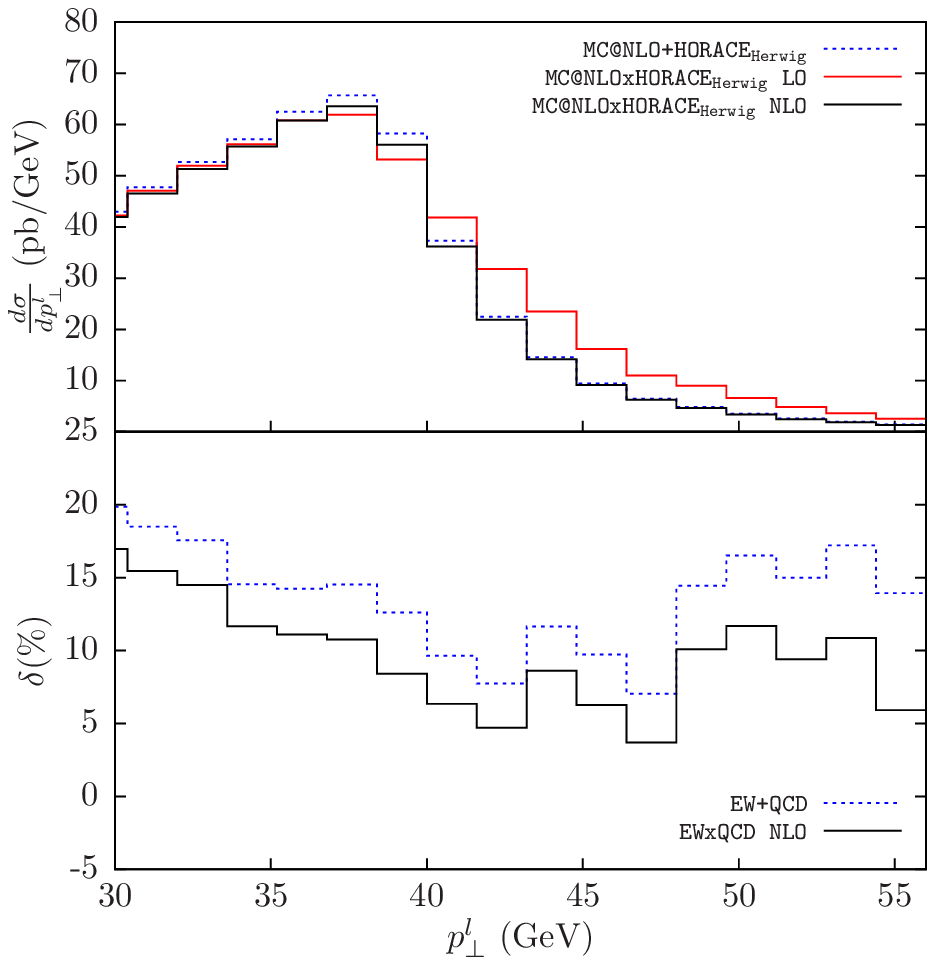}
\caption{
$W$ transverse mass (left plot) and muon transverse momentum (right plot) distributions according to 
the additive EW+QCD combination and the two EW$\otimes$QCD factorized prescriptions.  In the lower panels
the relative effects due to the different combinations are shown in units of ALPGEN~$S_0$. 
}
\label{mtw-ptl-tev-ewqcd-fvsa}
\end{center}
\end{figure}
Figure~\ref{mtw-ptl-tev-ewqcd-fvsa} shows the comparison between the additive formula of eq.~(\ref{eq:qcd-ew}) and the factorized prescriptions of (\ref{eq:qcd-ew-factor}), as normalized to the LO hadron level distributions convoluted with HERWIG PS.  Some comments are in order here. With respect to the case, already discussed, of $y_W$ and $\eta_l$, the differences between the two factorized prescriptions are sizable.  In particular, in the case of $p_\perp^l$, the factorized formula with LO normalization leads to pathological results. This can be understood as due to the fact that NLO corrections are dominated, in the hard $p_\perp$ tail, by $2 \to 3$ QCD subprocesses not included in a LO calculation. For this kind of observables, the NLO calculation is the first perturbative order able to cover all the relevant phase space regions; hence it is the ``lowest order'' approximation w.r.t. which evaluate the size of radiative corrections.

Figure~\ref{mtw-ptl-tev-ewqcd-fvsa-comp} shows the comparison between the additive and factorized formulae on the one hand and the ResBos-A prediction on the other one. 
\begin{figure}[h]
\begin{center}
\includegraphics[height=5.5cm]{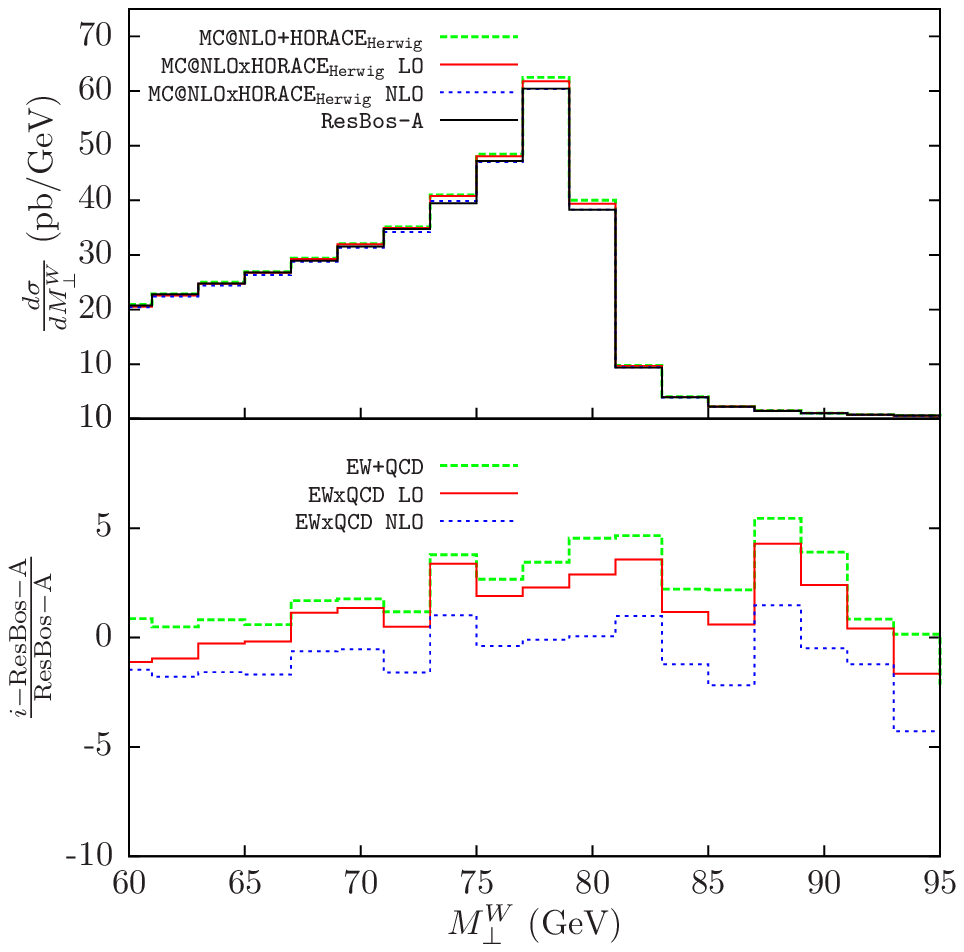}~\hskip 24pt\includegraphics[height=5.5cm]{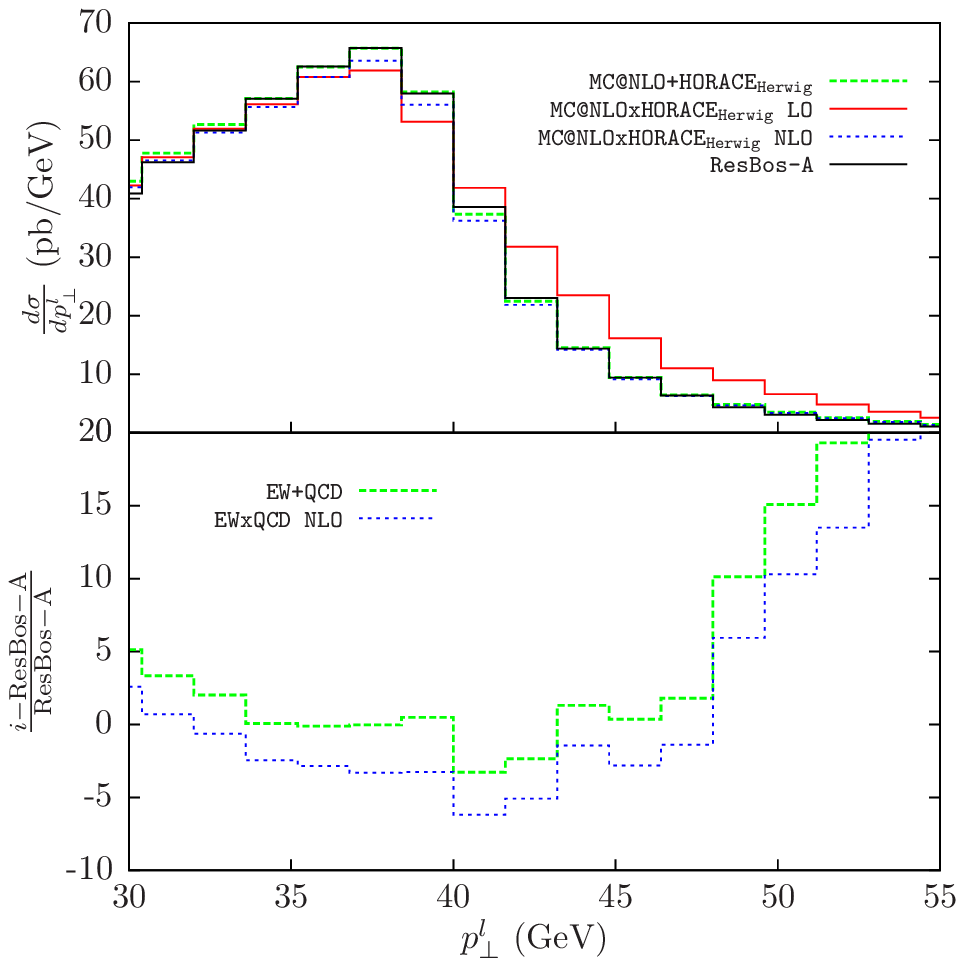}
\caption{
$W$ transverse mass (left plot) and muon transverse momentum (right plot) distributions according to 
the additive EW+QCD combination, the two EW$\otimes$QCD factorized prescriptions and ResBos-A.  In the lower panels
the relative differences of the various combinations w.r.t. ResBos-A are shown. 
}
\label{mtw-ptl-tev-ewqcd-fvsa-comp}
\end{center}
\end{figure}
From 
the comparison with Figure~\ref{mtw-ptl-tev} referring to QCD shape differences only, it can be
seen that the trend of the relative deviations  is very similar to that
observed between the results of MC@NLO and ResBos-A, while the size of the differences
increases of some per cents in the tails, as a consequence of the inclusion of the 
normalization in the codes predictions. It can also be noticed that the differences between 
the two approaches are largely dominated by QCD effects.  Therefore, as already remarked about Figure \ref{mtw-ptl-tev}, the differences below the peaks are probably due to soft-gluon resummation effects. On the other hand, the discrepancies present above the peaks 
have to be presumably ascribed to hard collinear PS contributions taken into account in 
our calculation but not implemented in ResBos-A. To better understand the origin of
these differences, it would be interesting to perform comparisons with the predictions of
ResBos-A when taking into account the effect of finite order QCD contributions (the so-called $Y$ terms)  in its 
formulation. Unfortunately, no public grids are available for such theoretical configuration. Concerning the EW 
contributions, the two programs essentially predict  the same quantitative effects. This is not
unexpected because EW corrections in the vicinity of the $W$ peak are 
largely dominated by final-state QED radiation, as widely discussed in the literature, and therefore
the inclusion of exact NLO corrections in HORACE versus the approximation limited to the inclusion of NLO final-state QED corrections in ResBos-A translates into differences at the 0.1\% level, as
we explicitly checked. As a whole, it is worth stressing that closely around 
 the jacobian peak the additive recipe and ResBos-A show differences of a few per cent, 
for both $M_\perp^W$ and $p_\perp^l$, whereas the factorized NLO prescription exhibits smaller deviations in the $M_\perp^W$ case. 
Such  differences could be relevant in view of the
future measurements of the $W$ mass with $\sim 20$~MeV precision at the Tevatron and 
should be further scrutinized on the experimental side. In particular, the combination of MC@NLO with
HORACE, as adopted in this study, could be used to cross-check the precision measurement of 
the $W$-boson mass as currently obtained in terms of the  generators ResBos (for QCD 
effects) and HORACE, PHOTOS~\cite{GW} and  WGRAD (for EW corrections).

Results for photon-induced processes at the Tevatron are not shown since they are negligible because of the very small photon 
content inside the proton for the typical $x$ values probed by DY 
kinematics at the Tevatron.

\subsection{Numerical results for the LHC}
\label{lhc-results}

In this section, we present the results obtained for the process $p p \to W^\pm \to \mu^\pm + X$ at the
LHC  ($\sqrt{s}$ = 14~TeV), when imposing the cuts quoted 
in Table~\ref{tab:lhc}. As for the Tevatron analysis, we begin with a discussion
of QCD effects on the relevant observables (considering shape predictions only), to continue with the 
study of the  combination of EW and QCD corrections. Results 
taking into account higher-order leading log QED corrections and predictions for the 
$W$ transverse momentum observable will not be given in the following, for the same reasons already emphasized 
at the beginning of Section \ref{tev-results}.

\subsubsection{Observables for luminosity monitoring and PDF constraint}

\begin{figure}[h]
\begin{center}
\includegraphics[height=5.5cm]{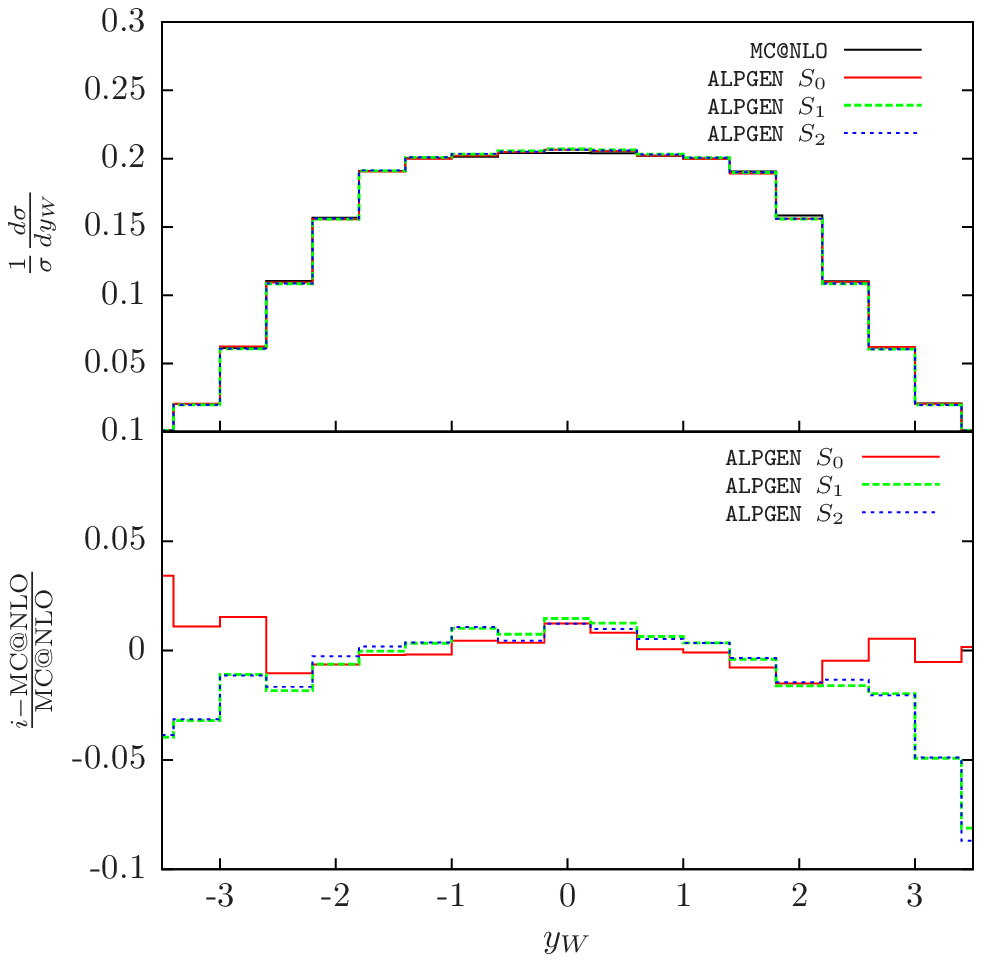}~\hskip 24pt\includegraphics[height=5.5cm]{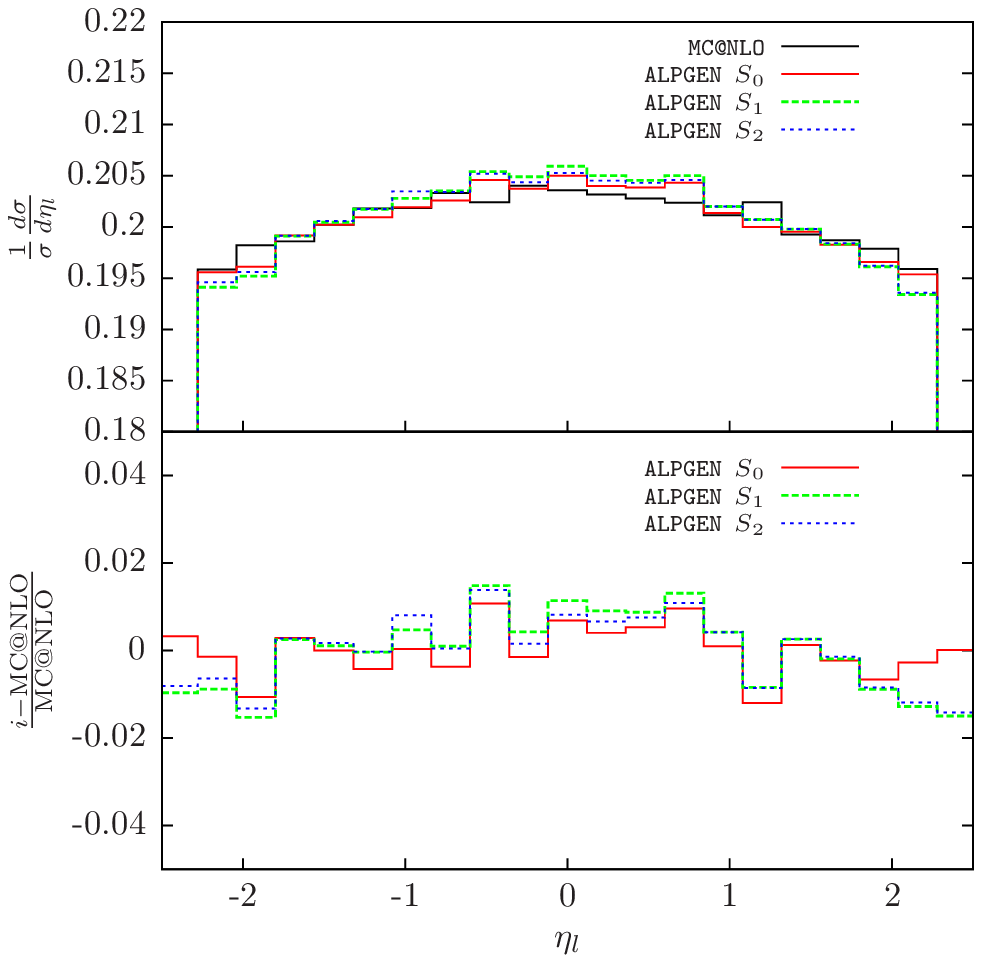}
\end{center}
\caption{$W$ rapidity (left plot) and muon pseudorapidity (right plot) distributions according to 
the QCD predictions of ALPGEN~S$_0$, ALPGEN~S$_1$, ALPGEN~S$_2$  and MC@NLO. In the lower panels
the relative deviations of each code w.r.t. MC@NLO are shown.}
\label{yw-etal-lhc}
\end{figure}
In Figure \ref{yw-etal-lhc} we show the results obtained for the $W$ 
rapidity (left plot) and muon pseudorapidity (right plot) with the QCD generators 
ALPGEN~S$_0$, S$_1$ and S$_2$, and MC@NLO, considering set up a. in 
Table~\ref{tab:lhc}. In the lower panels  we show the relative deviations of the ALPGEN variants w.r.t. 
 MC@NLO. It can be noticed that, while for $\eta_l$ the relative differences between
 ALPGEN and MC@NLO are at a few per cent level in the whole range, for the $W$ rapidity 
 the agreement in the shape predicted by the two generators is quite satisfactory 
 (at the 1\% level) in the central region, but it deteriorates in the very forward and very backward 
 limits, that anyway marginally contribute to the integrated cross section. In particular, in these regions, the pure PS predictions
 represented by ALPGEN~S$_0$ indicate a quite smooth excess of events w.r.t. MC@NLO, while the
 (practically indistinguishable) deviations of ALPGEN~S$_1$  and of ALPGEN~S$_2$  w.r.t. MC@NLO
 have a more pronounced opposite trend. 

\begin{figure}[h]
\begin{center}
\includegraphics[height=5.5cm]{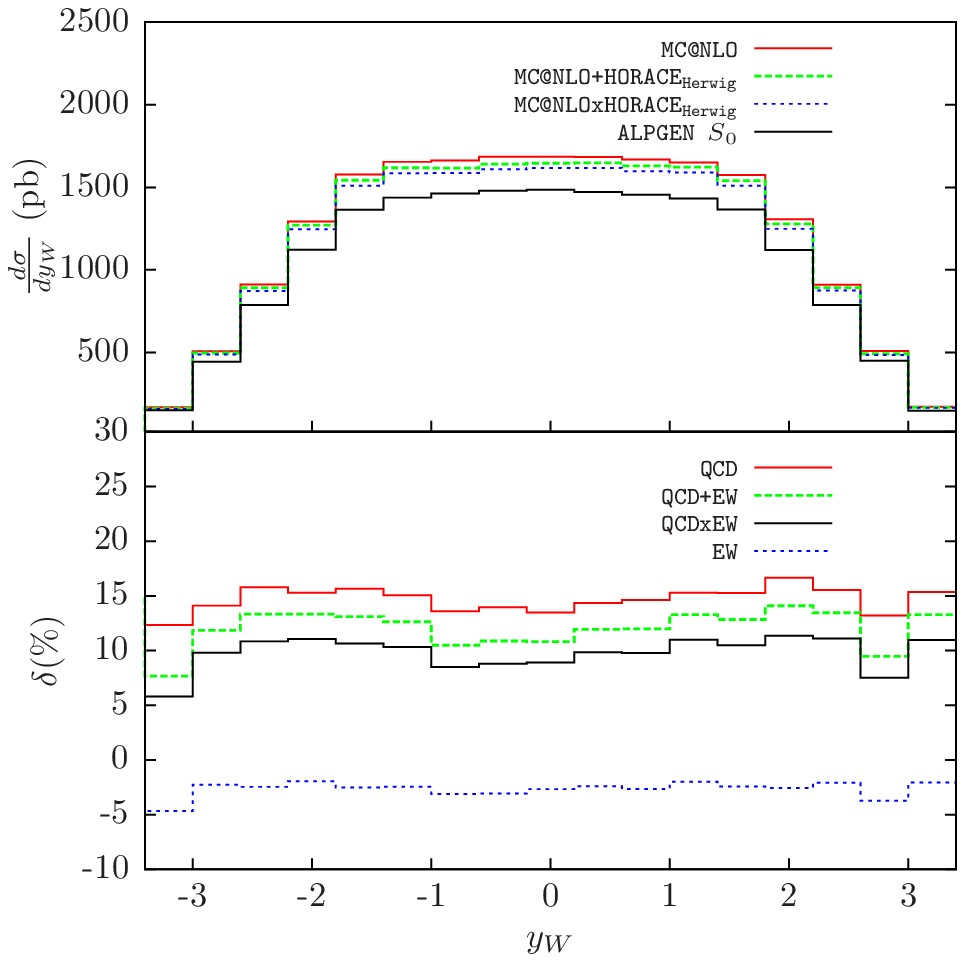}~\hskip 24pt\includegraphics[height=5.5cm]{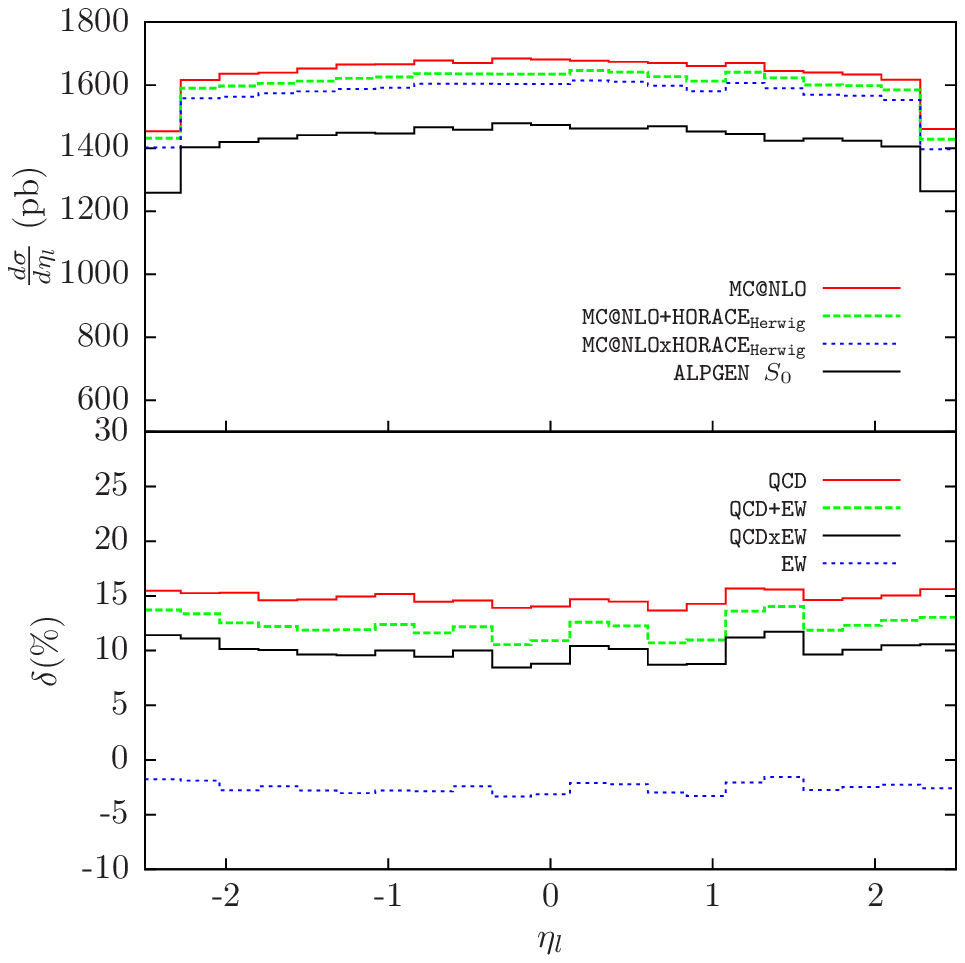}
\end{center}
\caption{$W$ rapidity (left plot) and muon pseudorapidity (right plot) distributions according to 
the predictions of MC@NLO, ALPGEN~$S_0$, MC@NLO+HORACE and  MC@NLO$\otimes$HORACE. In the lower panels 
the relative effects due to QCD, EW and combined  corrections are shown in units of ALPGEN~$S_0$.}
\label{yw-etal-lhc-ewqcd}
\end{figure}

The interplay between EW and QCD corrections at the LHC is shown
in Figure~\ref{yw-etal-lhc-ewqcd} for the $y_W$ and $\eta_l$ distributions. The upper 
panels show the {\it absolute} predictions of MC@NLO, ALPGEN~$S_0$, 
MC@NLO+HORACE (additive combination) and MC@NLO$\otimes$HORACE (factorized combination) interfaced to 
HERWIG PS. In the lower panels, the relative effects due to the combination of EW and QCD corrections, as well as for the EW and QCD contributions 
separately, are shown. The size of NLO EW and QCD corrections is
almost the same as the one seen for the Tevatron in Figure \ref{yw-etal-tev-ewqcd}. In particular, the negative 
NLO EW effects partially cancel the positive contribution due to NLO QCD corrections, resulting into
combined EW and QCD corrections at the 10\% level, for both  $y_W$ and $\eta_l$ and for both additive and factorized prescriptions. Hence, the 
interplay between EW and QCD contributions is crucial for precise simulations of the
observables relevant for luminosity monitoring and PDF constraint at the LHC.

\begin{figure}[h]
\begin{center}
\includegraphics[height=5.5cm]{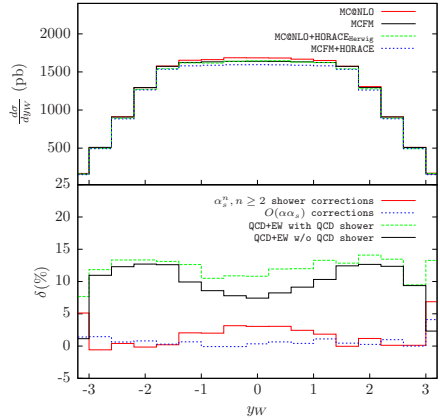} 
\end{center}
\caption{$W$ rapidity  distribution according to 
the predictions of MCFM, MC@NLO,  MCFM+HORACE and  MC@NLO+HORACE$_{\rm HERWIG}$. In the lower panel  
the relative effects due to $O(\alpha_s^2)$ QCD corrections, $O(\alpha \alpha_s)$ mixed corrections and QCD+EW corrections with and without QCD showering. }
\label{yw-lhca-aas}
\end{figure}

As remarked in Sect. 2.4, both the additive and factorized combination of EW
and QCD corrections contain, beyond the very same $O(\alpha)$ and $O(\alpha_s)$ 
content, higher-order contributions dominated by $O(\alpha_s^2)$ and mixed 
$O(\alpha \alpha_s)$ corrections. It is therefore interesting to analyze how
these higher-order corrections compare to each other. Since such higher-order 
effects originate, in both the additive and factorized recipe, from the QCD PS
and the convolution of the QCD PS with the NLO predictions, a simple
strategy to disentangle the relative effect of $O(\alpha_s^2)$ and $O(\alpha \alpha_s)$ 
corrections is to compare the predictions of the combination of pure NLO codes 
(lacking, by construction, the contribution of the QCD shower evolution) with those
of one of the two recipes described in Sect. 2.4 or their variants. An example of such
a study is given for $y_W$ in Fig.~\ref{yw-lhca-aas}, where, for definiteness, the results of the additive
recipe MC@NLO + HORACE$_{\rm HERWIG}$ are compared with those of the pure NLO 
combination MCFM + HORACE, as well as with the predictions of the "intermediate"
theoretical formulations given by MC@NLO + HORACE (missing mixed $O(\alpha \alpha_s)$
corrections) and MCFM + HORACE$_{\rm HERWIG}$ (missing $O(\alpha_s^2)$ shower effects).
As it can be seen from the lower panel, the fixed NLO predictions of MCFM + HORACE 
differ from those of MC@NLO + HORACE$_{\rm HERWIG}$ at the some per cent level, showing 
the overall importance of QCD shower and mixed EW-QCD corrections for a precise 
simulation of the $y_W$ distribution. It can be also noticed that $O(\alpha \alpha_s)$
corrections are very small,  while QCD 
shower corrections amount to a few per cent, thus dominating the EW-QCD combination
for this observable.

\subsubsection{Observables for $W$ precision physics}

\begin{figure}[h]
\begin{center}
\hskip 8pt~\includegraphics[height=5.5cm]{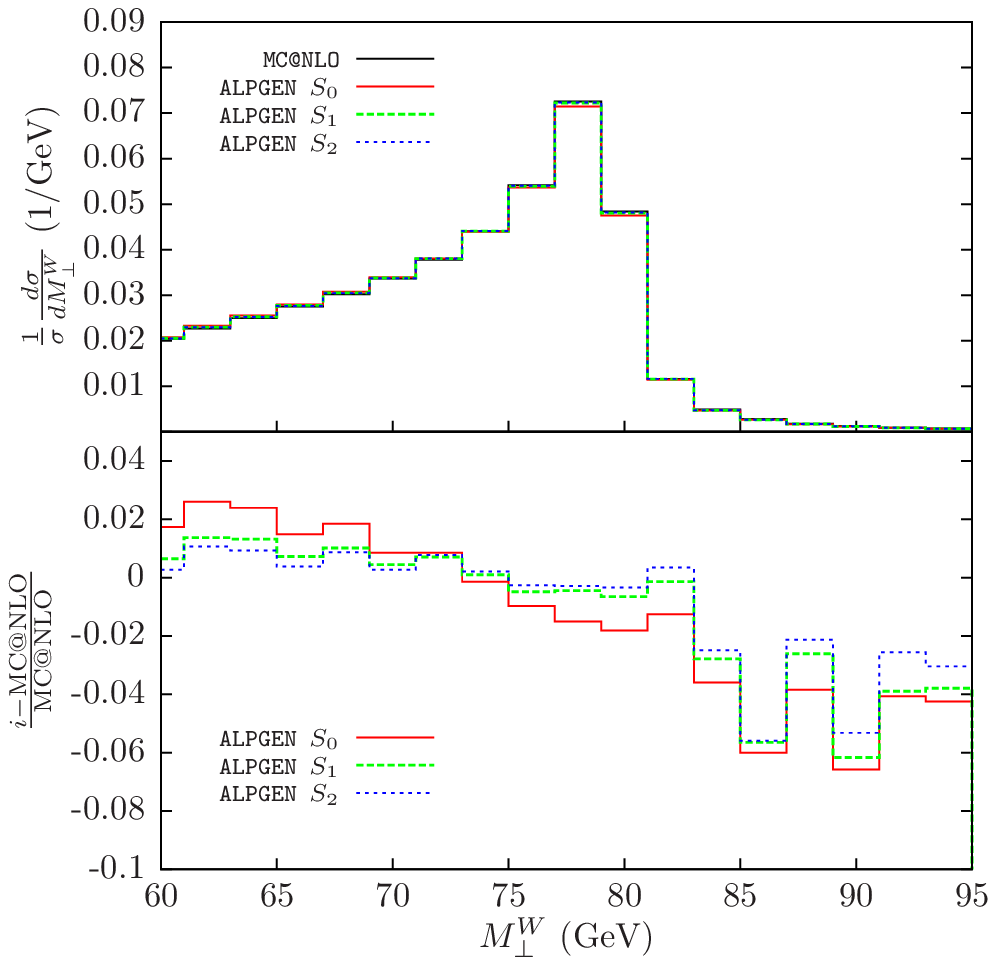}~\hskip 24pt\includegraphics[height=5.5cm]{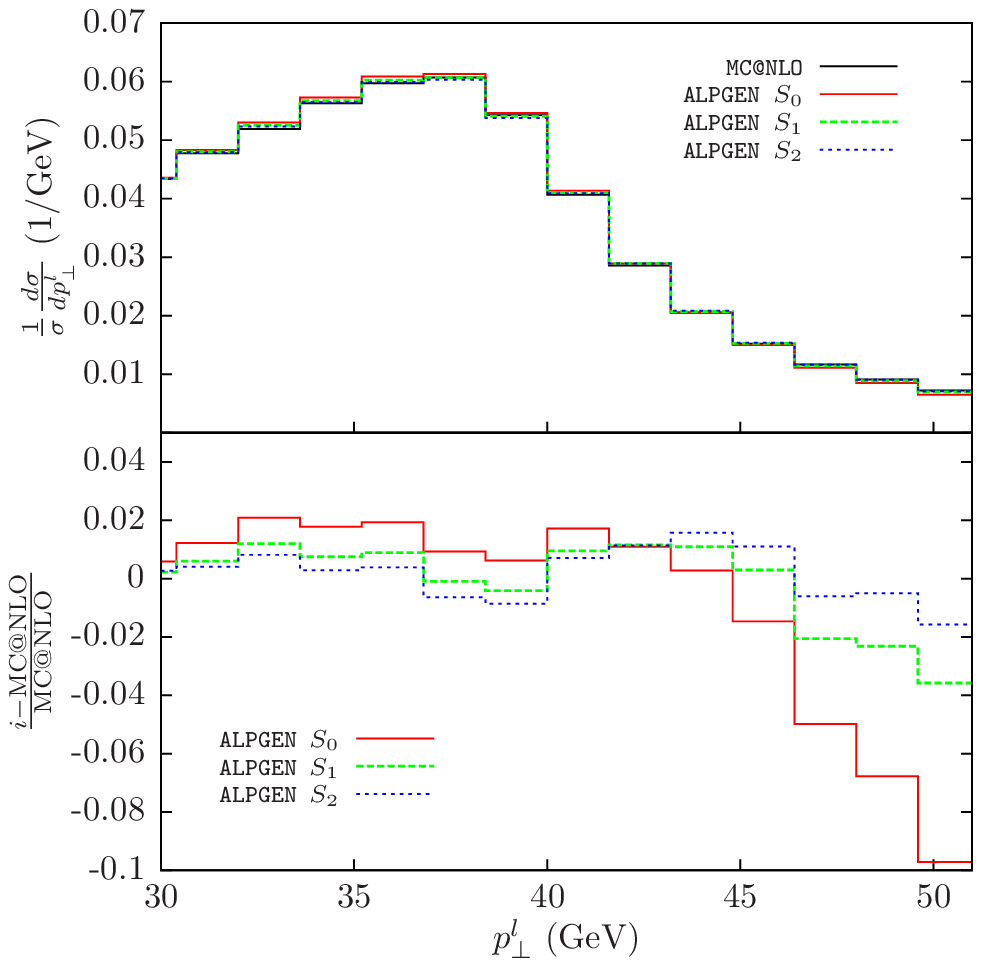}
\end{center}
\caption{$W$ transverse mass (left plot) and muon transverse momentum (right plot) distributions according to 
the QCD predictions of ALPGEN~S$_0$, ALPGEN~S$_1$, ALPGEN~S$_2$  and MC@NLO. In the lower panels
the relative deviations of each code w.r.t. MC@NLO are shown.}
\label{mtw-ptl-lhc-qcd}
\end{figure}

The QCD predictions for the $W$ transverse mass and muon transverse momentum according to 
 set up a. of Table~\ref{tab:lhc} is shown in Figure \ref{mtw-ptl-lhc-qcd}. The results of different variants of the 
 ALPGEN generator are compared with the predictions of MC@NLO. As can be seen from the lower panels, 
 the agreement  in the shape predicted by the two generators is very satisfactory (at the 1\% level) around the
 jacobian peak, while the relative deviations reach the 5-10\% level 
in the hard tail of  both $M_\perp^W$ and  $p_\perp^l$ spectrum. In particular, it can be noticed that for $M_\perp^W$ the 
 relative difference of ALPGEN~S$_0$  w.r.t. MC@NLO has the same size and shape of the deviations observed 
 for the predictions of ALPGEN~S$_1$  and of ALPGEN~S$_2$, while this is not the case for the hard 
 $p_\perp^l$ tail, where the MC@NLO predictions are in better agreement with the ALPGEN versions including matched matrix-elements corrections rather than with ALPGEN~S$_0$. This can be easily understood in terms of the rather 
 smooth dependence of the $M_\perp^W$ distribution from QCD corrections, at variance of the pronounced 
 sensitivity of $p_\perp^l$ from hard QCD radiation. The importance of the latter is particularly visible in 
 Figure \ref{ptl2-lhc-qcd}, showing the predictions of the QCD generators for the muon transverse momentum 
 distribution up to 80 GeV. Actually, one can see the well-known fact \cite{fm} that in the high tail of such 
 distribution, i.e. above $\sim 50$~GeV, there is an important enhancement due to NLO and matched matrix elements
 corrections w.r.t. the QCD PS approximation represented by ALPGEN~S$_0$. Furthermore, a substantial agreement 
 between the shape predicted by MC@NLO and the ones obtained with ALPGEN~S$_1$  and ALPGEN~S$_2$  
 is observed, although a closer inspection reveals discrepancies above $\sim 50$~GeV in the 10\% range, as illustrated
 in the lower panel of Figure \ref{ptl2-lhc-qcd}, where however the cross section is quite small and the relative statistics quite limited.   As a whole, the deviations between MC@NLO and ALPGEN
 in its different flavors shown in Figure \ref{mtw-ptl-lhc-qcd} and Figure \ref{ptl2-lhc-qcd} can be ascribed to the different matching procedures between real and virtual QCD 
 radiation implemented in the two generators. 

\begin{figure}[h]
\begin{center}
\includegraphics[height=5.5cm]{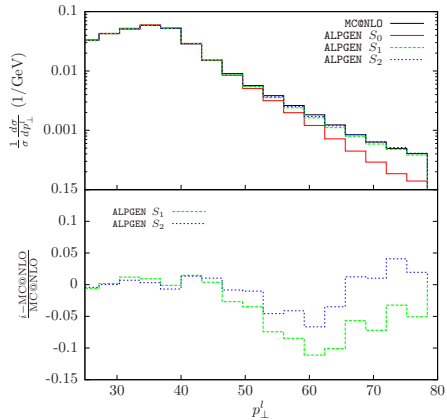}
\caption{The muon transverse momentum  distribution up to 80~GeV according to 
the QCD predictions of ALPGEN~S$_0$, ALPGEN~S$_1$, ALPGEN~S$_2$  and MC@NLO. In the lower panels
the relative deviations of ALPGEN~S$_1$  and ALPGEN~S$_2$   w.r.t. MC@NLO are shown.}
\label{ptl2-lhc-qcd}
\end{center}
\end{figure}

 \begin{figure}[h]
 \begin{center}
\hskip 8pt~\includegraphics[height=5.5cm]{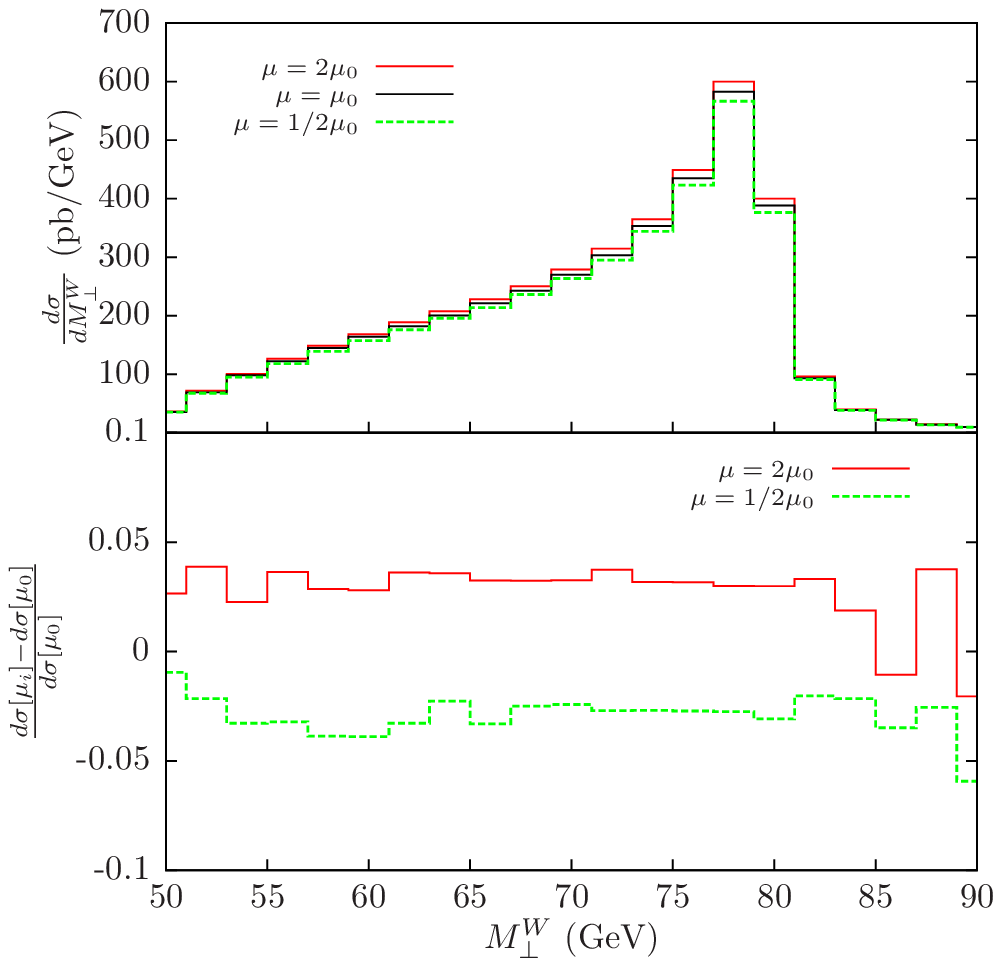}~\hskip 24pt\includegraphics[height=5.5cm]{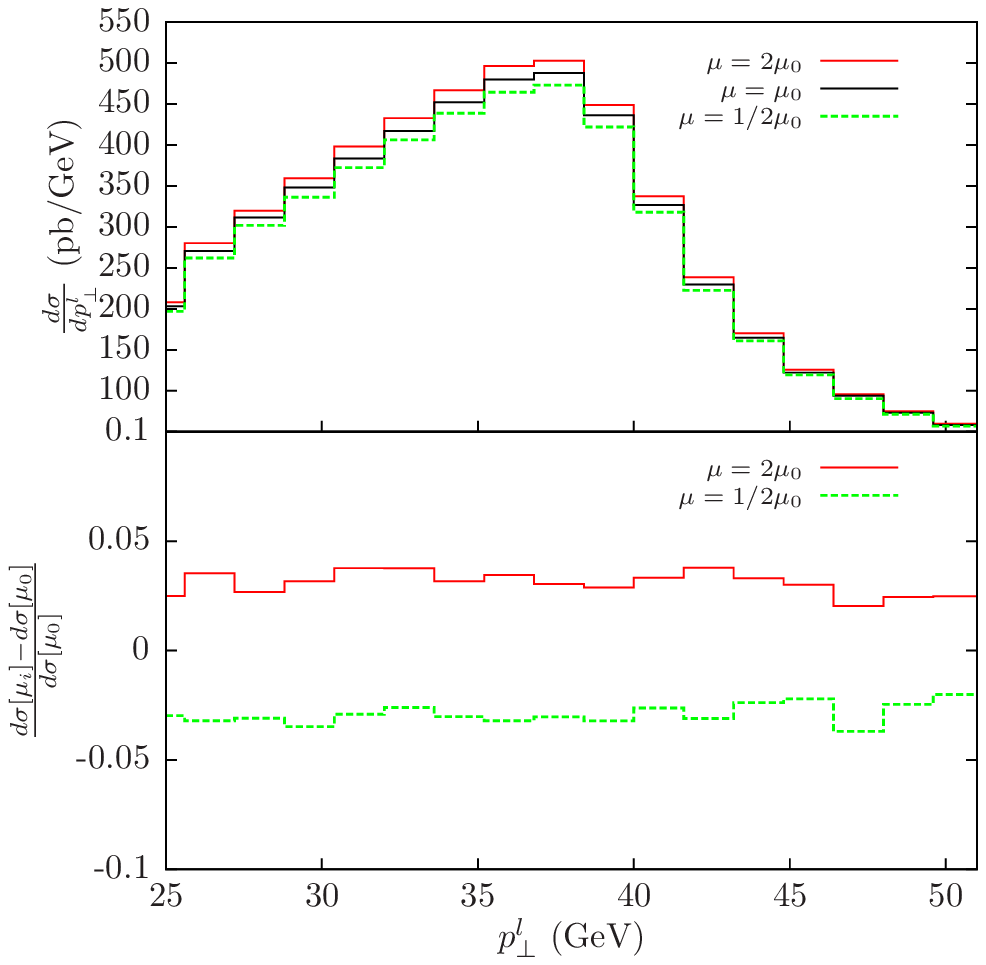}
\end{center}
\caption{$W$ transverse mass (left plot) and muon transverse momentum (right plot) distributions according to 
the QCD predictions of MC@NLO for renormalization/factorization scale $\mu$ at its default value $\mu_0$ 
given in the text and for $\mu = 1/2 \mu_0, 2 \mu_0$. In the lower panels
the relative deviations due to scale variations w.r.t. the default choice are shown.}
\label{mtw-ptl-lhc-qcd-scale}
\end{figure}

Because of the importance of precise predictions for the $M_W$ determination from fits to $M_\perp^W$
and  $p_\perp^l$, we show in Figure \ref{mtw-ptl-lhc-qcd-scale} the $W$ transverse mass (left plot) and muon transverse momentum (right plot) distributions according to the QCD predictions of MC@NLO, when varying the 
renormalization/factorization scale from its default value $\mu_0 = \mu_R = \mu_F = \sqrt{p_{\perp W}^2 + M_W^2}$
to $\mu_0/2 $ and $2 \mu_0$. As it can be seen from the lower panels in Figure \ref{mtw-ptl-lhc-qcd-scale}, the scale
variations induce relative differences w.r.t. the default choice at a few per cent level for both the distributions, without
modifying the shape of the distributions themselves. These deviations reflect the $\sim \pm 3\%$ variations obtained for the cross sections by integrating the above distributions 
 by MC@NLO, given by 8026(5) pb for $\mu =  \mu_0/2$, 8254(5) pb for $\mu =  \mu_0$
and 8502(5) pb for $\mu = 2 \mu_0$.

\begin{figure}[h]
\begin{center}
\hskip 8pt~\includegraphics[height=5.5cm]{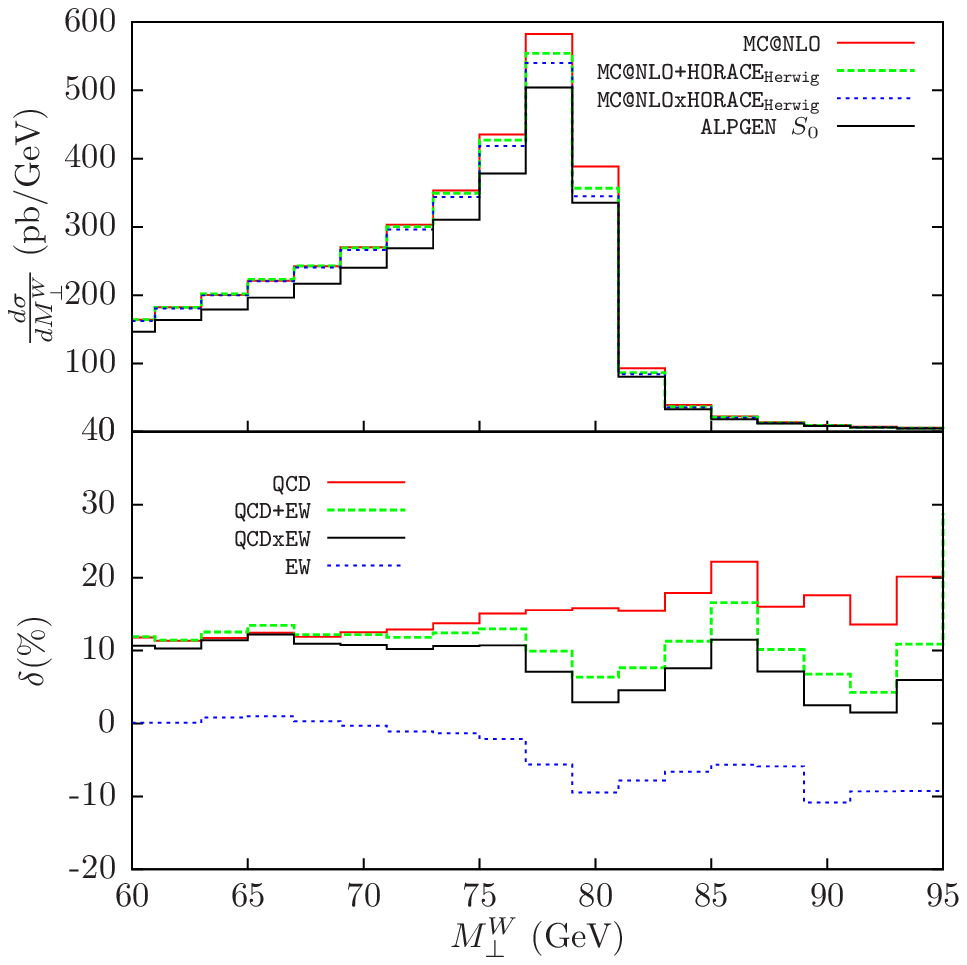}~\hskip 24pt\includegraphics[height=5.5cm]{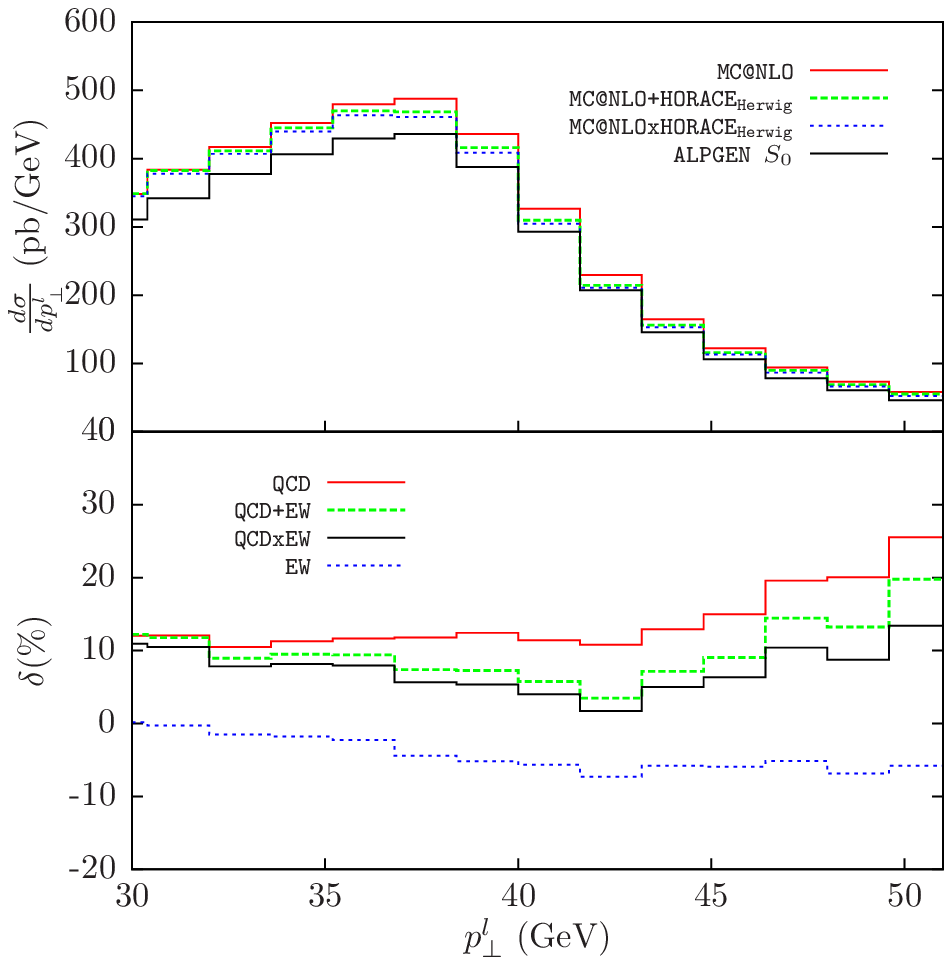}
\end{center}
\caption{The same as Figure 16 for the $W$ transverse mass distribution (left plot) and 
muon transverse momentum (right plot).}
\label{mtw-ptl-lhc-ewqcd}
\end{figure}

The combined effect of EW and QCD 
contribution at the LHC is illustrated in Figure \ref{mtw-ptl-lhc-ewqcd}.  
The upper panels show the 
predictions of the generators ALPGEN~$S_0$, MC@NLO, MC@NLO + HORACE and MC@NLO$\otimes$HORACE. The lower panels illustrate the relative effects of NLO QCD and 
EW corrections, as well as their combination. From fig. \ref{mtw-ptl-lhc-ewqcd} it can be seen that the NLO QCD corrections are positive  and tend to compensate the effect due to EW corrections. Therefore, 
their interplay is unavoidable for a precise $M_W$ extraction at the LHC, yielding an overall correction of about 5-10\% close to the peaks for both the additive and factorized prescriptions. 
\begin{figure}[h]
\begin{center}
\hskip 8pt~\includegraphics[height=5.5cm]{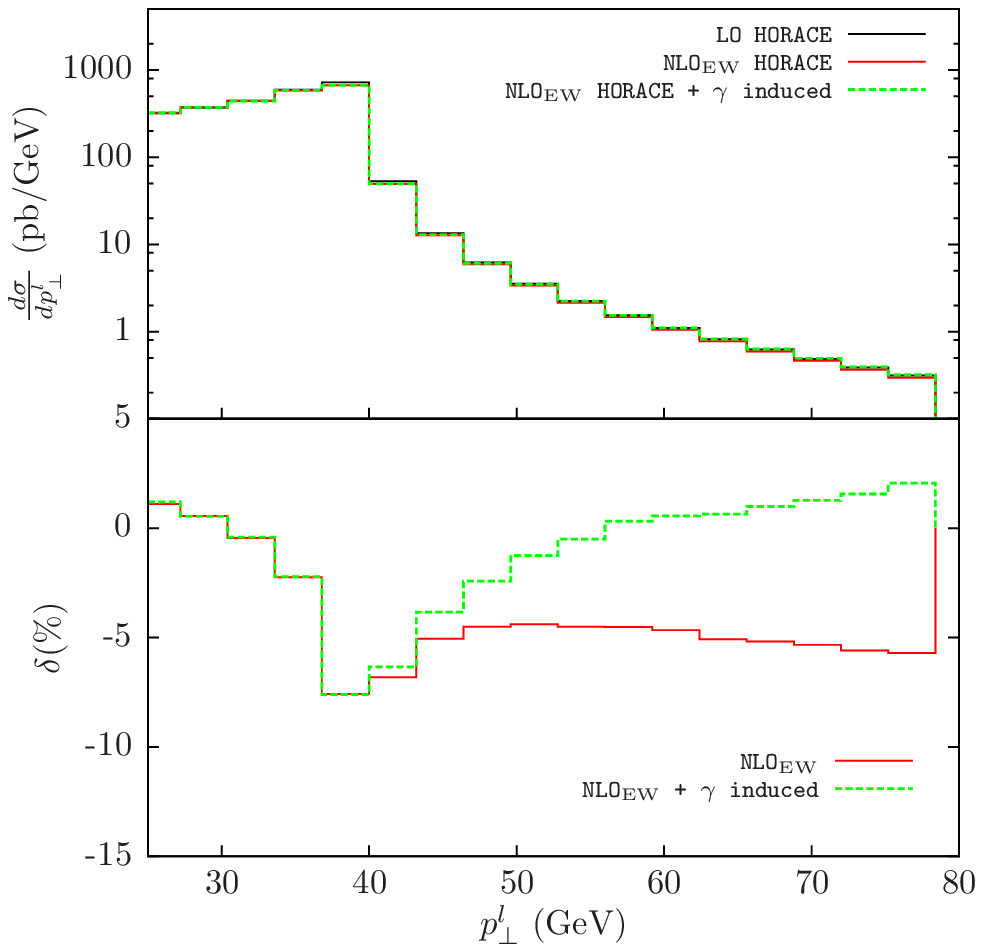}~\hskip 24pt\includegraphics[height=5.5cm]{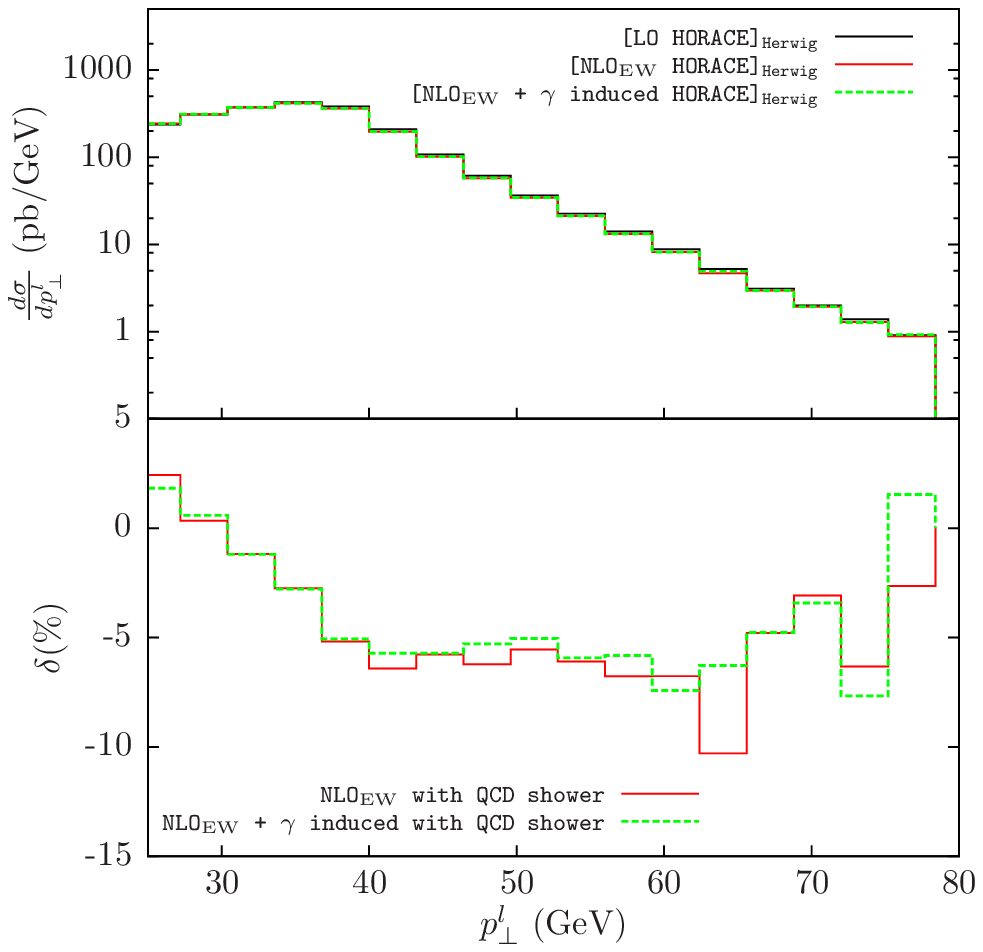}
\end{center}
\caption{Electroweak and $\gamma$-induced corrections without (left) and with (right) 
QCD shower evolution to the lepton transverse momentum distribution.}
\label{ewqcd-shower}
\end{figure}
As already stressed for such distributions at the Tevatron, it is also worth noticing that the relative size and shape of EW corrections to $M_\perp^W$ and 
$p_\perp^l$ is modified by the convolution with 
QCD shower evolution, as can be inferred from comparison with the results for pure EW  
contributions existing in the literature. This indicates that a naive
combination of QCD and EW corrections, in the absence of QCD shower 
evolution on top of EW effects, is rather inadequate for precise simulation of
physical observables. 
This feature is remarked in Figure \ref{ewqcd-shower}, showing the 
relative effect of EW corrections and photon-induced processes for the lepton 
$p_\perp$ distribution in the absence of QCD parton shower convolution (left plot) and in the presence of QCD
shower evolution (right plot). In particular, it can be noticed that the some per cent 
enhancement due to the $\gamma$-induced processes in the hard tail of the $p_\perp^l$ 
distribution without QCD shower largely disappears when the EW effects 
are convoluted with QCD shower radiation.
\begin{figure}[h]
\begin{center}
\includegraphics[height=5.5cm]{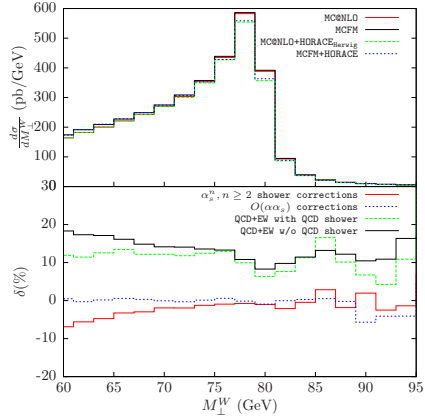} 
\end{center}
\caption{The same as Fig. 17 for the $W$ transverse mass distribution. }
\label{mtw-lhca-aas}
\end{figure}

An inspection of the relative impact due to $O(\alpha_s^2)$ and mixed 
$O(\alpha \alpha_s)$ corrections analogous to that shown in Fig.~\ref{yw-lhca-aas} for $y_W$ is
shown in Fig.~\ref{mtw-lhca-aas} for the transverse mass distribution. For this observable, it can
be seen that the QCD shower evolution is particularly important below the Jacobian
peak, where it introduces corrections of some per cent, while mixed EW-QCD 
contributions are not negligible above it. More importantly, a naive combination of
QCD and EW corrections, omitting the contribution of QCD shower, appears to be inadequate
for a precision calculation of $M_T^W$, as the clearly visible difference between 
the predictions of MC@NLO + HORACE$_{\rm HERWIG}$ and of MCFM + HORACE points out. 
It follows immediately that a measurement of the $W$ mass at the LHC with an aimed 
accuracy of a few MeV must necessarily rely upon MC generators including 
combined QCD and EW corrections according to the highest theoretical standards.\footnote{It is worth noticing that for the first bins close to the kinematical boundary ($M_T^W = 50$~GeV, not shown in the figure) the prediction by MCFM is negative. This is a well known effect already discussed in~\cite{as2, Catani:1997xc}, and due to perturbative instabilities of the NLO calculation. As a consequence, the results by MCFM around the jacobian peak appear to be slightly larger than the corresponding ones by MC@NLO, without contradicting the results for the integrated cross sections shown in Tab.~\ref{comparison}. }

\subsubsection{Observables for new-physics searches}

The numerical results corresponding to the distributions interesting for the search 
for new physics at the LHC
are shown in Figure \ref{mtw-ptl-lhc-qcd-cut} and Figure \ref{mtw-ptl-lhc-qcd-cut-scalevar}
 for what concerns  QCD predictions, and in 
Figure \ref{mtw-ptl-lhc-ewqcd-cut} and 
Figure \ref{mtw-ptl-lhc-ewqcd-cut-jetveto}, for the combination of EW and QCD corrections. The simulations have been performed using  the cuts referring to set up b. of Table \ref{tab:lhc} and imposing additional jet veto conditions for what concerns the results shown in 
Figure \ref{mtw-ptl-lhc-ewqcd-cut-jetveto}. 

Concerning Figure \ref{mtw-ptl-lhc-qcd-cut}, it can be seen that the shape predicted by the different variants 
of ALPGEN w.r.t. the one obtained with MC@NLO has a very similar trend for both $M_\perp^W$
in the range 1-2~TeV and $p_\perp^l$ in the region 0.5-1~TeV. The predictions of all the ALPGEN versions, both without and with matrix element corrections, nicely agree with the MC@NLO results within a 5\% band.

\begin{figure}[h]
\begin{center}
\includegraphics[height=5.5cm]{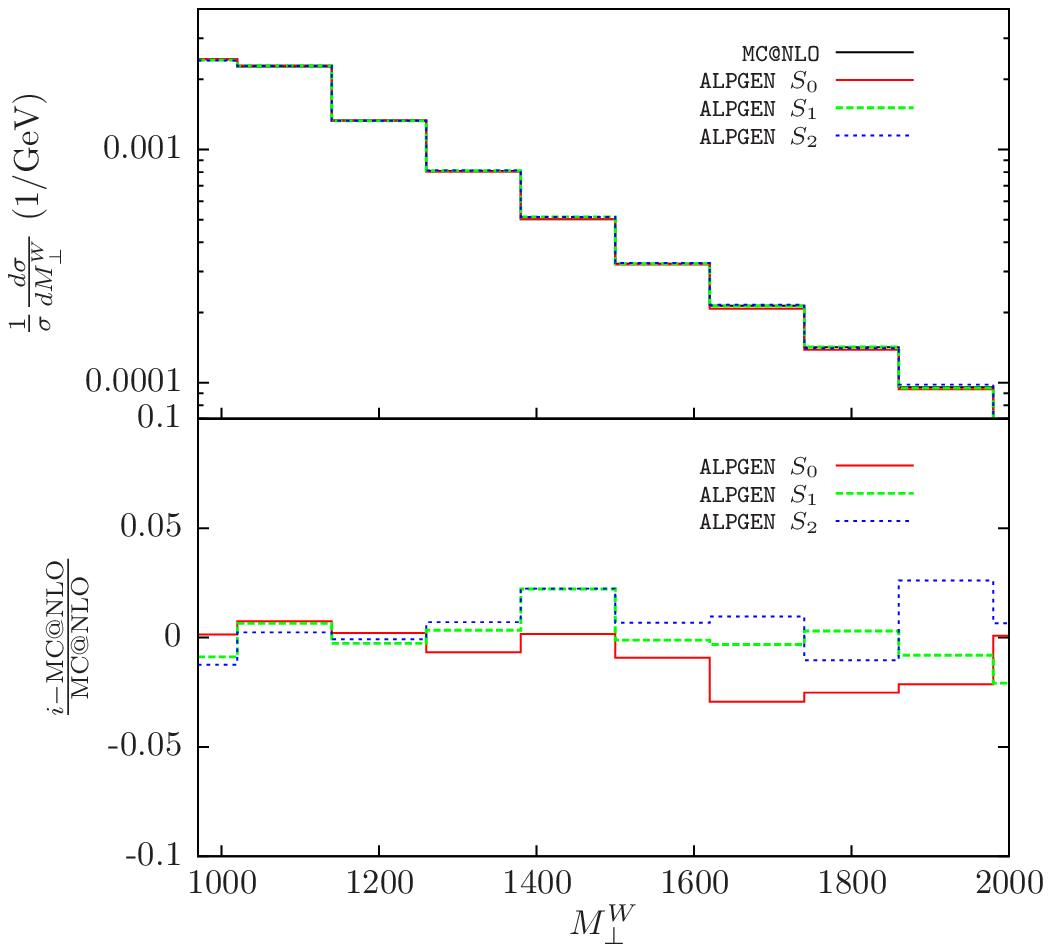}~\hskip 24pt\includegraphics[height=5.5cm]{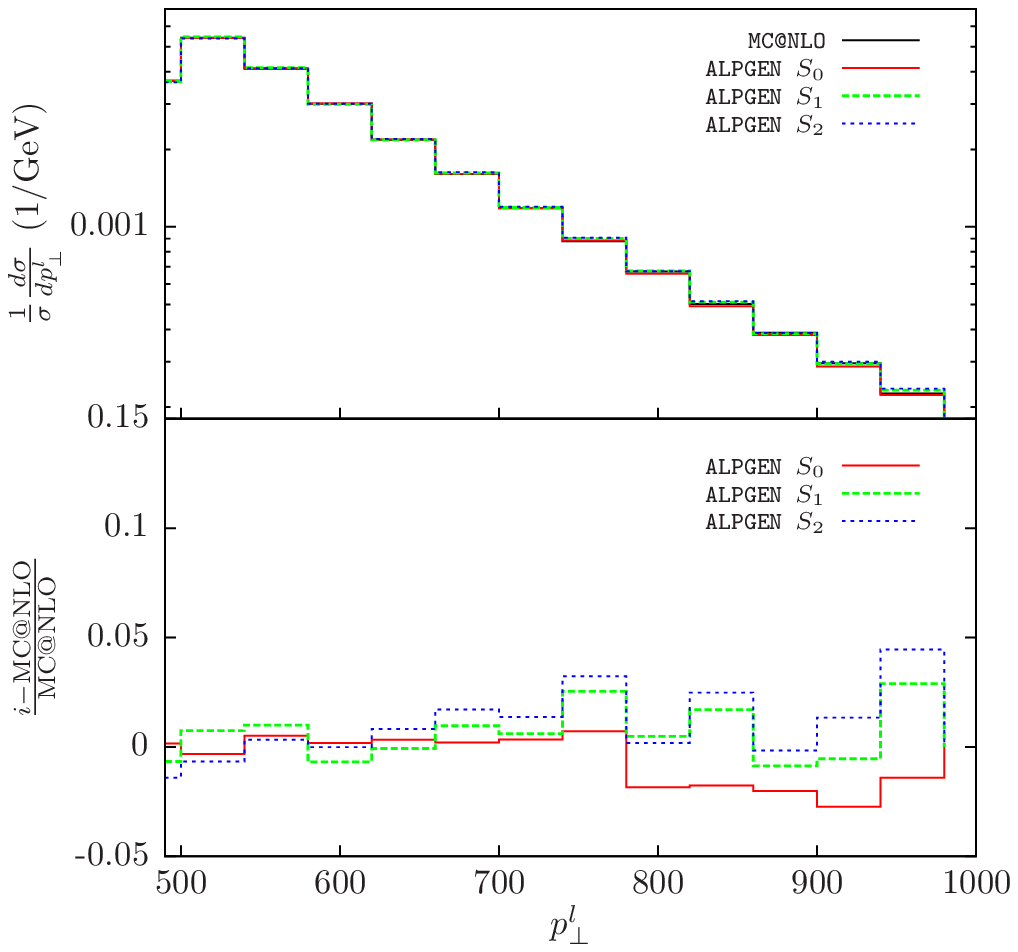}
\caption{The same as Figure 18 for the transverse mass (left plot) and lepton transverse momentum (right plot) 
distributions in the very high tails, according to set up b. at the LHC.}
\label{mtw-ptl-lhc-qcd-cut}
\end{center}
\end{figure}

Figure \ref{mtw-ptl-lhc-qcd-cut-scalevar} shows the $W$ transverse mass (left plot) and muon transverse momentum (right plot) distributions according to the QCD predictions of MC@NLO, when varying the 
renormalization/factorization scale from its default value $\mu_0 = \mu_R = \mu_F = \sqrt{p_{\perp W}^2 + M_W^2}$
to $\mu_0/2 $ and $2 \mu_0$. As can be seen from the lower panels, 
the scale variations induce relative differences w.r.t. the default choice 
of a few per cent around 1~TeV for $M_\perp^W$, reaching O(-10\%) at 3~TeV, but where the cross section is two orders of magnitude smaller. For $p_\perp^l$ the scale variations are milder, always within  $\pm 5\%$.

\begin{figure}[h]
\begin{center}
\includegraphics[height=5.5cm]{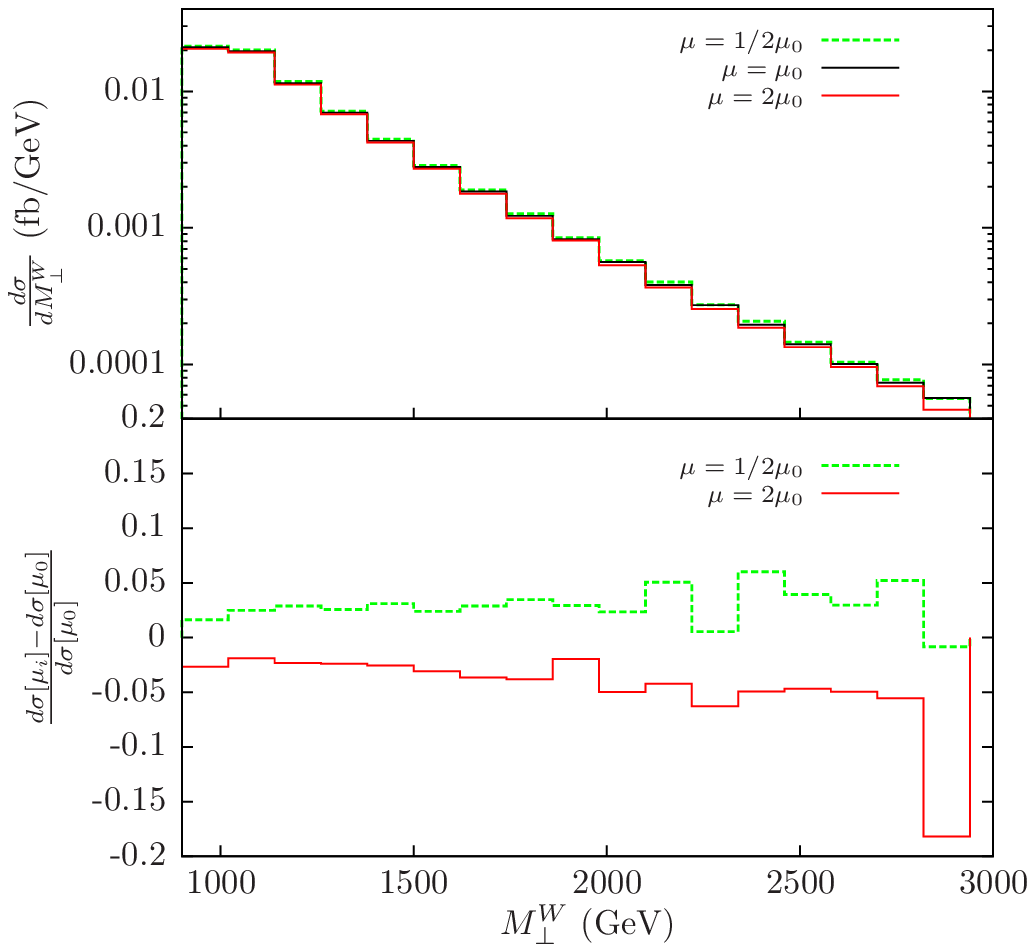}~\hskip 24pt\includegraphics[height=5.5cm]{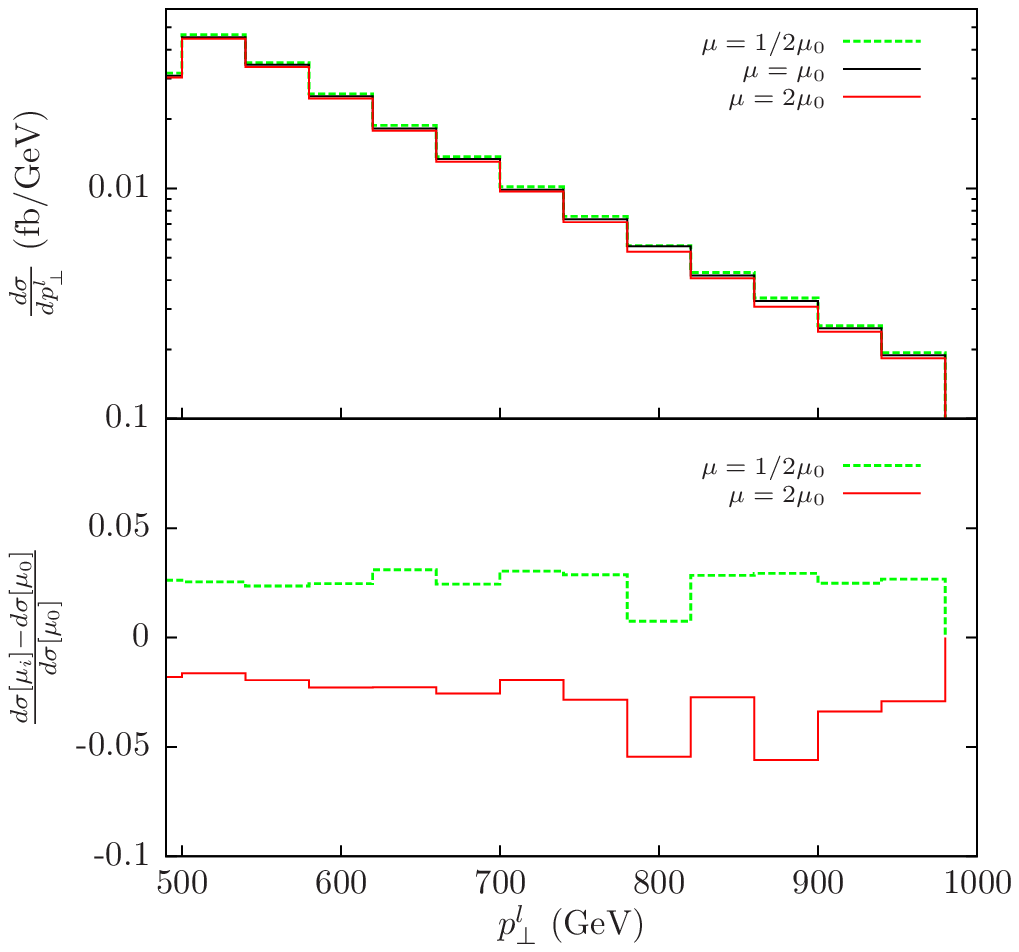}
\caption{The same as Figure 20 for the transverse mass (left plot) and lepton transverse momentum (right plot) 
distributions in the very high tails, according to set up b. at the LHC.}
\label{mtw-ptl-lhc-qcd-cut-scalevar}
\end{center}
\end{figure}

The interplay between QCD and EW corrections in the high tail of $M_\perp^W$ and $p_\perp^l$ distributions
is shown in Figure \ref{mtw-ptl-lhc-ewqcd-cut} (for the standard cuts of set up b. in Table \ref{tab:lhc}) and 
Figure \ref{mtw-ptl-lhc-ewqcd-cut-jetveto} (when including additional jet veto conditions). 

\begin{figure}[h]
\begin{center}
\includegraphics[height=5.5cm]{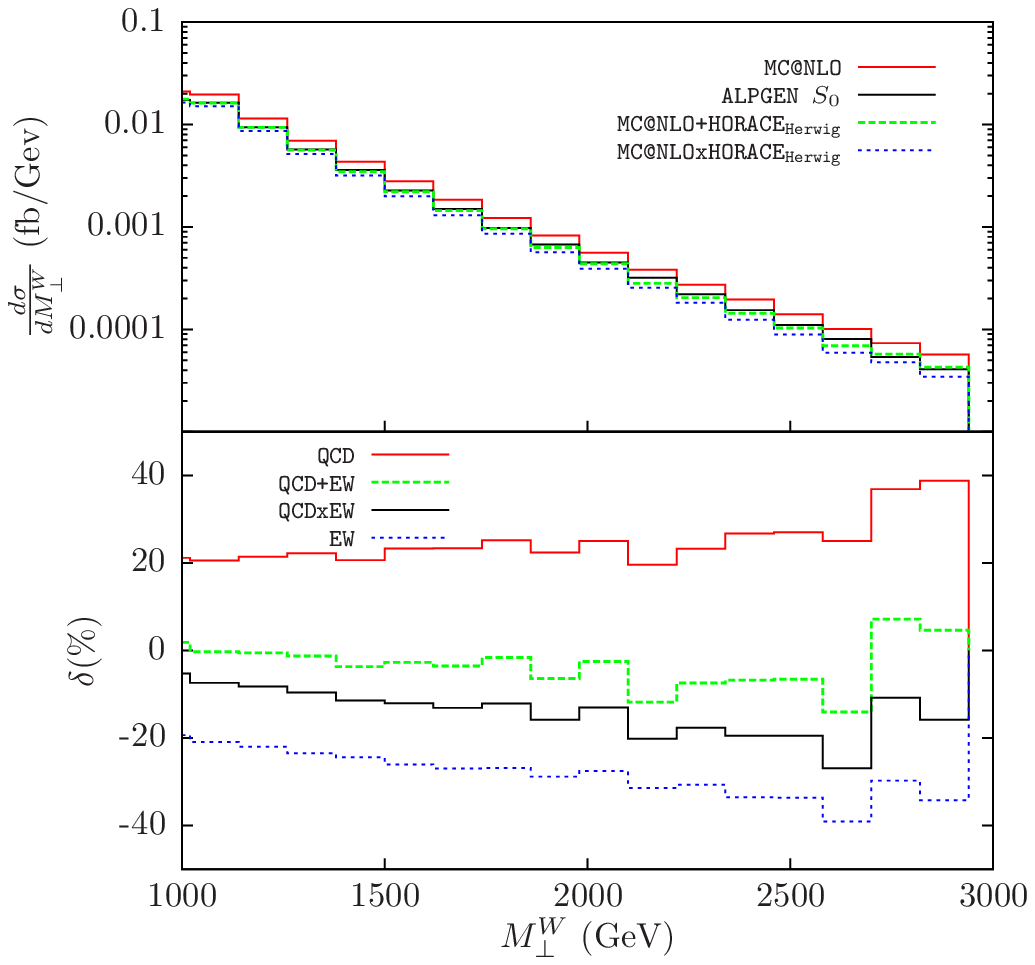}~\hskip 24pt\includegraphics[height=5.5cm]{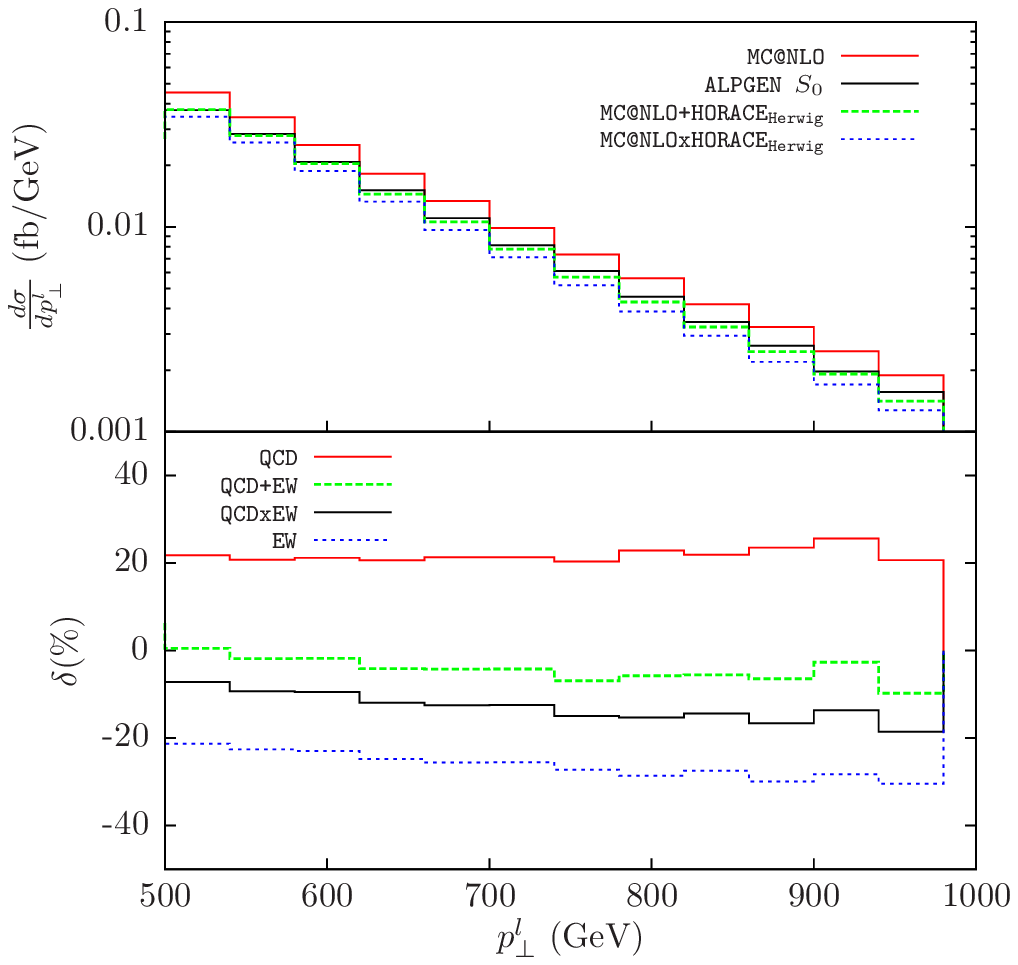}
\caption{The same as Figure 21 for the transverse mass (left plot) and lepton transverse momentum (right plot) 
distributions in the high tails, according to set up b. at the LHC.}
\label{mtw-ptl-lhc-ewqcd-cut}
\end{center}
\end{figure}

The 
jet veto conditions have been implemented by vetoing any hard jet with $p_\perp > 30$~GeV
in the central rapidity region ($|y_{\rm jet}| < $ 2.5).
In both cases, and for both $M_\perp^W$ and $p_\perp^l$, NLO QCD corrections are positive  (of the order of 20-40\% for standard cuts and between 20-60\% in the presence of jet veto 
requirements) and combine with  very large negative EW Sudakov logarithms. 
When imposing standard selection criteria, their factorized combination is  about $- 10(+10)\%$ for $M_\perp^W \simeq 1.5(3)$~TeV
and  about $- 5(-20)\%$ for $p_\perp^l \simeq 0.5(1)$~TeV, the additive combination yielding smaller effects. 
Therefore, a precise
normalization of the SM background to new physics searches 
necessarily requires the simultaneous control of QCD and 
EW corrections, as also pointed out  in \cite{LH2007}  for the NC DY 
process in the high invariant mass region, where a similar compensation between 
strong and EW contributions takes place. It is worth noticing that in such a region the large negative Sudakov virtual logarithms can be compensated by real weak boson emission, as pointed out in~\cite{UBaur}. 
Concerning the inclusion of two-loop EW Sudakov 
logarithms, whose calculation is available in the literature~ \cite{Denner:2006jr,Jantzen:2005az,Jantzen:2005xi}, 
it should be considered in view of a theoretical accuracy at the
percent level. For the same reason it should be accompanied by the
simultaneous control of the NNLO QCD corrections. 

\begin{figure}[h]
\begin{center}
\includegraphics[height=5.5cm]{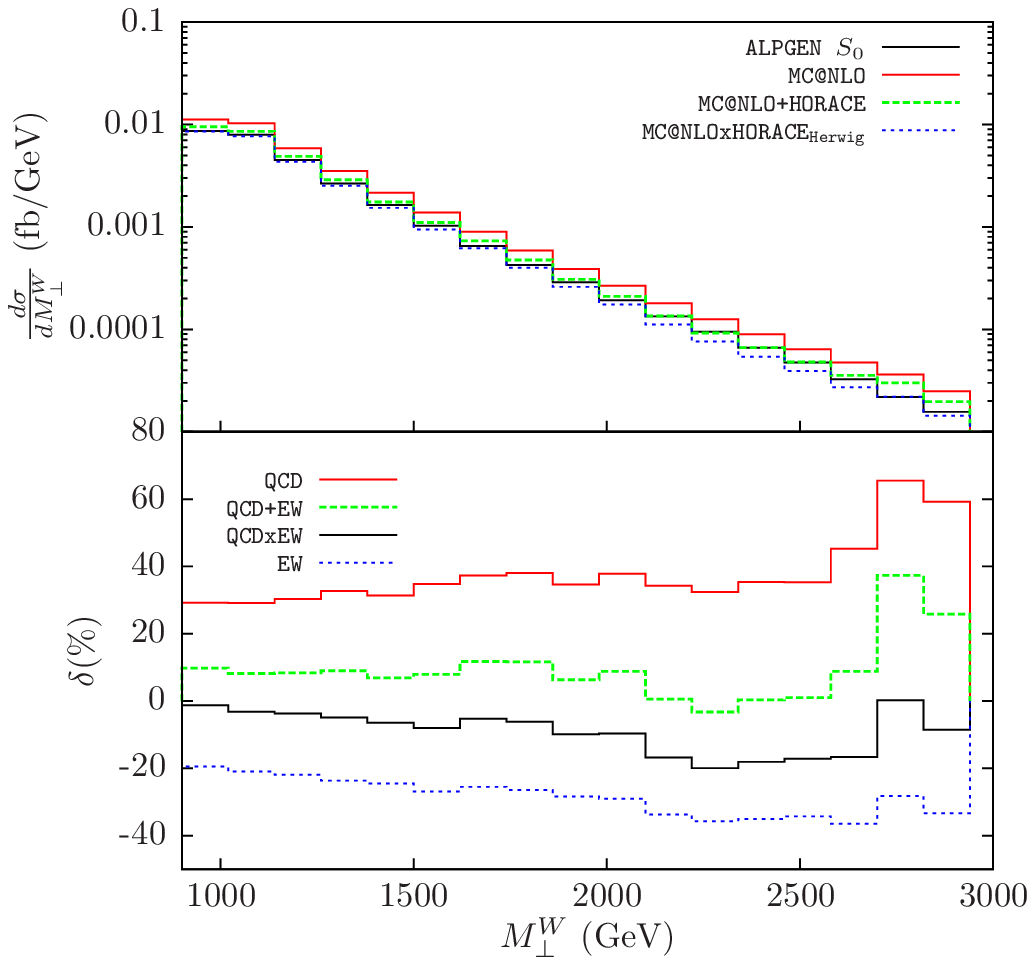}~\hskip 24pt\includegraphics[height=5.5cm]{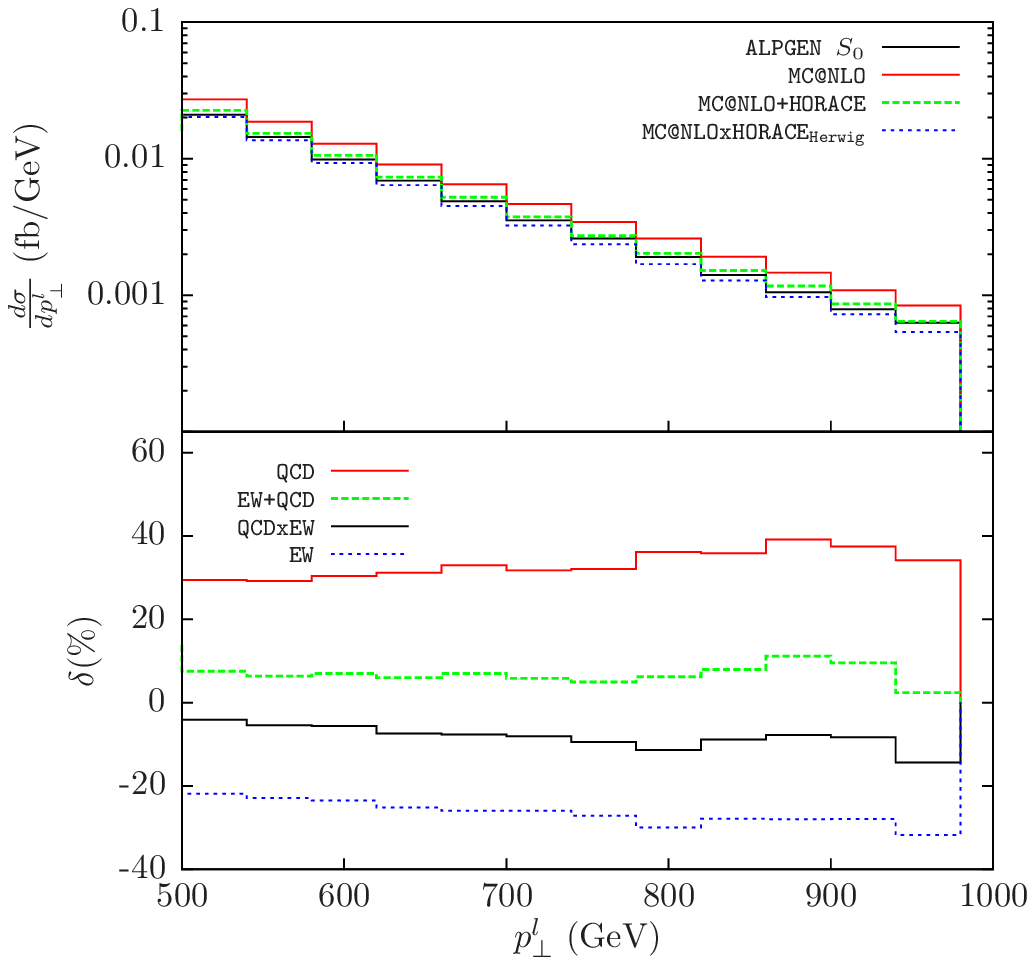}
\caption{The same as Figure 21,  according to set up b. at the LHC and in the 
presence of additional jet veto conditions.}
\label{mtw-ptl-lhc-ewqcd-cut-jetveto}
\end{center}
\end{figure}

\begin{figure}[h]
\begin{center}
\includegraphics[height=5.5cm]{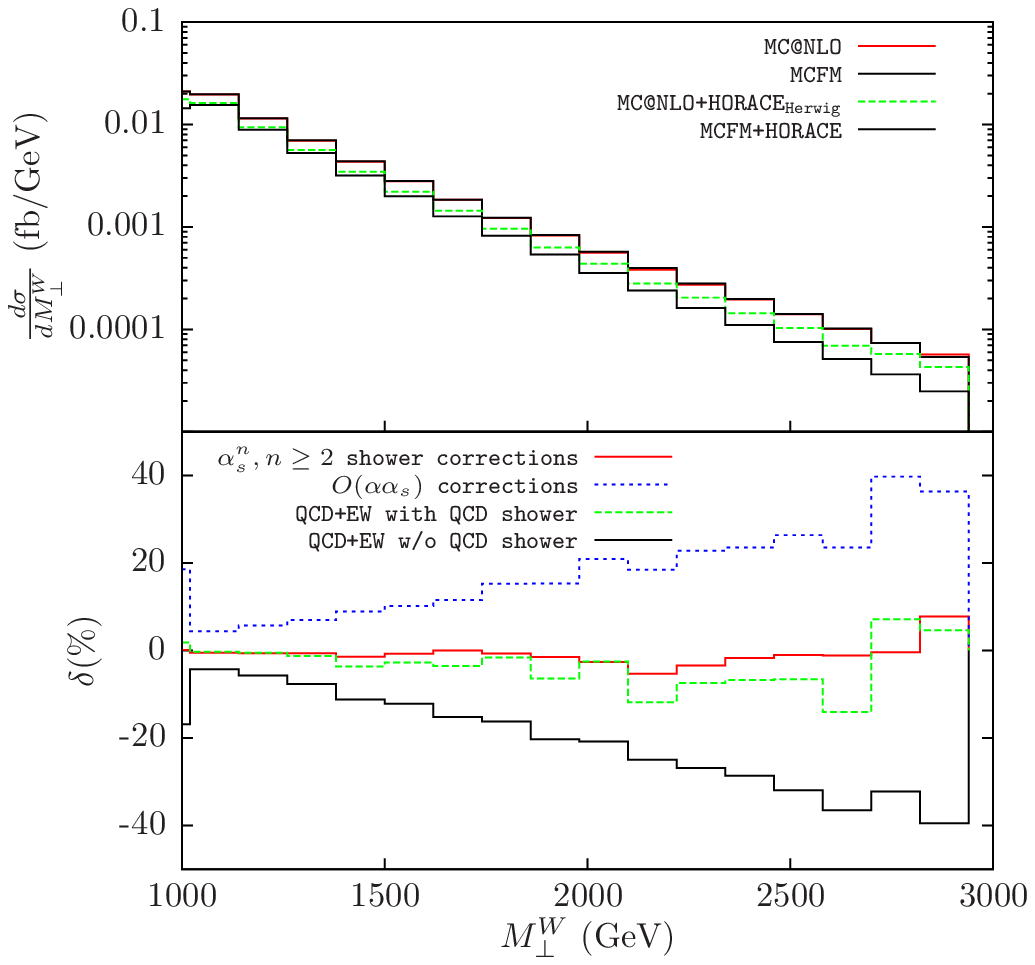}~\hskip 24pt\includegraphics[height=5.5cm]{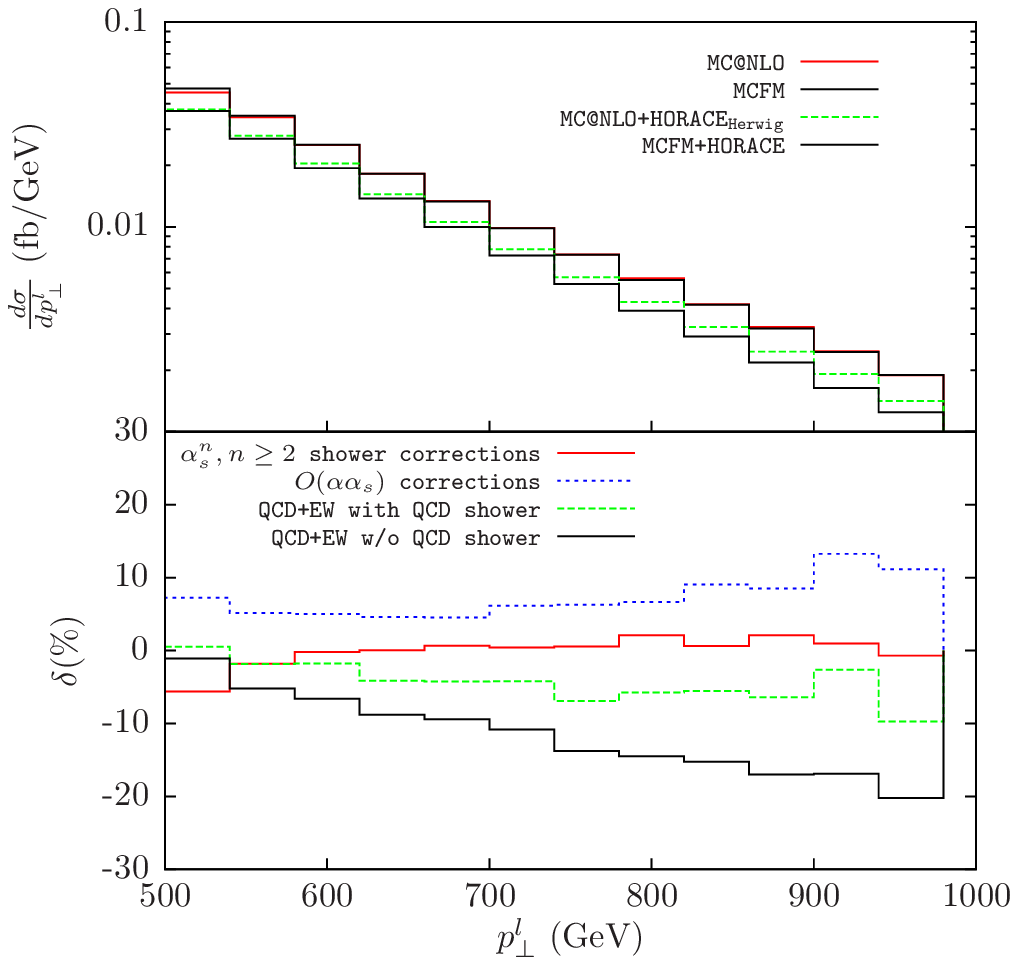}
\caption{The same as Figure 17 for the transverse mass (left plot) and lepton transverse momentum (right plot) 
distributions in the high tails, according to set up b. at the LHC.}
\label{mtw-ptl-lhc-ewqcd-aas}
\end{center}
\end{figure}
As done for $y_W$ and $M_T^W$ in the presence of the standard analysis cuts denoted as LHCa,
we also investigated how the predictions of a combination of pure NLO codes for EW and QCD corrections 
compare with those in which QCD shower effects are taken into account when
considering $M_T^W$ and $p_T^l$ in the set up LHC b. The results of this study are shown in 
Fig.~\ref{mtw-ptl-lhc-ewqcd-aas} for $M_T^W$ (left plot) and $p_T^l$ (right plot). It can be seen that the pure NLO
results differ from those including QCD shower contributions at the 10\% level for 
$p_T^l$ and even more, between 20-30\%, for $M_T^W$ in the very hard tail. This important difference
is due to the presence of large EW Sudakov logarithms in the NLO EW calculation, giving rise
to sizable $O(\alpha \alpha_s)$ corrections at the NNLO level. Hence, the evaluation of the DY
background to new physics searches in this region should take carefully under control the 
combined effect of QCD-EW corrections, especially in the event new physics would manifest 
as a moderate modification of the DY continuum, rather than as a clearly visible resonance.

\begin{figure}[h]
\begin{center}
\hskip 8pt~\includegraphics[height=5.5cm]{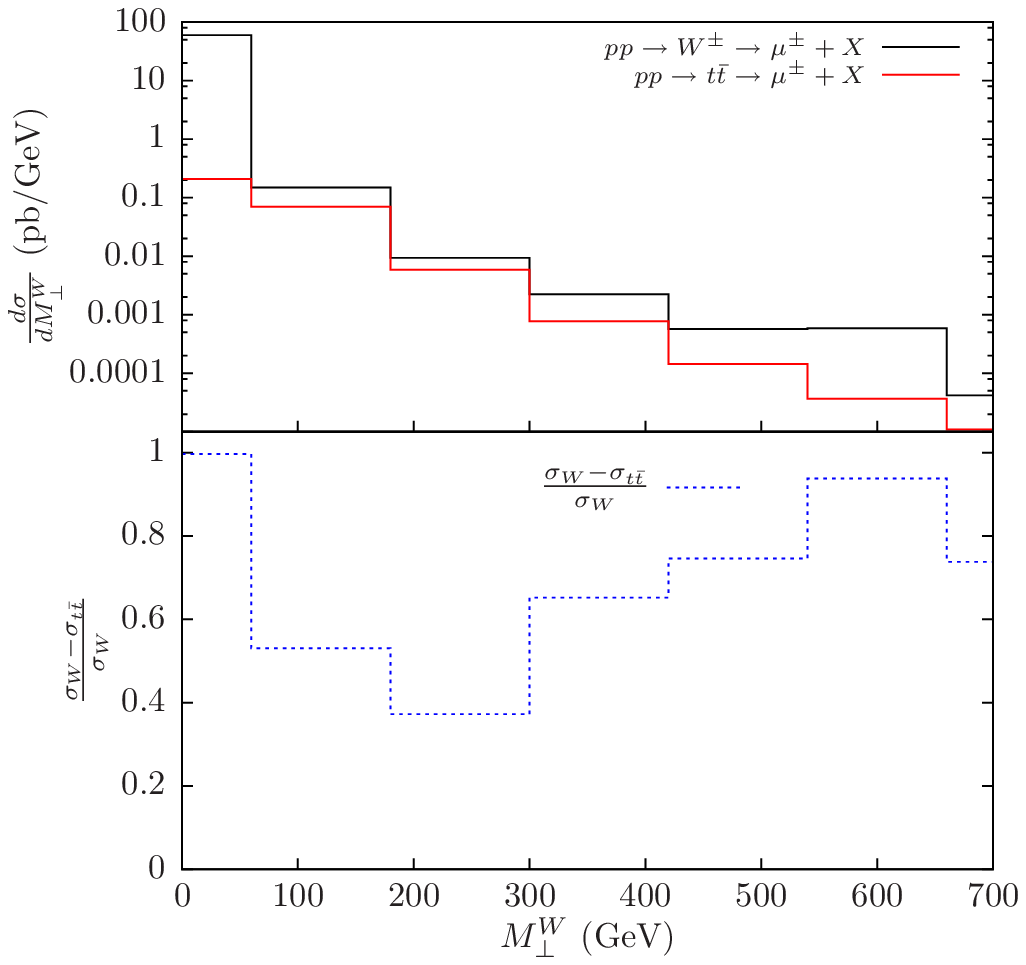}~\hskip 24pt\includegraphics[height=5.5cm]{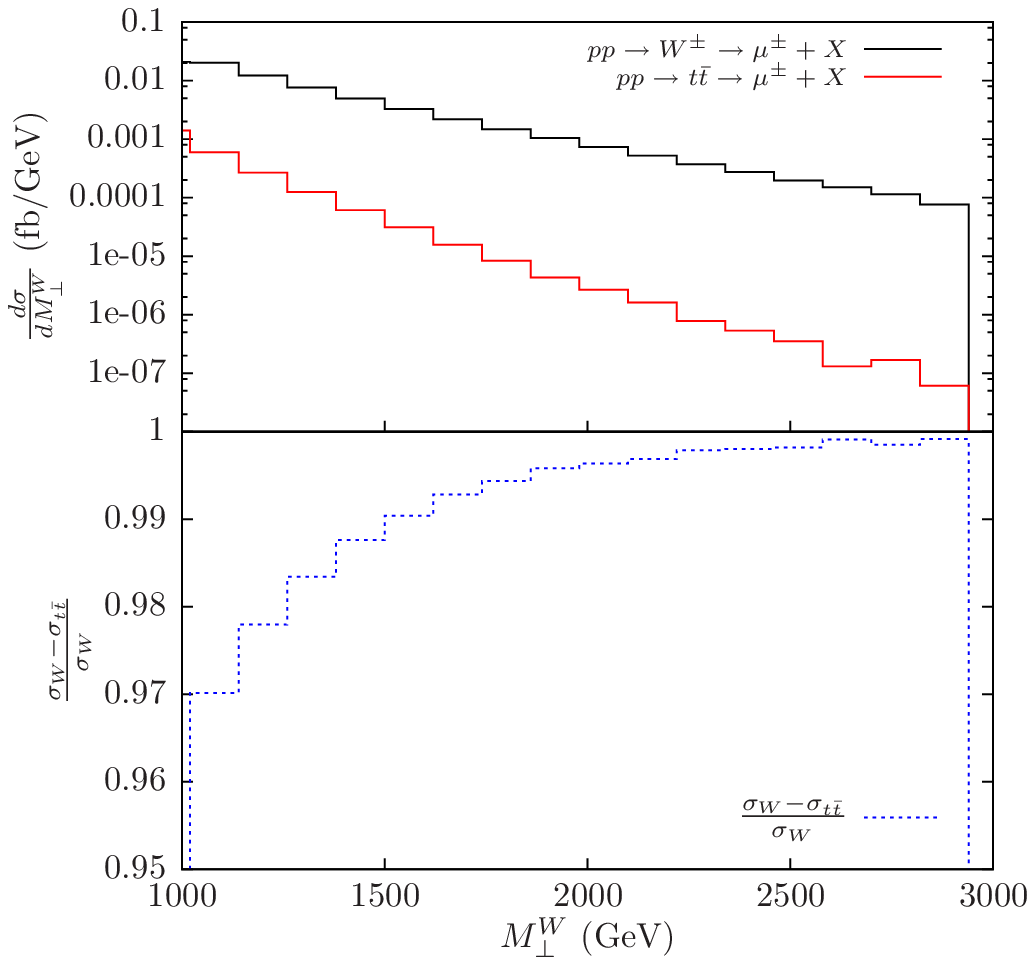}
\caption{Upper panels: transverse mass distribution of the single $W$ production process (black histogram) and of the $t \bar{t}$ process (red histogram), according to the event selection 
specified in the text. Lower panels: relative difference of the
two contributions, for two transverse mass windows. }
\label{fig:w-tt}
\end{center}
\end{figure}

To complete the phenomenological analysis, we also performed an investigation of the 
contribution to the transverse mass due to the $top$-pair  production process $pp \to t \bar{t} \to \mu^{\pm} + X$ 
(simulated with ALPGEN) in comparison with $pp \to W^\pm \to \mu^\pm + X$. The obtained
results are shown in Figure \ref{fig:w-tt}, for a moderate transverse mass range up to 700 GeV
(left plot) and in the high tail up to 3 TeV (right plot). For the $ t \bar{t}$ process, all the $t$
leptonic decays are simulated and the events are considered as contributions to the signature 
when at least a muon is present in the final state. Whenever just a single muon is part of the final state 
products, the $W$ transverse mass is reconstructed in terms of its transverse momentum, whereas
in the case two muons are present the $W$ transverse mass is calculated in terms of the leading-$p_t$ 
muon. The unobserved $X$ state consists of the two $b$-quarks and two neutrinos coming from 
the top and $W$ decays, respectively. The missing transverse momentum entering the definition of $W$
 transverse mass is obtained by summing over the missing $p_t$ of the two neutrinos and requiring 
 for each of the two the condition $\rlap{\slash}{\! E_T}  \geq$~25~GeV, consistently with the cuts imposed 
 in the analysis of $W$ production events. As can be seen from Figure \ref{fig:w-tt}, the $ t \bar{t}$ process gives, for the event selection assumed,  a sizable (of the order of several tens per cent) 
 contribution to the $\mu^\pm + X$ 
 signature in the transverse mass range between about 100 GeV and 600 GeV, while it is at a
 few per cent level in the high tail.

\subsection{Results for jet multiplicity (Tevatron and LHC)} 

We conclude our phenomenological study with the presentation of the results 
concerning the distributions 
of the number of jets, showing the predictions obtained for the 
Tevatron in comparison with those valid for the LHC.  The jets are required to satisfy the 
following cuts: $E_{Tj}>$ 20 GeV, $|\eta_j|<$ 5, $\Delta R_{ij} >$ 0.7, 
where $\Delta R_{ij}$ is the separation between the $i$-th and $j$-th jet in the $\eta-\phi$ plane. 
Jets are reconstructed by means the cone algorithm 
          provided by the GETJET package~\cite{getjet}, 
          which represents a simplified jet cone algorithm a la UA1 (for the sake of simplicity, we stop the evolution at the 
           shower level).

 \begin{figure}[h]
 \begin{center}
\includegraphics[height=5.5cm]{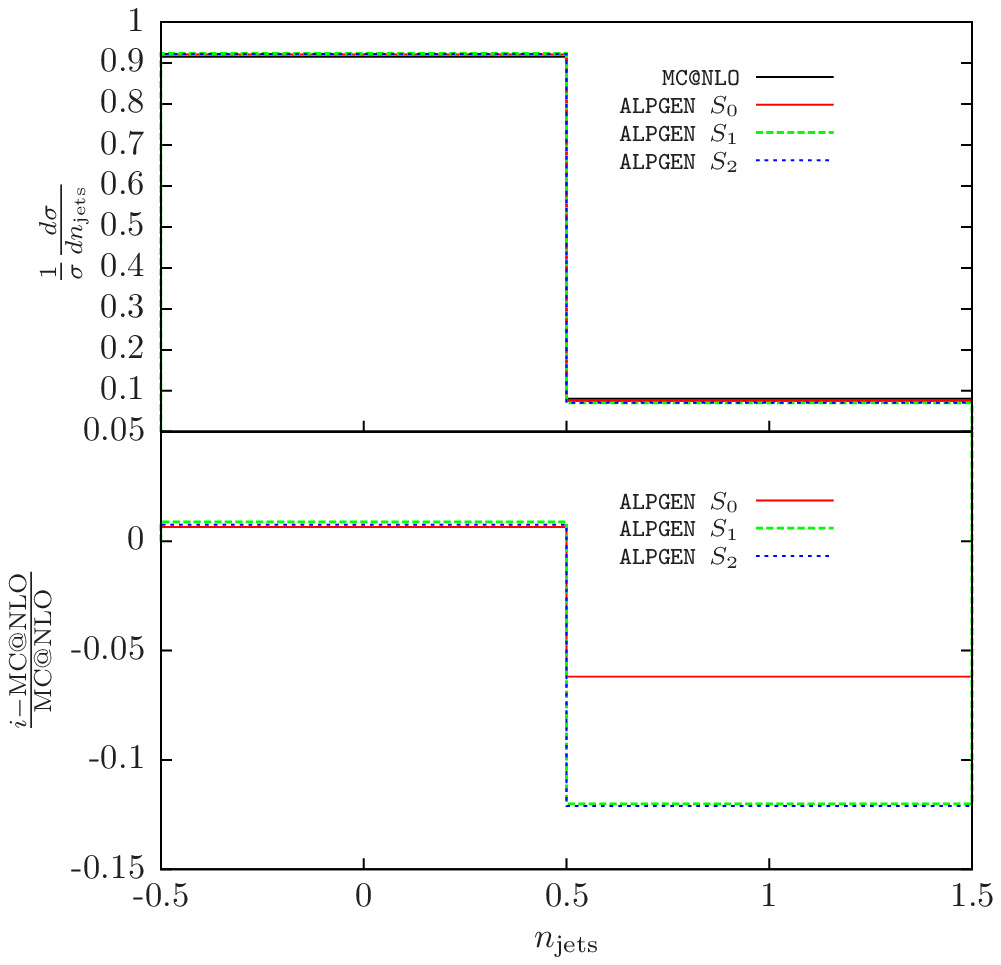}~\hskip 24pt\includegraphics[height=5.5cm]{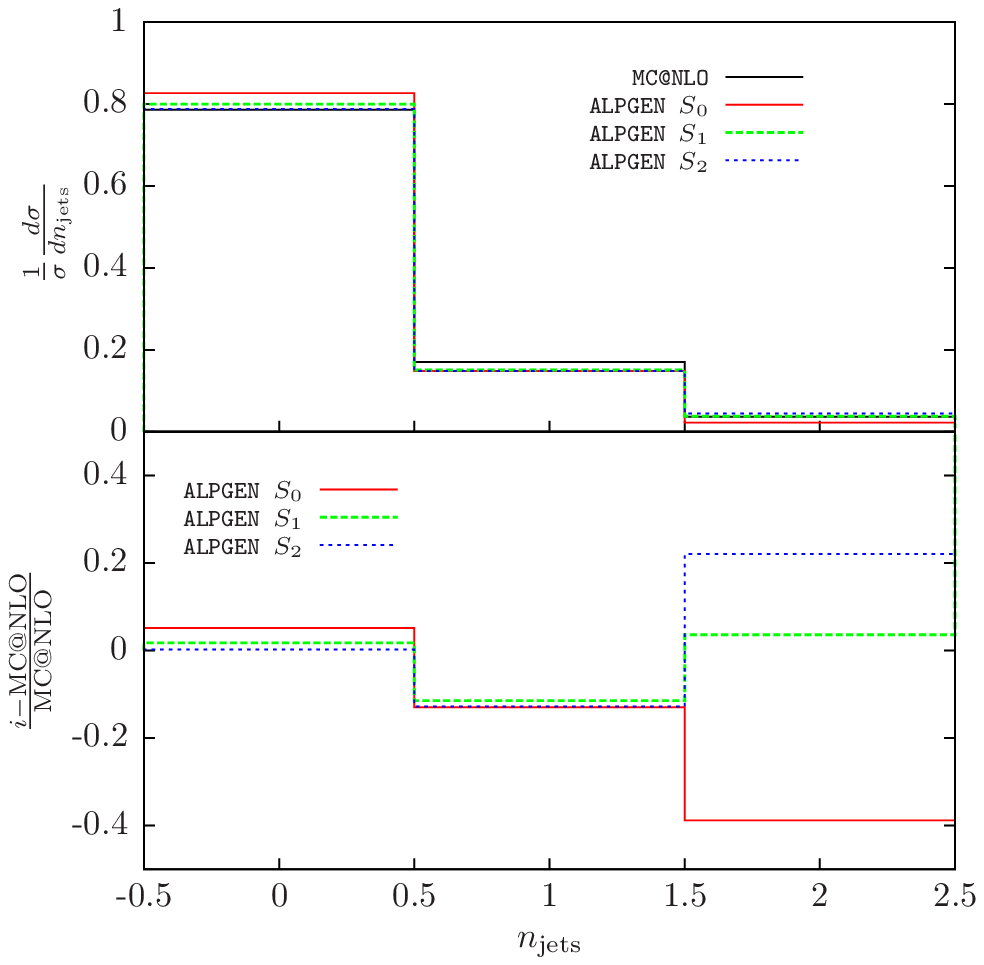}
\caption{QCD predictions of ALPGEN~S$_0$, ALPGEN~S$_1$, ALPGEN~S$_2$  and MC@NLO 
for the number of jets distribution at the Tevatron (left plot) and at the LHC, set up a. (right plot). In the lower panels
the relative deviations of each code w.r.t. MC@NLO are shown.}
\label{njets-tev-lhc}
\end{center}
\end{figure}
 
From the QCD generators used in our analysis, the predictions for the number of jets
distribution are shown in Figure \ref{njets-tev-lhc} for the Tevatron (left plot) and the LHC (right plot), set up a. At the 
Tevatron energies, about 90\% of single $W$ production events are without any extra real hard QCD radiation, while
the fraction of events with one additional jet is about 10\%.  At the LHC, the sharing of the events is, not unexpectedly, 
quite different: about  80\% of the events do not contain extra jets, the fraction of events with one additional jet is slightly smaller than  
20\%, while events with al least two additional jets amount to a few per cent. 
As it can be seen from the lower panels
in Figure \ref{njets-tev-lhc}, there is good agreement between the predictions of MC@NLO and ALPGEN 
variants for the $W$ production rate without any extra jet both at the Tevatron and LHC. At the Tevatron, 
for the events with one additional jet, MC@NLO predictions differ from ALPGEN~S$_0$  results (where the additional jet is generated through the PS cascade) at the 5\% level and from ALPGEN~S$_1$  and ALPGEN~S$_2$  predictions at the 
$\sim 10\%$ level. At the LHC, all the ALPGEN variants provide results differing from the MC@NLO ones at the 
$\sim 10\%$ level for events with one extra jet, while for the fraction of two additional jet events there is good 
agreement between MC@NLO and ALPGEN~S$_1$  but significant deviations are present between MC@NLO and either
ALPGEN~S$_0$  or ALPGEN~S$_2$, as expected. 

\begin{figure}[h]
\begin{center}
\includegraphics[height=5.5cm]{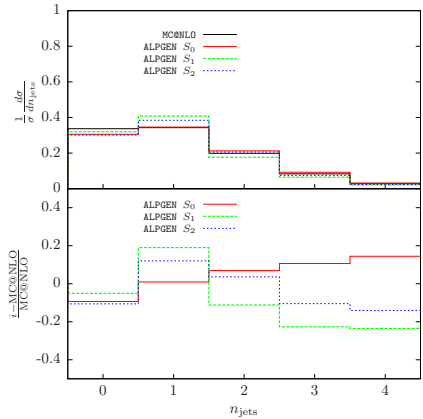}
\caption{QCD predictions of ALPGEN~S$_0$, ALPGEN~S$_1$, ALPGEN~S$_2$  and MC@NLO 
for the number of jets distribution at the LHC, set up b. In the lower panels
the relative deviations of each code w.r.t. MC@NLO are shown.}
\label{njets-lhc}
\end{center}
\end{figure}

In Figure \ref{njets-lhc} we show the number of jets distribution corresponding to set up b. 
at the LHC. As can be seen, the requirement $M_\perp^W > 1$~TeV changes very significantly
the number of jets distribution w.r.t. the situation of the more inclusive set up a. shown in 
fig. \ref{njets-tev-lhc}. In particular, the fraction of events without extra jets is about 30\% and it's
exceeded by the fraction of events with one additional jet, at the 40\% level, at least for what
concerns the ALPGEN~S$_1$  and ALPGEN~S$_2$  predictions. There are important fractions 
of events containing two and three extra jets, at the 20\% and 10\% level, respectively. From the
lower panel, it can be noticed that the relative differences between the results of 
ALPGEN~S$_1$  and ALPGEN~S$_2$  and those of MC@NLO lie in a $\pm 20\%$ range
for the predictions relative to the fraction of events with at least one extra jet.

\section{Conclusions and perspectives}
\label{concl}

In this paper we presented a detailed phenomenological analysis of the EW and QCD 
contributions, as well as of their combination, to single $W$ boson production in hadronic collisions, considering
the energies of interest for the experiments at the Fermilab Tevatron and the CERN LHC and in the 
presence of realistic selection criteria. 
In our study, we made use of MC tools which include the
recent advances in EW and QCD calculations.

We analyzed all the quantities of interest for the
many facets of $W$ physics programme at hadron colliders and showed that a far from trivial interplay 
between EW and QCD effects is present for most of the observables. We noticed, 
in particular, partial cancellations between EW and QCD corrections  for the experimental distributions
of interest for both precision studies (such as the luminosity monitoring or the measurement of the $W$ mass)
and searches for new physics. Remarkably, in the region of the high transverse mass tail above 1~TeV, important for the search for new gauge bosons, these cancellation occurs almost completely between huge positive QCD corrections and very large negative contributions due to EW Sudakov logarithms. 
This emphasizes the need for a careful combination of strong and
EW contributions in present and future analyses of the CC DY process, as
also pointed out  in~\cite{LH2007} for the NC DY channel in the high invariant mass region 
at the LHC. We also remarked that the convolution of the EW effects with QCD shower evolution 
is definitely relevant for a correct simulation of the distributions, since the relative size and shape of 
EW contributions is considerably modified when compared with the same features in the absence of 
the combination with a QCD PS. This can be understood in terms of the modifications introduced by QCD 
PS on the kinematics of the final-state leptons w.r.t. a pure EW calculation. We also discussed in detail and evaluated the uncertainty inherent the combination of EW and QCD corrections by comparing different theoretical recipes (additive vs factorized) and showing that they can differ at a few per cent level.

In relation to the accuracy of the theoretical tools presently available and used by the experimental collaborations 
in comparison with the precision already reached at the Tevatron and foreseen at the LHC, some final
comments are in order. While the predictions of the EW programs are in very good 
agreement and the situation can be considered well under control  (also when taking into account
multiple photon emission), the same can not be said, strictly speaking, for the QCD generators currently used
in the experimental analyses at the Tevatron or for simulations by ATLAS and CMS collaborations.  Actually, we noticed, for example, differences of 2\% or more in 
predictions for the shape of the distributions
relevant for $W$ mass extraction between ResBos-A   and our combination of EW and QCD tools.  
 Therefore, it is our opinion that a cross-check of the Tevatron $W$ mass results, generally  obtained by means 
of  ResBos in association with  WGRAD or PHOTOS,   with the predictions of our recipe, based on the combination 
of MC@NLO with HORACE, would be highly desirable, in view of a robust, high-precision measurement of 
$M_W$, which is such an important input for indirect constraints on the Higgs boson mass. On the other hand, 
we concluded that the results of the different QCD generators are in very good agreement for what concern
the $W$ rapidity and lepton pseudorapidity distributions, and that EW corrections are well under control for
such distributions. This reinforces the need for deeper attempts to monitor the hadron collider luminosity 
with presently available tools in terms of such distributions, and this goal should be attainable at the Tevatron with 
a few per cent precision, a certainly useful ``exercise" to pave the way to LHC collaborations for future 
luminosity measurements along this direction.

Concerning the forthcoming 
data taking at the LHC, we agree with the conclusions of~\cite{Adam:2008pc,Adam:2008ge} about the feasibility of using the process-independent
module PHOTOS to simulate photonic corrections around the $W$ peak for early stages of data analysis, but we also provided clear evidences  that such an approach to the treatment of EW corrections will 
not be sufficient in the later stages of analysis at high integrated luminosity. This caveat applies to high-precision 
extraction of the $W$ boson properties  at the LHC, as well as to predict correctly the SM 
background to new-physics searches in the high tail, where pure EW Sudakov logarithms and not photonic
effects dominate within the full set of one-loop corrections. In particular, for such a region, two-loop 
Sudakov contributions and NNLO QCD corrections would be needed, together with the contribution of real
$W,Z$ emission. 

In general, for the LHC we remarked that available calculations and tools do not currently allow to reach a 
theoretical accuracy better than some per cent level, when excluding PDF uncertainties. If this could be acceptable
for earlier stages of analysis, future measurements at the LHC would probably require the calculation of 
complete ${\cal O} (\alpha\alpha_s)$ corrections. All the necessary theoretical ingredients are not for the time
being implemented into a single generator, which would represent the optimal solution for simulation and analysis of
DY process in hadronic collisions, in its many and quite different aspects. 

For readers' convenience, we summarize in a final Table  the relative effects of the different sources of corrections to the integrated cross section. We also include a further Table reporting an {\it estimate} of the theoretical accuracy 
of the best predictions that combine QCD and EW corrections according to the different recipes proposed.
\begin{table}[h]
\begin{center}
\begin{tabular}{|c|c|c|c|c|c|}
\hline
$\delta(\%)$ & NLO QCD & NLL QCD & NLO EW & Shower QCD & $O(\alpha \alpha_s)$  \\
\hline
Tevatron & 8 &16.8 & -2.6 & -1.3 & $\sim 0.5$  \\
\hline
LHC a & -2 & 12.4 & -2.6 & 1.4 &  $\sim 0.5$  \\
\hline
LHC b & 21.8 & 20.9 & -21.9 & -0.6 & $\sim 5$  \\
\hline
\end{tabular}
\caption{Relative effect of the main sources of QCD, EW and mixed radiative corrections to the integrated cross sections 
for the Tevatron, LHC a and LHC b.}
\label{tab:deltafinal}
\end{center}
\end{table}
In tab.~\ref{tab:deltafinal} NLO QCD is the complete $O(\alpha_s)$ correction, NLL QCD is the matrix element contribution of the NLO QCD correction, NLO EW is the full  $O(\alpha)$ correction, Shower QCD stands for the $O(\alpha_s^n), n \geq 2$ correction and $O(\alpha \alpha_s)$ represents the mixed EW-QCD corrections estimated by properly combining the additive and factorized cross sections. It is worth noticing in particular that the latter corrections remain below the 1\% level for typical event selections at the Tevatron and the LHC, while they can amount to some per cent in the region important for new physics searches at the LHC. 

\begin{table}[h]
\begin{center}
\begin{tabular}{|c|c|c|c|}
\hline
$\delta(\%)$ & $\delta \sigma / \sigma $~(scale) &  $\delta \sigma / \sigma$~(FA) & $\delta \sigma / \sigma$ \\
\hline
Tevatron & $\sim 1$ & $\sim 2$ & 2  \\
\hline
LHC a & $\sim 2.5$ & $\sim 2$ & 2.5  \\
\hline
LHC b & $\sim 1.5 $ & $\sim 5$ & 5   \\
\hline
\end{tabular}
\caption{Estimate of the present theoretical accuracy for the calculation of the integrated cross section 
at the Tevatron, LHC a and LHC b.}
\label{tab:deltath}
\end{center}
\end{table}
Concerning the theoretical accuracy of the integrated cross sections calculation shown in Tab.~\ref{tab:deltath}, it is a measure of the missing/incomplete higher order $\alpha_s^2$ and $\alpha \alpha_s$ contributions. It has been assessed by neglecting the effect of the PDF uncertainties and according to the following procedure: the relative scale variation of the additive cross section $\delta \sigma / \sigma $~(scale)  and the relative difference between the additive and factorized cross sections $\delta \sigma / \sigma$~(FA) have been computed and compared; as error estimate, the largest of the two entries has been taken. 

Some comments are in order here. The above error estimate are in agreement with phenomenological results available in the literature about the size of NNLO QCD corrections, that are known to contribute at the $\sim 2\%$ level for standard cuts at the Tevatron and the LHC. The errors quoted in Tab.~\ref{tab:deltath} do not include the contributions of the two-loop EW Sudakov logarithms, that amount to a few per cent limited to the very high $W$ transverse mass/lepton  transverse momentum tails. Last, the error estimate reported can be considered as rather conservative. Actually, the factorized cross sections used in the present analysis contain the bulk of NNLO QCD contributions, as can be inferred by comparing the relative difference between the additive and factorized predictions with the exact results for NNLO corrections given in~\cite{ADMP, ADMP1, mp, mp1} and finding them in pretty fine agreement. Hence, a more aggressive estimate of the theoretical error of the factorized formulae could be derived by taking the relative difference the factorized formulae, obtaining $O(1\%)$ at the Tevatron and the LHC a and $O(3\%)$ at LHC b. This estimate could be put on firmer grounds through detailed comparisons with the NNLO exact QCD calculations available in the literature. Furthermore, in the very hard tails of the distributions, the calculation of still unavailable $\alpha \alpha_s$ corrections should be performed. 

Possible perspectives of the present work would be a careful 
analysis of how the precision measurement of the $W$ boson mass and width is affected by the combination of EW and QCD corrections here proposed and 
an application of the same approach to the study of the NC DY channel.

\acknowledgments
We are indebted to all the colleagues who allowed us, with their invitations, to present preliminary results of
our work at various conferences and workshops: M. Czakon, H. Czyz, G. Degrassi, S. Ferrag, 
S. Forte, J. Gluza, M. Grazzini, S. Jadach, W.~Placzek, M. Srzypeck, D. Wackeroth, Z.~Was and G. Zanderighi. 
We thank K. Ellis  and J.M. Campbell for help with the use of the code MCFM and valuable correspondence, 
S. Frixione for helpful comments on the use of the code MC@NLO and 
M.L. Mangano for useful discussions and interest in our work. 
We acknowledge M.R. Whalley for help on LHAPDF. We are also
grateful to the colleagues of the common paper~\cite{LH2007} for fruitful collaboration.  We wish to thank 
several experimental colleagues, in particular 
M. Bellomo, G. Polesello, A. Tricoli and V. Vercesi, for useful discussions and interest in our work.
Useful discussions with  P. Nason  during the INFN Workshop 
on Monte Carlo's, Physics and Simulations at the LHC~\cite{Ambroglini:2009nz} are gratefully acknowledged. 

The work of A. V. was supported by the European Community's
Marie-Curie Research Training Network under contract MRTN-CT-2006-035505 
(HEP-TOOLS).

----------------------
\end{document}